\documentclass[12pt]{article}
\usepackage{amssymb,latexsym}
\usepackage[cmex10]{amsmath}
\usepackage[all]{xy}
\input{xy}
\xyoption{all}
\title{On the Annihilator Ideal  of an Inverse Form.
    \footnote{The published version is AAECC (2017) 28:31-78, DOI 10.1007/s00200-016-0295-6}}
\author{Graham H. Norton
\footnote{School of Mathematics and Physics, University of Queensland, Brisbane, Queensland 4072, Australia. (Email: ghn@maths.uq.edu.au)}
}

\newtheorem{theorem}{\bf Theorem}[section]
\newtheorem{corollary}[theorem]{\bf Corollary}
\newtheorem{proposition}[theorem]{\bf Proposition} 
\newtheorem{notation}[theorem]{\sc Notation} 
\newtheorem{definition}[theorem]{\bf Definition} 
\newtheorem{lemma}[theorem]{\bf Lemma} 
\newtheorem{example}[theorem]{\it Example}  
\newtheorem{examples}[theorem]{\it Examples}  
\newtheorem{algorithm}[theorem]{\bf Algorithm} 
\newtheorem{conjecture}[theorem]{\bf  Conjecture}
\newtheorem{remark}[theorem]{\it Remark}
\newtheorem{remarks}[theorem]{\it  Remarks}

\newenvironment{proof}{{\noindent\it Proof.\ }}{$\square$\par\vspace{4mm}} 
\def \bt{ \begin{theorem} }
\def \et{ \end  {theorem} }
\def \bl{ \begin{lemma} }
\def \el{ \end  {lemma} }
\def \bp{ \begin{proposition} }
\def \ep{ \end  {proposition} }
\def \bn{ \begin{notation} }
\def \en{ \end  {notation} }
\def \bq{ \begin {question} }
\def \eq{ \end {question} }
\def \bc{ \begin{corollary} }
\def \ec{ \end  {corollary} }
\def \bcj{ \begin{conjecture} }
\def \ecj{ \end  {conjecture} }
\def \bd{ \begin{definition} }
\def \ed{ \end  {definition} }
\def \bdp{ \begin{definitionprop} }
\def \edp{ \end  {definitionprop} }
\def \bdt{ \begin{definitiontheorem} }
\def \edt{ \end  {definitiontheorem} }
\def \bpr{ \begin  {proof} }
\def \epr{ \end  {proof} }
\def \ba{ \begin{algorithm} }
\def \ea{ \end{algorithm} }
\def \be{ \begin{example} }
\def \eex{ \end{example} }
\def \bes{ \begin{examples} }
\def \eexs{ \end{examples} }
\def \br{ \begin{remark} }
\def \er{ \end{remark} }
\def \brs{ \begin{remarks} }
\def \ers{ \end{remarks} }
\def \bpb{ \begin{problem} }
\def \epb{ \end{problem} }

\newcommand{\Ann} {\mathrm{Ann}}

\newcommand{\E}{\mathrm{E}}
\newcommand{ \GF} {\mathrm{GF}}

\newcommand{\Spol} {\mathrm{S}}
\newcommand{\Syz} {\mathrm{Syz}}

\newcommand{\lcm} {\mathrm{lcm}}

\newcommand{\la } {\leftarrow}

\newcommand{\ol} {\overline}

\newcommand{\ul} {\underline}
\newcommand{\ee} {\mathrm{E}}

\newcommand{\m} {\mathfrak{m}}

\newcommand{\sm} {\,\sharp\,}

\newcommand{\ra} {\rightarrow}

\newcommand{\rem} {\mathrm{rem\,}}

\newcommand{\vv} {\mathrm{v}}

\newcommand{\lc} {\mathrm{LC}}
\newcommand{\LC} {\ell}

\newcommand{\LM} {\mathrm{LM}}
\newcommand{\LT} {\mathrm{LT}}
\newcommand{\N} {\mathbb{N}}

\newcommand{\R} {\mathrm{R}}
\newcommand{\T} {\mathrm{T}}

\newcommand{\calE}{\mathcal{E}}
\newcommand{\F}{\mathcal{F}}
\newcommand{\sbbk}{\Bbbk}
\newcommand{\LL}{\mathrm{L}}

\newcommand{\Q}{\mathbb{Q}}
\newcommand{\Z}{\mathbb{Z}}

\newcommand{ \B} {\mathcal{B}}

\newcommand{\D} {\mathcal{D}}
\newcommand{ \G} {\mathcal{G}}
\newcommand{\I} {\mathcal{I}}
\newcommand{\J} {\mathcal{J}}

\newcommand{\M} {\mathrm{M}}

\begin{document}
\maketitle
\begin{abstract}
Let $\sbbk$ be a field. We simplify and extend work of Althaler \& D\"ur on finite sequences over  $\sbbk$ by regarding $\sbbk[x^{-1},z^{-1}]$ as a  $\sbbk[x,z]$ module, and studying forms in  $\sbbk[x^{-1},z^{-1}]$ from first principles. Then we apply our results to finite sequences. 

First we define the annihilator ideal $\I_F$ of a  form  $F\in\sbbk[x^{-1},z^{-1}]^\times$ of   finite total degree $m\leq0$. This is a homogeneous ideal.
We inductively construct an  ordered pair ($f_1$\,,\,$f_2$)  of forms in $\sbbk[x,z]$ which generate $\I_F$\,; our generators are special in that $z$ does not divide the leading grlex monomial of $f_1$ but  $z$  divides $f_2$\,, and the sum of their total degrees is always $2-m$.  The corresponding algorithm  is $\sim m^2/2$.   
 
We prove  that the row vector   obtained by accumulating  intermediate forms of the construction gives a minimal grlex Gr\"obner basis for $\I_F$ for no extra computational cost other than storage. (This is based on a closed-form description of a 'form vector' for $F$, an auxiliary  vector and   a syzygy triple derived from the construction. These imply that the  remainder of the S polynomial of $f_1,f_2$ is zero. Then we  inductively apply Buchberger's Criterion  to show that the form vector yields a minimal Gb for $\I_F$\,.) 
 We apply this to determining $\dim_\sbbk (\sbbk[x,z] /\I_F)$\,. 
 
 We show that  either the form vector  is  reduced or a monomial  of $f_1$ can be reduced by  $f_2$\,.  This enables us to efficiently construct the unique  reduced Gr\"obner basis for $\I_F$  from the vector extension of our algorithm. 

Then we specialise to the inverse form of a finite sequence, obtaining generator forms for its annihilator ideal   and a corresponding algorithm which does not use the last 'length change' of Massey.  We compute the intersection of two annihilator ideals using  syzygies in $\sbbk[x,z]^5$. This improves  a result of  Althaler \& D\"ur.

   Finally we show that dehomogenisation induces a one-to-one correspondence ($f_1$\,,$f_2$) $\mapsto$ (minimal polynomial,  auxiliary polynomial), the output of the author's variant of the Berlekamp-Massey algorithm. So we can also solve the LFSR synthesis problem via the corresponding algorithm for sequences. 
\end{abstract}


{\small \tableofcontents}
\section{Introduction}
\subsection{Background}
Let $\sbbk$ be a field. In  \cite{AD},  the authors addressed the following  problem: 'given several sequences in $\sbbk$ of finite but possibly different length, find the linear recurrence of least order that is satisfied by all the sequences'. For some history of this problem, see {\em loc. cit.} p. 378 and the references cited there. 

In \cite[Section 2]{AD} they defined a 'generating form' for a finite sequence $s$ using  negative Laurent  series in two variables, considered as a $\sbbk[x,z]$ module. It is  based on  Macaulay's inverse system.  They showed that their annihilator ideal is homogeneous and hence is generated by finitely many forms. In   \cite[Section 4]{AD} the authors gave an extension of  Berlekamp-Massey  (BM) algorithm and showed that it yields  
 a  minimal homogeneous graded-lexicographic (grlex) Gr\"obner basis  for the annihilator ideal. 
Their solution to the original problem was to then use Gr\"obner basis methods to compute the intersection of finitely many annihilator ideals.  The authors also showed that if $s$ has $n\geq 1$ terms, linear complexity $\LC_s$ and $2\LC_s\leq n$ then their  Gr\"obner basis for the annihilator ideal is its unique  reduced grlex Gr\"obner basis, \cite[Lemma 7]{AD}. 

The authors  related their ideal to the usual one for a linear recurring sequence,   \cite[Proposition 1]{AD}. Let $\m$ be  the maximal ideal of $\sbbk[x,z]$ generated by $x,z$. They also showed that  their ideal is $\m$ primary, \cite[Theorem 1]{AD}. In \cite[Section 3]{AD}, $\sbbk[[x^{-1},z^{-1}]]$ is regarded as a topological $\sbbk$ vector space and ideals of $\sbbk[x,z]$ correspond to closed $\sbbk[x,z]$ submodules of   $\sbbk[[x^{-1},z^{-1}]]$.  This is used to show that  any $\m$ primary homogeneous ideal $I$ of $\sbbk[x,z]$ is the annihilator ideal of at least $\dim_\sbbk(I:\m)/I$ sequences, \cite[Theorem 2]{AD}.

\subsection{Overview}
We simplify and extend \cite[Sections 2,4]{AD}.  Let $\R=\sbbk[x,z]$. First we recall the graded $\R$ module $\sbbk[x^{-1},z^{-1}]$ of inverse polynomials from \cite{Macaulay}, \cite{Northcott}.  Then we  define the  annihilator ideal $\I_F$ of a non-zero inverse form $F$. This ideal is homogeneous. We inductively work with a special 'viable ordered pair' $f$ of  forms which generate $\I_F$\,; our  basis is $F=x^m$ where $m\leq0$, giving $f=(x^{1-m},z)$ as viable ordered pair for $\I_F$\,.  Inductively, given  a viable ordered pair for $\I_F$\,, $a\in\sbbk$ and $G=ax^{m-1}+Fz^{-1}$, we  construct one  for $\I_G$\,.  This gives an effective Hilbert Basis theorem for $\I_F$\,.

The corresponding algorithm  requires $\sim m^2/2$ multiplications in $\sbbk$. We define an analogue $\lambda_F$ of the linear complexity of a finite sequence. Part I  concludes with 'essential' inverse forms, a prerequisite for Part II, and we apply essential inverse forms to characterise the  uniqueness/non-uniqueness of $f_1,f_2$\,.

Part II discusses Gr\"obner bases. Accumulating  intermediate forms (if any) gives a row vector headed by our viable ordered pair and length at most $\lambda_F+1$. We derive a syzygy for our pair and inductively apply Bucherger's Criterion to show that the form vector yields a minimal grlex Gr\"obner basis of $\I_F$\,. The $\sbbk$-dimension of the cyclic $\R$ module $\R/\I_F$ is $\lambda_F(2-m-\lambda_F)$.  We show how to modify the vector version of our algorithm to compute a reduced Gr\"obner basis for $\I_F$.

In the final Part III we apply our results to finite sequences. For example, the  coefficients of $x^m$  (in reverse order) constitute the {\em impulse response sequence  $0\ldots,0,1$} with $1-m$ terms.
The corresponding algorithm  has been implemented for sequences in COCOA. For completeness, we  recompute several ideal intersections from \cite{AD} using our algorithm and syzygies obtained via COCOA. We correct an error in \cite[Example 3]{AD}. 

Our approach is  simpler than \cite{N15} where we used Laurent polynomials  to derive a variant of the BM algorithm. (In fact \cite{N15} is closely related to $\sbbk[x^{-1},x]$ as standard $\sbbk[x]$ module; we were unaware of \cite{Macaulay}.) We obtained a 'minimal polynomial' $\mu$  and  an 'auxiliary polynomial' $\mu'$ of $s$.  Finally, dehomogenisation induces a one-to-one correspondence of the viable ordered pair for our annihilator ideal of  $s$ and $(\mu,\mu')$.

In the "Appendix", we  use evaluation homomorphisms to show that   if $F\in\sbbk[x^{-1},z^{-1}]$ is the generator form of a finite sequence then our ideal coincides with the ideal of \cite{AD}. The latter was defined using the $\R$ module of inverse power series $\sbbk[[x^{-1},z^{-1}]]$. 

\subsection{In More Detail}
This paper has three parts. First we give the necessary algebraic preliminaries and  recall the $\R$ module $\M=\sbbk[x^{-1},z^{-1}]$ of inverse polynomials from \cite{Northcott}. Part I  discusses the annihilator ideal of an inverse polynomial. Throughout the paper, $F\in\M^\times$ is an 'inverse  form' and $m=|F|\leq0$ unless stated otherwise.  We give a key characterisation of forms annihilating $F$. We have included a number of worked examples of determining $\I_F$\,, the annihilator ideal of $F$. We  give some properties of $\I_F$ e.g. $x^{1-m},z^{1-m}\in\I_F$\,, $\I_F$ is homogeneous  and  $\I_{x^m}=\langle x^{1-m},z\rangle$.  We say that an ordered pair of forms $f=(f_1,f_2)$ is 'viable for $\I_F$' if $f_1\not\in\langle z\rangle$, $f_2\in\langle z\rangle$, $|f_1|+|f_2|=2-m$ (where $|\ |$ denotes total degree) and $f_1,f_2$ generate $\I_F$\,. For example, the pair $(x^{1-m},z)$ is viable for $\I_{x^m}$. 
 
Next, we give preliminary results needed to obtain $\I_F$ inductively i.e. to obtain $\I_G$ where  $G=a\sm F=ax^{m-1}+Fz^{-1}$. We give our analogue of the discrepancy of \cite{Ma69} and show how it influences elements of  $\I_G$\,. We show that if $f$ is viable for $\I_F$ and the discrepancy is zero then $(f_1,f_2\,z)$ is viable for $\I_G$\,. The case of a non-zero discrepancy is more involved. We first define  parametrised operators $\curlyvee\,, \ominus:\R^2\ra\R$. Then we show that 
$$g=(f_1\ominus f_2, (f_1\,z)\curlyvee (f_2\,z)\,)\in\R^2$$ 
  is viable for $\I_G$ and how to update the parameters of $ \curlyvee$ and $\ominus$. This induction is simpler than \cite{N15} as we have the single inductive basis: $(x^{1-m},z)$ for $x^m$.  In this way, we  construct a viable pair.  The corresponding Algorithm \ref{calPA} requires $\sim m^2/2$ multiplications in $\sbbk$. 
  
 We conclude Part I with  'essential' inverse forms, an analogue of a notion used in \cite{N15}. It forms an important prerequisite for the construction of a minimal Gr\"obner basis in Part II. Since  $x^{1-m}\in\I_F$\,, $\lambda_F=\min\{|\varphi|\in\I_F^\times: z\nmid \varphi\}$ is a well-defined integer, and $0\leq\lambda_F=|f_1|\leq 1-m$. As an application of essential inverse forms, we show that  $2\lambda_F<2-m \Leftrightarrow f_1{\rm\ is\ unique}\Leftrightarrow  f_2 {\rm\ is\ not \ unique}$.

Part II concerns Gr\"obner bases and is more technical. Accumulating  intermediate forms (if any) for $F$ gives a row vector $\F$ of forms headed by our ordered pair --- at no extra computational cost other than storage --- and $\F$ has at most $\lambda_F+1$ components. We show how successive ideal generators are related and explicitly determine the leading terms of $\F$  in terms of $F$ and a syzygy of $(\F_1,\F_2,\F_3)$. The syzygy  enables us to show that the  $\F$ remainder of the  S polynomial of $f$ is  zero.   An inductive application of  Buchberger's Criterion  then shows that   $\F$ is a minimal Gr\"obner basis for $\I_F$. So if $\mathsf{G}$ is any minimal grlex Gb of $\I_F$ then $\mathsf{G}$ and $\F$ have the same  cardinality and the same set of leading terms. Also, $\dim_\sbbk(\sbbk[x,z]/\I_F)=\lambda_F(2-m-\lambda_F)$.
We adapt an inequality from the proof of \cite[Lemma 7]{AD} to show that  either $\F$ is already reduced or a monomial of $f_1$ reduces by  $f_2$\,. This yields an efficient reduced Gr\"obner basis (RGb) version; see Algorithm \ref{RGbA}. (A version of \cite[Algorithm 1]{AD} to compute the RGb of their ideal does not appear in {\em loc. cit.})

In Part III we apply our results to finite sequences. For the remainder of the introduction, let  $n\geq 1$, $s=s_0,\ldots,s_{n-1}$ and $F=F^{(s)}$   be the 'inverse form' of $s$. Two modifications of  Algorithm \ref{calPA} give Algorithm \ref{calPAs} which finds a viable ordered pair for $\I_F$. This algorithm does not use the last 'length change' or the variable $x=n-m$ of \cite{Ma69}.

 We relate $\I_F$ and the {\em set} of annihilating polynomials of $s$ from \cite{N15}.  The integer  $\lambda_F$ equals the linear complexity $\ell_s$ of $s$.   For completeness we  recompute the ideal intersections of two examples from \cite{AD} using syzygy modules in $\R^5$. 
(In \cite{AD},  syzygy modules are computed in $\R^p$ where $p-1$ is the sum of the cardinalities of the individual Gb's.)
Dehomogenisation induces a  one-to-one correspondence   $$f\mapsto (f_1^\vee,f_2^\vee)\mbox{ and }(\mu,\mu')\mapsto (\mu^\wedge, (\mu')^\wedge\,z^{n+1-|\mu|-|\mu'|}).$$ (Here $^\vee$ denotes dehomogenisation, $^\wedge$ is homogenisation, $\mu$ is a monic minimal polynomial  and $\mu'$ is a monic auxiliary polynomial  of $s$, as defined in \cite{N15}.) 

We conclude this outline with some remarks related to the Berlekamp-Massey  (BM) algorithm. Firstly, 'Thus the question arises whether the BM algorithm can be modified to always yield the reduced Gb', \cite[p. 387]{AD}. Whilst we do not  know the answer to this question, we have shown that Algorithm \ref{calPAs}  can be modified to  give  the  RGb of $\I_F$\,. 
 
 Secondly, the one-to-one correspondence extends to outputs of Algorithm \ref{calPAs}  and the first components of the outputs of \cite[Algorithm 4.12, normalised]{N15}. The latter yields a variant of the original BM algorithm (with different initialisation);
see \cite{N10a}. So if $f$ is viable for $\I_F$\,, $ |f_1|$ and the reciprocal $(f_1^\vee)^\ast$ are a solution of  the 'LFSR synthesis problem' of \cite{Ma69}.
Thus for sequences, with  Algorithm \ref{calPAs}$_{\,\mathrm{(R)Gb}}$ denoting the (R)Gb-vector extension of Algorithm \ref{calPAs}, we have the summary
$$
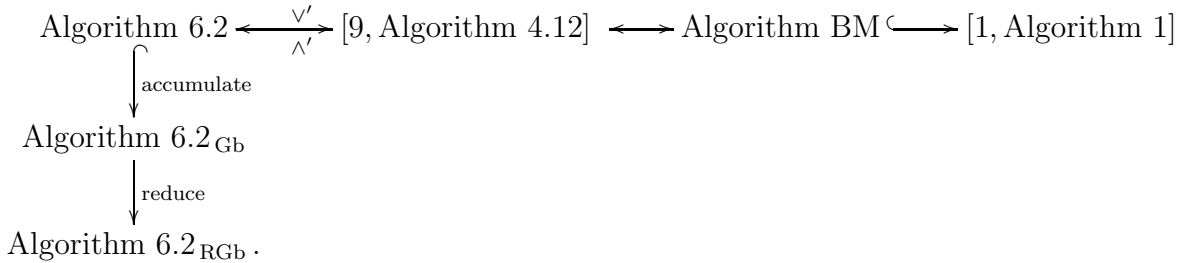
\xymatrix{
{\rm Algorithm}\ \ref{calPAs}\ar@{^{(}->}[d]^{\rm accumulate}     \ar@{<->}[r]^{\vee'}_{{\wedge '}}&{\rm \cite[Algorithm\ 4.12]{N15}}\ \ar@{<->}[r]&{\rm Algorithm\ BM\,}\ar@{^{(}->}[r]&{\rm \cite[Algorithm\ 1]{AD}}\\
{\rm Algorithm}\ \ref{calPAs}_{\,\mathrm{Gb}}\ar[d]^{\rm reduce}\\
{\rm Algorithm}\ \ref{calPAs}_{\,\mathrm{RGb}}\,.}
$$

It would be interesting to extend this work to inverse forms over more general rings such as finite chain rings and Ore rings.  

 {\em Acknowledgements.} We thank Anna Bigatti for useful conversations on {COCOA} \cite{COCOA} and the two reviewers for  helpful comments and suggestions.

\section{Algebraic Preliminaries}
\subsection{Notation}
 For any set $S$ containing 0, $S^\times=S\setminus\{0\}$ so that $\N^\times=\{1,2,\ldots\}$.  Throughout the paper, $\sbbk$ is an arbitrary field and $\R=\sbbk[x,z]$.  Multiplication in $\R$ is written as juxtaposition. For $\varphi,\varphi'\in\R$ and $k\in\N^\times$,  $x^k\,\varphi+\varphi'$ means $x^k\varphi(x,z)+\varphi'(x,z)$ and similarly for  $\varphi+\varphi'\,z^k$\,; for $p\in\N$, $z^p||\varphi$ if $z^p|\varphi$ but $z^{p+1}\nmid \varphi$.  The total degree of $\varphi\in\R^\times$ is $|\varphi|$,  with $|x|=|z|=1$. We put $|0|=-\infty$ and the usual rules for arithmetic involving $-\infty$ apply. 
We also use $|\cdot |$ for  degree  on $\sbbk[x]$.  The ideal  generated by $\varphi_1,\ldots ,\varphi_m\in\R$ is written $\langle \varphi_1,\ldots ,\varphi_m\rangle$ and $\m$ is the maximal ideal $\langle x,z\rangle$ of $\R$.

We will use the ring of Laurent polynomials in $x,z$ namely, $\LL=\sbbk[x^{-1},z^{-1},x,z]$ and $\cdot$ will denote multiplication;  $[\ \,]_i$ is the $i^{\rm{th}}$ coefficient of an element of $\LL^\times$.

We also include reference tables of commonly-used symbols for the aid of the reader.

\begin{center}
\begin{tabular}{|c|l|}\hline
Symbol & Meaning \\\hline\hline
$a\sm F$& the inverse form $a\,x^{m-1}+Fz^{-1}$ \\\hline
$d=d_f$ &$|f_2|-|f_1|$\\\hline
$f\in\R^2$ & the constructed viable ordered pair for $\I_F$\\\hline
$F$ & a non-zero form in $\M$\\\hline
$ F^{(i)}$& the 'subform' $F_i\,x^i+\cdots+x^\vv\,z^{i-\vv}$ of $F$, $m\leq i\leq\vv$\\\hline
$ F^{(s)}$& the inverse form of $s$\\\hline
$\F$ &  the constructed form vector  for $\I_F$\\\hline
$|\F|$ &  the vectors $\F$ and $\D$ have length $|\F|$  \\\hline
$\I_F$ & the annihilator ideal of $F$\\\hline
$\ell_s$& the linear complexity of $s$\\\hline
$m$& $m=|F|\leq0$, $F\neq0$\\\hline
$\LL$& the ring of Laurent polynomials $\sbbk[x^{-1},z^{-1},x,z]$\\\hline
$\M$& the $\R$ module $\sbbk[x^{-1},z^{-1}]$ of inverse polynomials\\\hline
$\D$ & a vector   with $z^{\D_i-m}||\F_i$\\\hline
$q=q_f$& $\Delta(f_1;a\sm F)/\Delta(f_2;a\sm F)$\\\hline
$r$& the common ratio $F_{-1}$ of a geometric $F$,   $|F|\leq-1$\\\hline
$s$ & the sequence $s_0,\ldots,s_{n-1}$, $n\geq 1$\\\hline
$\Spol(\F_i,\F_j)$& the syzygy polynomial of  $\F_i$ and $\F_j$\\\hline
$\vv=\vv(F)$ & the order of $F$.\\\hline
\end{tabular}
\end{center}
\begin{center}
\begin{tabular}{|c|l|}\hline
Greek Symbol & Meaning \\\hline\hline
$\Delta(f_i;a\sm F)$ &  the discrepancy of  $f_i$ and $a\sm F$, $i=1,2$\\\hline
$\lambda_F$ & a non-negative integer derived from $\I_F\cap \Phi$\\\hline
$\mu$& a monic minimal polynomial (MP) of $s$\\\hline
$\mu'$& a monic auxiliary polynomial of $s$\\\hline
$\varphi$ & a form in $\R^\times$\\\hline
$\varphi ^\vee$ & the dehomogenisation of $\varphi$\\\hline
$\varphi\circ F$ & $\varphi$ acting on $F$\\\hline
$\Phi$ & the set of non-zero, monic forms $\varphi$ with $z\nmid\LM(\varphi)$\\\hline
$\psi$ & an element of $\sbbk[x]$\\\hline
$\psi^\wedge$ & the homogenisation of $\psi\in\sbbk[x]$.\\\hline
\end{tabular}
\end{center}

\subsection{Grlex}
We adopt \cite{IVA} as a general reference. We write $\succ$ for graded-lexicographic order ({\em grlex}) on monomials of $\R^\times$, with $|x|=|z|=1$ and $x\succ z\succ1$. Recall that  $\succ$ is the linear  order defined on monomials of $\R^\times$ as follows: $M\succ M'$ if $|M|>|M'|$ or 
($|M|=|M'|$ and $M>_{\mathrm{lex}}M'$). 
 We write $\ee(\varphi)$ for the  grlex {\em exponent} or multidegree of $\varphi\in\R^\times$ :
 $$\ee(\varphi)=\max_\succ\{i\in\N^2:\varphi_i\neq 0\}$$
 and $\LM(\varphi)$ for its   $\succ$ leading  monomial; the leading coefficient of $\varphi$ is $\lc(\varphi)$ and its leading term is $\LT(\varphi)=\lc(\varphi)\LM(\varphi)$.
 If   $\varphi$ is also a   form and $d=|\varphi|\in\N$, it will be convenient to write  $\varphi=\sum_{j=0}^d\varphi_{d-j}x^{d-j}z^j$, on the understanding that $\varphi_{d-j}=0$ for $(d-j,j)\succ \ee(\varphi)$.  In this case  $z|\varphi$ if and only if $z|\LM(\varphi)$ and if $z\nmid \LM(\varphi)$ then $\ee(\varphi)=(|\varphi|,0)$.
  \subsection{Homogeneity}\label{homogsubsection}
The {\em homogenisation} of  $\psi\in\sbbk[x]^\times$ is $\psi^\wedge(x,z)=\psi(x/z)\,z^{|\psi|}$. Thus  if $d=|\psi|$ and $\psi=\sum_{i=0}^d \psi_{d-i}x^{d-i}$ then
$$\psi^\wedge(x,z)=\left(\sum_{i=0}^d \psi_{d-i}(x/z)^{d-i}\right)z^d=\sum_{i=0}^d \psi_{d-i}x^{d-i}z^i$$
 a non-zero form of total degree $d=|\psi|$ and $\LM(\psi^\wedge)=x^d$ since $\psi_d\neq0$. The {\em dehomogenisation} of $\varphi\in\R^\times$ is $\varphi^\vee(x)=\varphi(x,1)$.
If $\varphi$ is non-zero and $d=|\varphi|$, so that $\varphi=\sum_{i=0}^d \varphi_{d-i}\,x^{d-i}\,z^i$ then 
$$\varphi^\vee(x)=\sum_{i=0}^d \varphi_{d-i}\,x^{d-i}$$
$|\varphi^\vee|\leq |\varphi|$ and   if $z\nmid \LM(\varphi)$ then $|\varphi^\vee|=|\varphi|$. 
Thus if we let $$\Phi=\{\varphi\in\R^\times: \varphi\mbox{ is a monic form, } z\nmid\LM(\varphi)\}$$
then we have degree-preserving maps $^\wedge:\sbbk[x]^\times\rightleftarrows \Phi:\ ^\vee$ which are mutual inverses.
We will  need the following extension of division in $\sbbk[x]$,  also used in \cite{AD}.
\bl \label{Fdiv} If $g\in\Phi$  then for any form $\varphi\in\R^\times$ there are  forms $\alpha,\beta\in\R$ such that 

(i) $\varphi=\alpha\, g+\beta$ and if $\alpha\neq0$ then $|\alpha|=|\varphi|-|g|$ 

(ii) either (a) $\beta=0$ or  (b) $|\beta|=|\varphi|$ and $z|\beta$.
\el
\bpr Without loss of generality, we can assume that $z\nmid \varphi$. We can write $\varphi^\vee=\alpha' g^\vee +\beta'$ for some $\alpha',\beta'\in\sbbk[x]$ such that if $\alpha'\neq 0$, $d=|\varphi|=|\varphi^\vee|=|\alpha'|+| g^\vee|$ and $\beta'=0$ or $|\beta'|<d$. If $\alpha\neq0$, put $\alpha=(\alpha')^\wedge$. If $\beta'=0$ then we easily get 
$\varphi=(\varphi^\vee)^\wedge=\alpha'(x/z)\,  g^\vee(x/z)z^d=\alpha\,  g$\,. If $\beta'\neq0$  then $\varphi=\alpha\,  g+\beta'(x/z)z^d$. Put $\beta= (\beta')^\wedge\,z^{d-|\beta'|}\in\R$. Then $\varphi=\alpha g+\beta$ where $\beta$ is a form of total degree  $d$ and $z|\beta$ since  $|\beta'|<d$. 
\epr

\section{Inverse Forms}
 \subsection{Basics}
 We also order the  monomials of $\M^\times=\sbbk[x^{-1},z^{-1}]^\times$ using grlex, now written $\prec$\,, but with $|x^{-1}|=|z^{-1}|=-1$, $x^{-1}\prec z^{-1}\prec1$ and
  $$\ee(F)=\min_\prec\{i\in-\N^2:F_i\neq 0\}$$
  is the $\prec$ exponent of $F\in\M^\times$. If $F$ is also a form i.e. an {\em inverse form} and  $d=|F|\leq0$ is its total degree,    we write $F=\sum_{j=d}^0F_{j,d-j}x^jz^{d-j}$.  {\em When $F$ is understood, we write $F_j$ for $F_{j,d-j}$}  on the understanding that $F_j=0$ for $(j,d-j)\prec \ee(F)$. Note that for $d\leq i\leq 0$, $[x^{-1}F]_0=0$, $[x^{-1}F]_{i-1}=F_i$ and  $[Fz^{-1}]_{i,d-i}=F_{i,d-i+1}$.

 Throughout the paper, $F\in\M^\times$ denotes a  typical non-zero inverse form  and $m=|F|\leq0$.  
 We will use a restriction of the exponential valuation $\vv$ for (non-zero) inverse forms. The {\em order} of $F\neq0$ is $\vv=\vv(F)=\max\{j: |F|\leq j\leq 0, F_j\neq0\}$. 

  Let $F,G\in\M^\times$. We write 
  $F\sim G$ if for some $a\in\sbbk^\times$ we have  $F=a\,G$. In particular $|F|=|G|$ and for example, $F\sim F_\vv^{-1}F$. Then $\sim$ is an equivalence relation; let $\ol{F}$ denote the equivalence class of $F$. We will use $F_\vv^{-1}F\in\ol{F}$  for simplicity as $[F_\vv^{-1}F]_\vv=1$ and our annihilator ideal will only depend on $\ol{F}$ by linearity. 
  
  We will say that $F$ is {\em geometric} if   $F_0=1$, $m\leq-1$ and  $F=r^{-m}x^m+\cdots +z^m=\sum_{i=m}^0 r^{-i}x^iz^{m-i}$ or $F_i=rF_{i+1}$ for $m\leq i\leq -1$ where $r=F_{-1}$\,; e.g. if  $m\leq-1$, $F=z^m$ is geometric with $r=F_{-1}=0$. (We require $m\leq-1$ so that $F_{-1}$ is defined).

 We will often use  $a\in\sbbk$ to {\em augment an inverse form} $F\in\M^\times$ :   $a\sm F=ax^{m-1}+Fz^{-1}$, of total degree $m-1=|F|-1$\,. 
  For example, $a\sm z^m=a\,x^{m-1}+z^{m-1}$.  A form $F\in\M^\times$ defines  {\em inverse subforms} $\{F^{(i)}: m\leq i\leq \vv\}$ by $F^{(\vv)}=x^\vv$ and
  $$F^{(i)}=F_i\sm F^{(i+1)}=F_ix^i+F^{(i+1)}z^{-1}=F_ix^i+\cdots +F_{\vv-1}x^{\vv-1} z^{i-\vv+1}+x^\vv z^{i-\vv}\mbox{  for \ }m\leq i\leq \vv-1$$ 
 the last equality being an easy induction. If $F_i\neq0$ then $\E(F^{(i)})=(i,0)$ for $m\leq i\leq \vv$. 
  It is easy to check that $[a\sm F]_i=F_i$ and $(a\sm F)^{(i)}=F^{(i)}$ for $m\leq i\leq \vv$. 
  
{\em The following inductive principle will often be used to prove a result for an arbitrary (non-zero) inverse form}:
  
  (i)  prove the result for $x^\vv$ (the inductive basis)
  
  (ii) assuming the result is true for a subform $F^{(i)}$ of $F$, let $a\in\sbbk$ be arbitrary and   prove the result for $a\sm F^{(i)}$.

N.B. In applying this inductive principle to  $a\in\sbbk$ and a (generic) subform $F\in\M^\times$ with $\E(F)=(m,0)$ we have $\E(a\sm F)=(m-1,0)$ if $a\neq 0$ and $\E(0\sm F)=\E(F\,z^{-1})=(m,-1)$.

\subsection{The Module of Inverse Polynomials}
We recall the $\R$ module $\M=\sbbk[x^{-1},z^{-1}]$ of { inverse polynomials} from \cite[Introduction]{Northcott}. 
For  $x^i\in\R$ and $x^j\in\M$
\begin{eqnarray}\label{module}
x^i\circ x^j=\left\{\begin{array}{ll}
x^{i+j}&\mbox {if }x^{i+j}\in\M\\
0&\mbox{otherwise.}
\end{array}
\right.
\end{eqnarray}
The $\R$ module structure on $\M$ is obtained by linearly extending 
Equation (\ref{module})  to all of $\R$ and $\M$.
The next two basic results will be applied frequently. Recall that the ring of {\em Laurent polynomials} in $x,z$ is $\LL=\sbbk[x^{-1},z^{-1},x,z]$.
\bl \label{prebasic} If $\varphi\in\R^\times$, $F\in\M^\times$ are  forms  and $d=|\varphi|+|F|$ then (i) $$\varphi\circ F= \sum_{i=d}^0[\varphi\cdot F]_i\,x^i\,z^{d-i}$$

(ii) if $d>0$ then $\varphi\circ F=0$

(iii) if $\varphi\circ F\neq0$ then $\varphi\circ F$  is a form of total degree $d\leq0$. 
\el
\bpr (i) Since $\varphi,F\neq 0$, the integer $d\in\Z$ is well defined. Consider the  form $\varphi\cdot F\in\LL^\times$. It has total degree $d$ and so we can write $\varphi\cdot F$ as the finite sum
$$\sum_{i\in\Z} [\varphi\cdot F]_i\, x^iz^{d-i}.$$
By definition, the monomials of $\varphi\circ F$  are a subset of those of $\varphi\cdot F$\,: 
$$\varphi\circ F=\sum_{i\in\Z}\sum_{i\leq0\,,\,d-i\leq 0} [\varphi\cdot F]_i\, x^iz^{d-i}= \sum_{d\leq i\leq 0} [\varphi\cdot F]_i\, x^iz^{d-i}.$$ 
(ii) If $d>0$ then $\varphi\cdot F\not\in\M$ and  the sum of (i) is empty. (iii) From (i) and (ii),  any non-zero term of $\varphi\circ F$ has total degree $d\leq0$. 
\epr
\bc\label{basic} Let $F\in\M^\times$ be an inverse form.  Then for any form $\varphi\in\R^\times$
$$\varphi\circ F=0\mbox{ if and only if }[\varphi\cdot F]_i=0\mbox{ for }|\varphi|+|F|\leq i\leq 0.$$
\ec
\bpr Let $\varphi\circ F=0$. If $d=|\varphi|+|F|>0$ there is nothing to prove; otherwise the result follows from Lemma \ref{prebasic}. Conversely let $[\varphi\cdot F]_i=0\mbox{ for }d\leq i\leq 0.$ From Lemma \ref{prebasic}, if $d\leq0$ then $\varphi\circ F=0$ and if $d>0$ then $\varphi\circ F=0$.
\epr
\be {\rm Let $F=\sum_{j=m}^0r^{-j}x^jz^{m-j}$ be geometric. Direct evaluation gives $(x-rz)\circ F=0$. Alternatively, 
for $1+m\leq i\leq 0$, $zF\in\M$, $[rz\cdot F]_i=r[z\cdot F]_i=r\, F_i=r^{1-i}$ and
$$[(x-rz)\cdot F]_i=[x\cdot F]_i-[rz\cdot F]_i=F_{i-1}-r^{1-i}=0.$$ Note also that $z^{1-m}\circ F=0$.}
\eex
\be \label{firstIRS} {\rm Let  $m\leq0$ and $F=x^m$, so that $1-m>0$ and  $x^{1-m}\circ F=0$.  Now let  $a\neq0$ and $G=a\sm F=a\,x^{m-1}+Fz^{-1}=ax^{m-1}+x^mz^{-1}$. Then $x^{1-m}\circ G=a$ and $x^{-m}z\circ G=1$; in fact  $[x^{1-m}\circ G]_0=a$ and $[x^{-m}z\circ G]_0=1$. Let $\varphi=x^{1-m}-a\,x^{-m}z$. Then $\varphi\circ G=0$  by direct calculation; or    $1-m+|G|=0$ and $[\varphi\cdot G]_i=0$ for  $1-m+|G|\leq i\leq0$  so $\varphi\circ G=0$ by the previous corollary. It is trivial that $z^2\circ G=0$.} 
\eex
\be \label{preDelta}{\rm Let $\sbbk=\GF(2)$, $\varphi=x^4+x^3z+x^2z^2$ and $F=x^{-6}+x^{-4}z^{-2}+x^{-3}z^{-3}+z^{-6}$. One checks that $\varphi\circ F=0$ and trivially $z^7\circ F=0$. Now let $a\in\sbbk$ and $G=a \sm F$. The reader may verify that $\varphi\circ G=(1+a)x^{-3}$. As $|\varphi|+|G|=-3$, we have $\varphi\circ G=0$ if and only if $a=1$, so we may think of $1+a=[\varphi\cdot G]_{|\varphi|+{|G|}}$ as  the obstruction to $\varphi\circ G=0$. 
This more complicated example is derived from  Example \ref{ADex4} below.
}
\eex

   \subsection{The Ideal $\I_F$}
 Let $F$ be an inverse form. The {\em annihilator ideal} of $F$ is $$\I_F=(0:_{\,\circ}F)=\{\varphi\in\R: \varphi\circ F=0\}\subseteq\R.$$
The previous examples give (i) $x-r\,z, \ z^{1-m}\in\I_F$ (ii) $x^{1-m},z\in\I_F$ and (iii) $x^4+x^3z+x^2z^2,\, z^7\in\I_F$ respectively. We will revisit the above augmented forms $a\sm F$ below.
 
 If $F,G$ are non-zero forms and $F\sim G$ then $\I_F=\I_G$ by linearity and so we can without loss of generality assume that $F_\vv=1$ and $F=F_mx^m+\cdots +x^\vv z^{m-\vv}$.  It is clear that we always have  $x^{1-m} ,z^{1-m}\in \I_F$\,.  
For $F\neq0$ we have $1\not\in\I_F$ i.e. $\I_F$ is a proper ideal of $\R$.
If $m=0$ then  $\m=\langle x,z\rangle\subseteq\I_1$\,. As $\m$ is  maximal and $\I_1$ is proper, $\I_1=\m$. 
\be\label{firsta,b}{\rm Let $F=x^m$, so that $\vv=m$. Then  $\langle x^{1-m},z\rangle\subseteq\I_F$\,. We will shortly see that this is an equality.  For $a\in\sbbk^\times$ we have $ a\sm F=ax^{m-1}+x^mz^{-1}$ and we have seen that $\langle x^{1-m}-ax^{-m}z,z^2\rangle \subseteq\I_{ a\sm F}$\,.  We will see in Proposition \ref{seconda,b} that this inclusion is also an equality.}
\eex

In \cite{AD} the authors showed that their ideal is homogeneous.

\bp \label{homog} The ideal $\I_F$ is homogeneous.
\ep
\bpr  Since $\I_0=\R$ is homogeneous, we can assume that $F\neq0$ and $|F|=m<\infty$. Write an arbitrary $\varphi\in\I_F^\times$ as the sum of its non-zero homogeneous components   $\varphi=\sum_i\varphi^{(i)}$ say, with the $|\varphi^{(i)}|$ distinct. Then $0=\varphi\circ F=\sum_i\varphi^{(i)}\circ F$. If $\varphi^{(i)}\circ F\neq0$ then it has total degree $|\varphi^{(i)}|+m$ from Lemma \ref{prebasic}. In the latter case, $\varphi^{(i)}\circ F$ cannot cancel with any other  non-zero form  $\varphi^{(j)}\circ F$ of distinct total degree $|\varphi^{(j)}|+m$. Hence $\varphi^{(i)}\circ F=0$ for each $i$  and  $\I_F$ is homogeneous.  
\epr

If $I$ is a proper homogeneous ideal of $\R$, $I=\langle\varphi_1,\ldots,\varphi_k\rangle$ for some forms $\varphi_1,\ldots,\varphi_k\in I$ by \cite[Theorem 8.3.2]{IVA}.
Thus if  $F\neq0$ then $\I_F=\langle \varphi_1,\ldots,\varphi_k\rangle$ for some forms $\varphi_1,\ldots,\varphi_k\in \I_F$ by Proposition \ref{homog}.  If now $\J$ is an ideal of $\R$ and we want to show that $\I_F\subseteq \J$, it is enough to show that if $\varphi\in\I_F^\times$ is a form then $\varphi\in\J$. We will use this fact repeatedly without further mention.

\bp \label{001}(Inductive Basis) If $F=x^m$ then  $\I_F=\langle x^{1-m},z\rangle$.
 \ep
 \bpr   Firstly $x^{1-m},z\in\I_F$\,.  If $\varphi=\sum_{i=0}^d \varphi_{d-i}\,x^{d-i}\,z^{i}\in\I_F^\times$ is a form of total degree $d$ then $d\geq 1$. If $z|\varphi$ we are done, so let $z\nmid \LM(\varphi)$.  Then $\varphi_d\neq 0$ and $\varphi-\varphi_d x^d \in z\R$, so it suffices to show that $d\geq 1-m$. We have 
 $$0=\varphi\circ F=\left(\sum_{i=0}^d \varphi_{d-i}\,x^{d-i}\,z^{i}\right)\circ x^m=\varphi_d\,x^d\circ x^m$$
 since for $1\leq i\leq d$, $(x^{d-i}\,z^i)\circ x^m=x^{d-i}\circ (z^i\circ x^m)=0$. As $\varphi_d\neq 0$,  $d+m\geq 1$.  \epr
\be \label{100} {\rm  If $F=z^m$ we claim that $\I_F=\langle x,z^{1-m}\rangle$.  Firstly  $x,z^{1-m}\in\I_F$. Let $\varphi\in\I_F^\times$ be a form of total degree $d\geq 1$, $\varphi=\sum_{i=0}^d \varphi_{d-i}\,x^{d-i}\,z^{i}$ say. If $x|\varphi$ we are done, so suppose that $x\nmid \varphi$. Then $\varphi_0\neq 0$ for otherwise $x|\varphi$ and $\varphi-\varphi_0\,z^d\in \langle x\rangle$. So it suffices to show that $d\geq 1-m$. We have 
 $0=\varphi\circ F=\varphi_0\,z^d\circ z^m$
 since for $0\leq i\leq d-1$, $(x^{d-i}\,z^i)\circ z^m=0$. As $\varphi_0\neq 0$ we conclude that $d+m\geq 1$ as required.}
\eex

The following elementary lemma will be used repeatedly.
\bl \label{f/z} If $G=a\sm F$ then (i) $z\circ  G= F$ (ii) if $\varphi\in\R\,z$  then  $\varphi\in \I_G$ if and only if $\varphi/z\in \I_F$.
\el
\bpr  (i) We have $G=ax^{m-1}+F_m\,x^mz^{-1}+\cdots+\,z^{m-1}$ and so
$$z\circ  G=F_m x^m+F_{m+1} x^{m+1}z^{-1}+\cdots+z^{m}= F.$$  (ii) If $\varphi\in\R\,z$ then $\varphi\circ G=(\varphi/z)\circ (z\circ G)=(\varphi/z)\circ F$ from (i), whence the equivalence.
\epr

\bp \label{incl} For $F\neq0$, we have (i) 
$\I_{a\sm F}\subset \I_F$ and (ii)  there is a  filtration
$$\I_{a\sm F}\subset\I_F\subset\cdots\subset\I_{x^\vv}=\langle x^{1-\vv},z\rangle.$$
\ep
\bpr Let $G=a\sm F$.  
(i) If  $\varphi\in\I_G$ then $\varphi\,z\in\I_G\,$ and  $\varphi\circ F=\varphi\circ(z \circ G)=(\varphi\,z)\circ G=0$  by Lemma \ref{f/z}. Thus $\I_G\subseteq \I_F$\,. 
 To see that the inclusion is strict, write  $ F=F_mx^m+\cdots+x^\vv z^{m-\vv}$ where $m\leq \vv \leq 0$, so that $1-m+\vv\geq 1$ and $z^{1-m+\vv}\circ F=0$. But $ G=ax^{m-1}+\cdots+x^\vv z^{m-1-\vv}$ and $z^{1-m+\vv}\circ G=x^\vv$ i.e. $z^{1-m+\vv}\not\in\I_G$\,. (ii) The equality is Proposition \ref{001}. Now write $F=\lc(F)\sm\cdots \sm F^{(\vv)}$ and apply the first part inductively. 
 \epr

The following lemma will be a useful addition.
\bl \label{ally} Let $a\in\sbbk$, $F$ be an inverse form and $G=a\sm F$. If  $\I_F\,z\subseteq\J\subseteq \I_G$ then $\J=\I_G$.  
\el
\bpr We have to show that $\I_G\subseteq\J=\langle g_1,\ldots,g_l\rangle$ say.  Let $\varphi\in\I_G^\times$. If $z|\varphi$ then $\varphi/z\in\I_F$ by Lemma \ref{f/z}, so $\varphi\in\I_F\,z\subseteq \J$.
If $z\nmid \varphi$, we write $\varphi=\sum_k \alpha_k\, g_k+\beta$ for some forms $\alpha_k,\beta\in\R$ with $\beta=0$ or $z|\beta$ via Lemma \ref{Fdiv}. If $\beta=0$, we are done. Otherwise $\beta=\varphi-\sum_k \alpha_k\, g_k\in\I_G$ since $\J\subseteq \I_G$. As $z|\beta$, $\beta/z\in\I_F$ by Lemma \ref{f/z} and so $\beta\in \I_F\,z\subseteq \J$. Hence  $\varphi=\sum_k \alpha_k\, g_k+\beta\in\J$ and we are done. 
\epr

The next result completes Example \ref{firsta,b}; its proof will be generalised below.
\bp  \label{seconda,b}  If $F=x^m$ and  $a\in\sbbk^\times$ then $\I_{ a\sm F}=\langle x^{1-m}-a\,x^{-m}\,z,\,z^2\rangle$.
\ep
\bpr Let $G= a\sm F$, $g_1=x^{1-m}-a\,x^{-m}\,z$, $g_2=z^2$ and $\J=\langle g_1,g_2\rangle$. From Example \ref{firsta,b}, it is enough to show that $\I_G\subseteq\J$. We  have 
$\I_F\,z=\langle x^{1-m}\,z, z^2\rangle$ and $x^{1-m}z=g_1\,z+ax^{-m}\,z^2\in\J$. Thus Lemma \ref{ally} implies that  $\I_G=\J$.
\epr

\bd  \label{viable}  {\rm Let $F\neq0$. We will call  $f\in\R^2$   {\em a viable ordered pair  for}  $\I_F$  if $f_1,f_2$ are monic non-zero forms, $\I_F=\langle f_1,f_2\rangle$, $f_1\not\in\langle z\rangle$, $f_2\in\langle z\rangle$ and  $|f_1|+|f_2|=2-m$. }
\ed 

From Proposition \ref{001}, if $F=x^m$ then $(x^{1-m},z)$ is viable for  $\I_F$. From Example \ref{100}, if $F=z^m$ then $(x,z^{1-m})$ is viable for $\I_F$. In  Proposition \ref{seconda,b}, $(x^{1-m}-a\,x^{-m}z,z^2)$ is viable for  $\I_{a\sm F}$.
For  $m\leq-1$, $f=(x^{-m},xz)$  satisfies $f_1\not\in\langle z\rangle$, $f_2\in\langle z\rangle$ and $|f_1|+|f_2|=2-m$, but $\langle f_1,f_2\rangle\neq \I_F$ for any $F$ since  $z^{1-m}\in\I_F\setminus\langle f_1,f_2\rangle$.

If $f$ is viable for $\I_F$ then $f_1\neq f_2$ since $f_1\not\in\langle z\rangle$. As $f_1,f_2$ are non-zero, $ |f_1|,|f_2|$ are finite and $ |f_1|,|f_2|\geq 1$ since $F\neq0$. Hence $|f_1|=2-m-|f_2|\leq 1-m$ and similarly $|f_2|\leq 1-m$.

We develop our main result  Theorem \ref{indthm} in several stages. First we need to discuss our version of 'linear complexity' for an inverse form.
 Then we give the two  inductive steps (depending on a zero or non-zero discrepancy) to obtain a viable ordered pair  for  $\I_{ a\sm F}$ from one for $\I_F$\,.
\subsection{The  Integer $\lambda_F$}
The following definition makes sense since  $x^{1-|F|}\in\I_F$\,.
\bd  For an inverse form $F$, we define $\lambda_F\in\N$ by $$\lambda_F=\min\{|\varphi|:\ \varphi\in\I_F\cap\Phi\}$$
and  $\Lambda_F=\sum_{i=m}^\vv\lambda_{F^{(i)}}$. When $F$ is understood, we put  $\lambda_i=\lambda_{F^{(i)}}$ for $m\leq i\leq\vv$. 
\ed
Thus  $\lambda_F\leq 1-|F|$. We have   $\lambda_F\leq\lambda_{a\sm F}$ from Proposition \ref{incl}, which we paraphrase as  '$\lambda$ is non-decreasing'. 
\bp \label{Lambda} If $f$ is viable for $\I_F$ then  $\lambda_F=|f_1|$.
\ep
\bpr We have $\lambda_F\leq |f_1|$ as $f_1\in\I_F\cap\Phi$. If $\varphi\in\I_F\cap\Phi$ then $\varphi=a_1f_1+a_2f_2$ for forms $a_1,a_2\in\R$ and $a_1\neq 0$ since $z\nmid \varphi$. Thus $|\varphi|=|a_1|+|f_1|\geq |f_1|$, so $\lambda_F\geq |f_1|$. \epr

For example, if $F=x^m$ then $\lambda_F=1-m$  since $(x^{1-m},z)$ is viable for $\I_F$ from Proposition \ref{001} and if $F=z^m$  then $\lambda_F=1$ as $(x,z^{1-m})$ is viable for $\I_F$ by Example \ref{100}.

   \bp\label{Lambdageom} The inverse form $F$ is geometric if and only if $m\leq-1$, $F_0=1$ and $\lambda_m=\cdots=\lambda_0=1$. 
\ep
\bpr Suppose that $F$ is geometric. Then $F_0=1$, so $\lambda_0=1$. We have seen that  $x-rz\in\I_F\cap\Phi$ and so $\lambda_m\leq1$ by definition. As $\lambda_i\geq \lambda_{i+1}$, we therefore have $1\geq\lambda_m\geq  \cdots\geq \lambda_0=1$ i.e. $\lambda_m=\cdots=\lambda_0=1$.  Conversely, if $m\leq-1$ then $\lambda_0=1$ implies that $F_0$ is non-zero i.e. $F_0=1$ and $\lambda_m=1$ implies that  $\varphi=x-\rho\,z\in\I_F$ for some $\rho\in\sbbk$. By Corollary \ref{basic}  we have
$$0=[(x-\rho z)\cdot F]_j=F_{j-1}-\rho F_j$$
for $1+m\leq j\leq 0$ i.e. $F_0=1$ and $F_i=\rho F_{i+1}$ for $m\leq i\leq -1$, so $F$ is geometric. 
\epr
\section{The Main Theorem}
We develop our main result Theorem \ref{indthm} in several stages. First we discuss a 'discrepancy' which shows how $a$ and $\I_F$ affect $\I_{a\sm F}$. Then two inductive steps (depending on a zero or a non-zero discrepancy) to obtain a viable ordered pair for $\I_{a\sm F}$ from one for $\I_F$.
\subsection{The Discrepancy}
Here  we discuss  a  'discrepancy' which shows how $a$ and $\I_F$ affect $\I_{a\sm F}$\,.
\bp\label{Delta} Let  $\varphi\in\I_F^\times$ be a form. If $G= a\sm F$ and $d=|\varphi|+|G|$ then we have  $\varphi\circ G=[\varphi\circ G]_d\,x^d$. In particular, $\varphi\in\I_G$ if and only if $[\varphi\circ G]_d=0$. 
\ep
\bpr If $d>0$ there is nothing to prove, so assume that $d\leq 0$. Let $\varphi\in\I_F^\times$ and $e=|\varphi|\geq 0$. Then $d+1=e+m$ and from Lemma \ref{prebasic} 
\begin{eqnarray*}
\varphi\circ G&=&\sum_{i=d}^0 [\varphi\cdot G]_i \,x^i\,z^{d-i}
=[\varphi\cdot G]_d\,x^d+\sum_{i=e+m}^0 [\varphi\cdot G]_i\, x^i\,z^{e+m-1-i}.\\
\end{eqnarray*}
Now $G=ax^{m-1}+F\,z^{-1}$ and $[\varphi\cdot x^{m-1}]_i=\varphi_{i-m+1}=0$ since if $e+m\leq i\leq 0$ then $e+1\leq i-m+1$. Since $\varphi\cdot F$ is a form and $i\leq 0$,  $[\varphi\cdot F\, z^{-1}]_i=[\varphi\cdot F]_i$. By Lemma \ref{prebasic},  $(\sum_{i=e+m}^0 [\varphi\cdot F]_i\, x^i\,z^{e+m-i})\,z^{-1}=(\varphi\circ F)z^{-1}=0$ since $\varphi\in\I_F$.  Thus $\varphi\circ G=[\varphi\cdot G]_d\,x^d=[\varphi\circ G]_d\,x^d$ as claimed.
\epr

The next definition is our analogue of  'discrepancy' as introduced in \cite{Ma69}.
\bd {\rm If     $\varphi\in\I_F^\times$ is a form and $G= a\sm F$ then the {\em discrepancy} of $\varphi$ and $G$ is
 $$\Delta(\varphi;G)=[\varphi\cdot G]_{|\varphi|+|G|}\in\sbbk\mbox{\ \ \ \  if }|\varphi|+|G|\leq0$$ and $\Delta(\varphi;G)=0$ otherwise.}
\ed

For $G=a\sm F$, putting $\Delta(\varphi; G)=0$ if $|\varphi|+|G|>0$ is consistent with Lemma \ref{prebasic}, so that $\varphi\in\I_G$ if and only if $\Delta(\varphi; G)=0$, the first step being to check if $|\varphi|+|G|\leq0$. Also, if $\Delta(\varphi; G)\neq0$ then $|\varphi|+|G|\leq0$.

If $f$ is viable for $\I_F$ and $G=a\sm F$ then  $|f_i|+|G|\leq0$ and so $\Delta(f_i;G)=[f_i\cdot G]_{|f_i|+|G|}$ for $i=1,2$.
 Next some examples of determining discrepancies given $a\in\sbbk$ and $F$. 
\be\label{IRS+1}{\rm  Let  $F=x^m$, so that $(x^{1-m},z)$ is viable for $\I_F$ by Proposition \ref{001}. If $G= a\sm F$ then $ G=ax^{m-1}+x^mz^{-1}$ and $\Delta(x^{1-m};G)=[x^{1-m}\circ G]_{1-m+m-1}=G_{m-1}=a$. So $x^{1-m}\in\I_G$ if and only if $a=0$ by Proposition \ref{Delta}. Let $b\in\sbbk$ and $H=b\sm G$. Then  $H=bx^{m-2}+Gz^{-1}=bx^{m-2}+ax^{m-1}z^{-1}+x^mz^{-2}$. The discrepancy of $x^{2-m}$ and $H$ is $$\Delta(x^{2-m};H)=[x^{2-m}\circ H]_{2-m+m-2}=H_{m-2}=b$$
so that $x^{2-m}\in\I_H$ if and only if $b=0$. The discrepancy of $z$ and $H$ is 
$$\Delta(z;H)=[z\circ H]_{1+m-2}=G_{m-1}=a$$
since $z\circ H=G$. Thus $z\in\I_H$ if and only if $a=0$.}
\eex
\be
{\rm Let $\sbbk=\mathrm{GF}(2)$ and  $ F=x^{-6}+x^{-4}z^{-2}+x^{-3}z^{-3}+z^{-6}$ of Example \ref{preDelta}. We saw that   $\varphi=x^4+x^3z+x^2z^2\in\I_F$\,. Now let $G= a\sm F$ where $a\in\sbbk$. Then $ G=a x^{-7}+F\,z^{-1}$. We saw that $[\varphi\cdot G]_{|\varphi|+{|G|}}=1+a$ was the obstruction to $\varphi\in\I_G$\,. In fact $[\varphi\cdot G]_{|\varphi|+{|G|}}$ is none other than $\Delta(\varphi;G)$.}
\eex
Our final example is not altogether unexpected.
\bl \label{geomdelta} If  $F$ is geometric  and $G= a\sm F$ then $\Delta(x-r\,z;G)=a-r^{1-m}$.
\el
\bpr We have $F_i= r^{-i}$ for $m\leq i\leq 0$, $ F=\sum_{i=m}^0 r^{-i}\,x^iz^{m-i}$ and
\begin{eqnarray*}\Delta(x-r\,z;G)&=&[(x-r\,z)\circ G]_{1+|G|}=\left[(x-r z)\circ\left(ax^{m-1}+\sum_{i=m}^0 r^{-i}\,x^iz^{m-i-1}\right)\right]_m\\
&=& a+\sum_{i=m}^0 r^{-i}\,[x\circ x^iz^{m-1-i}]_m-\sum_{i=m}^0r^{1-i}\,[z\circ x^iz^{m-i-1}]_m\\
&=&a+B-C
\end{eqnarray*} 
say, for some $B,C\in\sbbk$. To obtain $B$ we require $1+i=m$, but $m\leq i\leq0$ so $B=0$, and to obtain $C$ we require $i=m$ so $C= r^{1-m}\,[z\circ x^mz^{-1}]_m= r^{1-m}\,[x^m]_m= r^{1-m}$ giving $\Delta(x-r\,z;G)=a- r^{1-m}$.
\epr

 From Proposition \ref{Delta}, if $\varphi\circ F\neq0$ then  $\varphi\circ (a\sm F)\neq0$ so we obtain another proof that $\I_{a\sm F}\subseteq \I_F$. 

 \subsection{The Inductive Step, I}
Here we treat the zero-discrepancy case.
\bp \label{Delta=0}(Inductive Step, I) If  $f$ is viable for $\I_F$, $G= a\sm F$ and  $\Delta(f_1;G)=0$  then $g=(f_1, f_2\,z)$  is viable  for $\I_G$\,. Iteratively, $\Delta(g_2;b\sm G)=\Delta(f_2;G)$ for any $b\in\sbbk$. 
\ep
\bpr Let $\J= \langle g_1,g_2\rangle$. Then $f_1\,z=g_1\,z$, $f_2\,z=g_2$ so $\I_F\,z\subseteq \J$. We have $g_1=f_1\in\I_G$ by Proposition \ref{Delta} and $g_2\circ G=f_2\,z\circ G=f_2\circ F=0$, so $\J\subseteq \I_G$\,. It follows from Lemma \ref{ally} that $\I_G=\J$.
 It is now immediate that $g$ is viable for $\I_G$\,. Finally  $g_2\circ (b\sm G)= f_2\circ G$ by Lemma \ref{f/z} and $|g_2|+|b\sm G|=|f_2|+|G|$.
\epr
\be\label{IRS+0} {\rm Let $F=x^m$ and $G=0\sm F=x^mz^{-1}$. We claim that $(x^{1-m},z^2)$ is viable for $\I_G$\,. For $(x^{1-m},z)$ is viable for $\I_F$ by Proposition \ref{001} and
$$\Delta(x^{1-m};G)=[x^{1-m}\circ  x^mz^{-1}]_{1-m+m-1}=0_0=0$$
so the claim follows from Proposition \ref{Delta=0}. Note that if $H=a\sm G=ax^{m-2}+ x^{m}z^{-2}$ 
 $$\Delta(x^{2-m}; H)=[x^{2-m}\circ H]_{2-m+|H|}=H_{m-2}=a\mbox{\ and\ }\Delta(z^2;H)=[z^2\circ H]_{2+|H|}=F_m=1$$
so  $z^2\not\in\I_H$ and if $a\neq0$ then $x^{2-m}\not\in\I_H$. At this stage,  we have not determined $\I_H$\,. } 
\eex 
\subsection{The Inductive Step, II}

We begin with the analogue for inverse forms of a well-known result \cite[Theorem 2]{Ma69}.
 
\bl\label{onemore2}   If    $f$ is viable for $\I_F$, $G= a\sm F$ and $f_1\not\in\I_G$ then for any $\varphi\in\I_G\cap\Phi$ we have $|\varphi|\geq 2-|F|-|f_1|$.
\el
\bpr  Since $f_1\not\in\I_G$  we know that  $\Delta=\Delta(f_1;G)\neq0$. Write  $\varphi=\sum_{i=0}^d \varphi_{d-i}\,x^{d-i}z^i$ where $d=|\varphi|$ and $\varphi_d\neq 0$ since $z\nmid \varphi$. As $f$ is viable for $\I_F$, $|f_1|\leq 1-m=1-|F|$. Since $\Delta\neq 0$, $f_1\circ G\neq 0$, so $e=|f_1\circ G|=|f_1|+m-1$ is well-defined.  Proposition \ref{Delta}  implies that $f_1\circ G=\Delta\cdot x^e$. As $\I_G$ is an ideal, $\varphi f_1\in\I_G$ and $z^i\circ x^e=0$ for $1\leq i\leq d$ so 
$$ 0=(\varphi f_1)\circ  G=\varphi \circ (f_1\circ  G)
=\varphi \circ(\Delta\, x^e)=\Delta\cdot\left(\sum_{i=0}^d \varphi _{d-i}\,x^{d-i}z^i\right)\circ x^e=\Delta\cdot  \varphi _d\, x^d\circ x^e
$$ 
 and $\Delta\cdot \varphi _d\in\sbbk^\times$ so $x^d\circ x^e=0$ i.e.  $|\varphi |+|f_1|+m-1=d+e\geq 1$.
 \epr
 \bc \label{onemore2cor}If $f$ is viable for $\I_F$\,, $G=a\sm F$ and $f_1\not\in\I_G$ then $\lambda_G\geq\max\{|f_1|,|f_2|\}$.
 \ec
 \bpr By definition $\lambda_F=\min\{|\varphi|: \varphi\in\I_F\cap\Phi\}$ and $\lambda_F=|f_1|$  from  Proposition \ref{Lambda}. Proposition \ref{incl} implies that  $\lambda_G\geq\lambda_F=|f_1|$. We also have $\lambda_G\geq 2-m-|f_1|$ from Lemma \ref{onemore2}, and $2-m-|f_1|=|f_2|$ since $f$ is viable for $\I_F$. 
 \epr

The examples developed so far suggest the following two definitions. For $d\in\Z$, let us define $\ol{d},\ul{d}\in\Z$ by $\ol{d}=\max\{d,0\}$ and $\ul{d}=\min\{d,0\}$. 
Then  $\ol{d}+\ul{d}=d$. Secondly
\bd {\rm Let $f$ be viable for $\I_F$ and $d=d_f=|f_2|-|f_1|$ be the difference of their total degrees. If  $G= a\sm F$ and $f_1,f_2\not\in\I_G$ let $q$ be the quotient $q=q_f=\Delta(f_1;G)/\Delta(f_2;G)$. Then  
$$f_1\ominus f_2=f_1\ominus_{d,\,q} f_2=x^{\ol{d}}f_1-q\, x^{-\ul{d}}f_2\,.$$
We put $(f_1z)\curlyvee_{d}(f_2z)=f_1z$ if $d>0$ and    $(f_1z)\curlyvee_{d}(f_2z)=f_2z$ if $d\leq 0$. Similarly, $\Delta(f_1;G)\curlyvee_{d}\Delta(f_2;G)=\Delta(f_1;G)$ if $d>0$ and    $\Delta(f_1;G)\curlyvee_{d}\Delta(f_2;G)=\Delta(f_2;G)$ if $d\leq 0$. }
\ed
We will omit $d,q$ when they are clear from the context. Note that $d$ is well-defined since if $f$ is viable for $\I_F$ then $f_1,f_2\neq 0$ and $q\in\sbbk^\times$ is well-defined since $f_i\not\in\I_G$\,. Of course if $\sbbk=\mathrm{GF}(2)$ then $q=1$ for any $f_1\not\in\I_G$.
\bl \label{newg_1} If $f$ is viable for $\I_F$, $G= a\sharp F$ and $f_1,f_2\not\in\I_G$ then  $|f_1\ominus f_2|=\ol{d}_f+|f_1|=\max\{|f_1|,|f_2|\}$ and $f_1\ominus f_2\in\I_G\cap\Phi$.
\el
\bpr    Put  $g_1=f_1\ominus f_2$\,. We have $\ol{d}+|f_1|=d-\ul{d}+|f_1|= |f_2|-\ul{d}$. As $f_1\in\Phi$ and $z|f_2$, $g_1$ is a non-zero form with $\LM(g_1)=x^{\,\ol{d}}\,\LM(f_1)$. In particular $g_1\in\Phi$ and $|g_1|=\ol{d}+|f_1|$. If $d\leq 0$ then  $\ol{d}=0$ and $|g_1|=|f_1|\geq|f_2|$; otherwise $d>0$, $|f_2|>|f_1|$ and $|g_1|=d+|f_1|=|f_2|>|f_1|$.  Thus $|g_1|=\max\{|f_1|,|f_2|\}$.

Since $\I_F$ is an ideal and $f_1,f_2\in\I_F$ we have $g_1=f_1\ominus f_2\in\I_F$\,. Also $f$ is viable for $\I_F$ so $|f_1|,|f_2|\leq 1-m$ and therefore $|g_1|=\max\{|f_1|,|f_2|\}+|G|\leq 0$. Thus $g_1\in\I_G$ if and only if $\Delta(g_1;G)=0$. But
\begin{eqnarray*}\Delta(g_1;G)&=&[g_1\circ G]_{|g_1|+|G|}
=[(x^{\,\ol{d}}f_1-q\,x^{-\ul{d}}f_2)\circ G]_{\,\ol{d}+|f_1|+|G|}\\
&=&[f_1\circ G]_{|f_1|+|G|}-q\,[f_2\circ G]_{\,\ol{d}+|f_1|+|G|+\ul{d}}=
\Delta(f_1;G)-q\,[f_2\circ G]_{|f_1|+d+|G|}\\
&=&\Delta(f_1;G)-q\,[f_2\circ G]_{|f_2|+|G|}=\Delta(f_1;G)-q\,\Delta(f_2;G)=0.
\end{eqnarray*}
 Thus $f_1\ominus f_2\in\I_G\cap\Phi$.
\epr
Next our second inductive step of a viable ordered pair. 
The significance of  Theorem \ref{template}(ii) is that we can apply either the first or second inductive step to  $g$ and $b\sm G$.
\bt \label{template}(Inductive Step, II) Let     $f$ be viable for $\I_F$,  $G= a\sm F$ and $f_1,f_2\not\in\I_G$\,. If  $d=|f_2|-|f_1|$ and $q=\Delta(f_1;G)/\Delta(f_2;G)$ then (i) 
$$g=(f_1\ominus_{d,\,q} f_2\,, (f_1z)\curlyvee_{d}(f_2z)\,)$$
  is viable for $\I_G$,  
$|g_1|=\max\{|f_1|,|f_2|\}=\max\{\lambda_F,2-m-\lambda_F\}$ 

(ii) iteratively, $d_g=1-|d|$ and   $\Delta(g_2;b\sm G)=\Delta(f_1;G)\curlyvee_{d} \Delta(f_2;G)$ for any $b\in\sbbk$.
\et
\bpr  
 (i)   Since $f_1,f_2\not\in\I_G$, $q\in\sbbk^\times $ is well-defined  and $g_1\in\I_G\cap\Phi$ from Lemma \ref{newg_1}. Next we apply Lemma \ref{ally} to prove that $\I_G=\langle g_1,g_2\rangle=\J$ say.
First we show that  $\I_F\,z=\langle f_1,f_2\rangle\,z\subseteq \J$.  If $d\leq 0$ then by definition, $g_1=f_1-q\,x^{-d}f_2$ and $g_2=f_2\,z$ so $f_1\,z=(g_1+q\,x^{-d}f_2)z=g_1\,z+q\,x^{-d}g_2\in\J$, and trivially $f_2\,z=g_2\in\J$.  Now suppose that $d>0$. Then $g_1=x^df_1-q\,f_2$ and $g_2=f_1\,z$. So  $q\, f_2\,z=(x^df_1-g_1)z=x^d\,g_2-g_1\,z\in\J$. As $q\neq 0$ this implies that  $f_2\,z\in\J$, and trivially $f_1\,z\in \J$, so $\I_F\,z\subseteq \J$. Secondly, for any form $\varphi\in\I_F$ we have  $(\varphi\,z)\circ  G=\varphi\circ F=0$,  so $g_2\in\I_G$ and $\J\subseteq\I_G$. Hence by Lemma \ref{ally}, $\I_G=\J$.  

It is trivial that $z|g_2$\,. If $d\leq0$ then $|g_1|+|g_2|=|f_1|+|f_2|+1=3-m$ and  otherwise $|g_1|+|g_2|=(d+|f_1|)+|f_1|+1=|f_2|+|f_1|+1=3-m$. Thus $|g_1|+|g_2|=2-|G|$ and $g$ is viable for $\I_G$\,.   As $f$ is viable for $\I_F$, $|f_1|=\lambda_F$ by Proposition \ref{Lambda} and from Lemma \ref{newg_1}, $|g_1|=\max\{|f_1|,|f_2|\}=\max\{\lambda_F,2-m-\lambda_F\}$.
 
(ii)    By definition,  $d_g=|g_2|-|g_1|$ which  is $|g_2|-(\ol{d}+|f_1|)$ from Lemma  \ref{newg_1}. If $d\leq 0$, $d_g=|g_2|-|f_1|=|f_2z|-|f_1|=1+d=1-|d|$ and if $d>0$, $d_g=|g_2|-|g_1|=(|f_1|+1)-(d+|f_1|)=1-d=1-|d|$. Finally, let $H=b\sm G$. We claim  that $\Delta(g_2;H)=\Delta(f_1;G)\curlyvee_{d} \Delta(f_2;G)$. Firstly  $z\circ H=G$ by Lemma \ref{f/z} and if  $d> 0$, $g_2=f_1z$  so $|g_2|+|H|=|f_1|+|G|\leq 0$ since $f$ is viable for $\I_F$ and 
$$\Delta(g_2;H)=[g_2\circ H]_{|g_2|+|H|}=
[(f_1z)\circ H]_{|g_2|+|G|-1}=[f_1\circ G]_{|f_1|+|G|}=\Delta(f_1;G).$$
Similarly, if $d\leq0$, $g_2=f_2z$ and $\Delta(g_2;H)=[g_2\circ H]_{|g_2|+|H|}=
[(f_2z)\circ H]_{|g_2|+|G|-1}=[f_2\circ G]_{|f_2|+|G|}=\Delta(f_2;G)$. 
\epr

 We again see that $\I_{a\sm F}\subseteq \I_F$\,, for the two  generators of $\I_{a\sm F}$ (as constructed in Proposition \ref{Delta=0} or Theorem \ref{template}) are in $\I_F$. 

 \subsection{Examples and Applications}
 First we reconsider Example \ref{firsta,b}.
  \be\label{IRSdisc} {\rm If   $a\in\sbbk^\times$ and $G= a\sm x^m=ax^{m-1}+x^mz^{-1}$ then $(x^{1-m}-a\,x^{-m}z,z^2)$ is viable for $\I_G$\,.  For  $f=(x^{1-m},z)$ is viable for $\I_{x^m}$ by Proposition \ref{001}, $d=m\leq 0$ and
 $q=a$ as
 $$\Delta(f_1;G)=[x^{1-m}\circ G]_{1-m+|G|}=a \mbox{\ and\ } \Delta(z;G)=[z\circ G]_{1+|G|}=1.$$
 }\eex
 
Now apply Theorem \ref{template}. Together with Example \ref{IRS+0} where $a=0$, this completes the inductive step for  $x^m$.  

Next we apply the construction to geometric inverse forms. The first result generalises Example \ref{100} where $F=z^m$ and $r=F_{-1}=0$.
 \bc \label{geomviable} If  $F$ is geometric then $(x-r z,z^{1-m})$ is viable for $\I_F$\,.
\ec 
\bpr We have $F_0=1$, so that  $f=(x,z)$ is   viable for $\I_{F_0}$\,. Consider  $G=a\sm 1=ax^{-1}+z^{-1}$; $\Delta(f_1;G)=[f_1\circ  G]_0=a$ and $\Delta(f_2;G)=[z\circ  G]_0=1$.  Hence if $a=0$, Proposition \ref{Delta=0} implies that $(x,z^2)=(x-r,z^2)$ is viable for $\I_G$ (where $r=0$). But if $a\neq 0$, $d=0$ and $q=\Delta(f_1;G)/\Delta(f_2;G)=a$ so that Theorem \ref{template} implies that $(x-a\,z,z^2)$ is viable for $\I_G$\,. Suppose now that $m\leq -1$,  the result is true for $F$ and $G= a\sm F$ is geometric.  Then $F$ is also geometric, so $f=(x-r z,z^{1-m})$ is viable for $\I_F$\,. Since $G$ is geometric, $a=r^{1-m}$ and hence $\Delta(f_1;G)=0$ by Lemma \ref{geomdelta}.  Thus by Proposition \ref{Delta=0} $(f_1,f_2z)=(f_1,z^{2-m})$ is viable for $ a\sm F$ and the induction is complete.
\epr

\bc \label{geom} Let $F$ be geometric and $f=(x-r\,z, z^{1-m})$. If $G= a\sm F$ is not geometric   then  $d=-m>0$, $q=a-r^{1-m}\neq0$ and $(x^{-m}f_1- q\,z^{1-m},f_1 z)$ is viable for $\I_G$\,.
\ec
\bpr   From Corollary \ref{geomviable},  $(f_1,z^{1-m})$ is viable for $\I_F$, $d=-m>0$, so $\ol{d}=d$ and $\ul{d}=0$. Suppose that $G= a\sm F=ax^{1-m}+\sum_{i=m}^0 r^{-i}x^iz^{m-i-1}$ is not geometric. Then $\Delta(f_1;G)=a-r^{1-m}\neq0 $ from Lemma \ref{geomdelta}. Further 
$$\Delta(f_2;G)=[z^{1-m}\circ  G]_{1-m+|G|}=\left[z^{1-m}\circ \left(ax^{1-m}+\sum_{i=m}^0 r^{-i}x^iz^{m-i-1}\right)\right]_0=1$$ 
so $q=a-r^{1-m}$ and
 $(x^{-m}f_1- q z^{1-m},f_1\,z)$ is viable for $\I_G$ by Theorem \ref{template}.
\epr

\be {\rm For $m\leq-1$, if $F=z^m$ then $f=(x,z^{1-m})$  is viable for $\I_F$ by Example \ref{100}. Consider $G= a\sm F=ax^{m-1}+z^{m-1}$ where $a\in\sbbk^\times$. We have $d=-m>0$. Next, $x\circ x^{m-1}=x^{m}$ since $m\leq-1$. Hence
\begin{eqnarray*}
\Delta(x;G)&=&[x\circ (a\,x^{m-1}+z^{m-1})]_{1+|G|}=a[x\circ x^{m-1}]_{m}+[x\circ z^{m-1}]_{m}
=a[x^m]_{m}=a\neq 0
\end{eqnarray*}
and $\Delta(z^{1-m};G)=[z^{1-m}\circ G]_{1-m+|G|}=[z^{1-m}\circ(a\,x^{m-1}+z^{m-1})]_0=1$. Hence $q=a$ and Theorem \ref{template}  implies that  $(x^{1-m}-az^{1-m}, x\,z)$ is  viable for $\I_G$.}
\eex

Corollary \ref{geom} also provides another characterisation of geometric inverse forms.
 \bp \label{isgeom} An inverse form $F$ is geometric  if and only if   $m\leq-1$, $F_0=1$ and $(x-r\,z,z^{1-m})$ is  viable for $\I_F$\,.
\ep
\bpr From Corollary \ref{geomviable}, it suffices to show that if $F_0=1$, $m\leq-1$ and $(x-r\,z,z^{1-m})$ is  viable for $\I_F$ then $F$ is geometric.
From Proposition \ref{Lambda} we have $\lambda_F=|f_1|=1$ since $f$ is viable for $\I_F$ and  $\lambda_0=1$ since $F_0=1$. Thus $1=\lambda_m\geq \cdots\geq\lambda_0=1$ i.e.  $\lambda_m=\cdots=\lambda_0=1$, so by Proposition \ref{Lambdageom}, $F$ is geometric.
 \epr

\bt\label{indthm} If $F\neq0$   we can construct a viable $f$ for $\I_F$\,. 
\et
\bpr    If  $F=x^\vv$,  Proposition \ref{001} implies that $(x^{1-\vv},z)$ is viable for $\I_F$\,. So we continue inductively as long as we have not exhausted the inverse form: if  $f$ is viable for $F$ and $G= a\sm F$, apply either Proposition \ref{Delta=0}  or Theorem \ref{template} to $f$ and $G= a\sm F$ according as $\Delta(f_1;G)=0$ or not.
 \epr

\bc If $F\neq0$ then $\I_F$ has two generators.
\ec
\bpr We know that  $\I_F$ is proper and $x^{1-m},\,z^{1-m}\in\I_F$\,, so $\I_F$ has at least two generators and by Theorem \ref{indthm}, two generators suffice.
\epr

The next result is  essentially \cite[Proposition 4.17]{N15}  reproved for an inverse form.
 \bp\label{stable} If $F\neq0$, $\Lambda_F=\lambda_F(2-m-\lambda_F)\leq  (1-\frac{m}{2})^2$ with equality if and only if $\lambda_F=1-\frac{m}{2}$ if and only if $\Lambda_F=\lambda_F^2$.
\ep
\bpr First we prove the equality by induction. If $F=x^m$ then $\Lambda_F=\lambda_F=1-m$ by Proposition \ref{Lambda}  and $2-m-\lambda_F=1$. Suppose inductively that the result is true for $F$, $f$ is viable for $\I_F$ and $G=a\sm F$. We have $\lambda_{G^{(i)}}=\lambda_{F^{(i)}}$ for $m\leq i\leq \vv$. If $\Delta(f_1;G)=0$ then $\lambda_G=\lambda_F$ by Proposition \ref{Delta=0} and from the inductive hypothesis
$$\Lambda_G=\sum_{i=m-1}^0\lambda_{G^{(i)}}=\lambda_F+\lambda_F(2-m-\lambda_F)=\lambda_F(3-m-\lambda_F)=\lambda_G(2-(m-1)-\lambda_G)$$
as required. For $\Delta(f_1;G)\neq 0$ we apply Theorem \ref{template}. If $d\leq0$ then $\Lambda_G=\Lambda_F$ and we have just seen that the result is true in this case. But if $d>0$ then $\lambda_G=2-m-\lambda_F$ and by the inductive hypothesis, 
$$\Lambda_G=\sum_{i=m-1}^0\lambda_{G^{(i)}}=
\lambda_G+\lambda_F(2-m-\lambda_F)=(\lambda_F+1)(2-m-\lambda_F)$$
Now $\lambda_G(3-m-\lambda_G)=(2-m-\lambda_F)(3-m-(2-m-\lambda_F))=(2-m-\lambda_F)(\lambda_F+1)=\Lambda_G$\,. 

For the inequality, we show that
$4\lambda_F(2-m-\lambda_F)\leq (2-m)^2$ with equality if and only if $2\lambda_F=2-m$.  For  integers $a,b$ we have $4ab\leq (a+b)^2$, with equality if and only if $a=b$. Put $a=\lambda_F$ and $b=2-m-\lambda_F$\,. Then $a+b=2-m$, so that $4ab\leq (2-m)^2$, with equality if and only if $\lambda_F=2-m-\lambda_F$\,. 

Finally, if $\lambda_F=1-\frac{m}{2}$ then $\Lambda_F=\lambda_F^2$ by the first part. Conversely, if $\Lambda_F=\lambda_F^2$ then $2-m-\lambda_F=\lambda_F$ since $\lambda_F\neq0$ and so $\lambda_F=1-\frac{m}{2}$.
\epr
\bc \label{ha}  If $f$  is viable for $\I_F$ then 

(i)  $|f_1|\,|f_2|=\Lambda_F$
 
(ii) $\Phi\cap\I_F=\{a_1 f_1+a_2f_2  {\rm\  monic}, \ a_i\ {\rm forms}, z\nmid a_1, a_2=0 \mbox{ or }|a_1|-|a_2|= d\}\,$.
\ec
\bpr     (i) Since $f$ is viable for $\I_F$\,, $\lambda_F=|f_1|$  from Proposition \ref{Lambda}  and $|f_2|=2-m-\lambda_F$\,. Hence  $|f_1|\,|f_2|=\lambda_F(2-m-\lambda_F)=\Lambda_F$ by Proposition \ref{stable}. 

(ii)  It is elementary  that  the right-hand side is a subset of $\I_F\cap\Phi$ so let $\varphi\in\I_F\cap\Phi$ be a form. We can write $\varphi=a_1f_1+a_2f_2$ for forms  $a_1,a_2\in\R$ where $z\nmid a_1$ since  $z\nmid\varphi$.   If  $a_2\neq 0$ then $|a_1|+|f_1|=|\varphi|=|a_2|+|f_2|$ so $|a_1|-|a_2|=
|f_2|-|f_1|$. 
 \epr
 \subsection{The Corresponding Algorithm}
This follows from the inductive proof of Theorem \ref{indthm}. First the inductive basis Proposition \ref{001} via a repeat loop giving $f=(x^{1-\vv},z)$ and $d=|f_2|-|f_1|=\vv$. Then we initialise for  Proposition \ref{Delta=0} and Theorem \ref{template}. In the main loop we compute $\Delta(f_1;F^{(i)})$ iteratively. We use $f$ throughout for the viable pair and if Theorem \ref{template} applies, expand  
 $f_1\ominus _{d,q}f_2$ according to $d,q$ and then update $f_2,\,d$. It is easy to check that  Algorithm \ref{calPA} also incorporates Proposition \ref{Delta=0}.
\begin{algorithm}\label{calPA}(Viable ordered pair for $\I_F$)
\begin{tabbing}
\noindent {\tt Input}: \ \ \=  Inverse form $F\in\M^\times$.\\

\noindent {\tt Output}: \> Viable ordered pair $f$ for $ \I_F$\,.\\\\

$\lceil$ $i\la 1$; {\tt repeat}  $i\la i-1$  {\tt until} $(F_i\neq0)$; $\vv\la i$; $ f\la (x^{1-\vv},z)$; $\Delta_2\la 1$;\ $d \la  \vv$; \\\\

$G\la x^\vv$;\\
{\tt for} \= $i\la \vv-1$ {\tt downto }$|F|$ {\tt do}\\\\

   \>$\lceil$  $G\la F_i\sm G$; $\Delta_1\la \Delta(f_1; G); q\la \Delta_1/\Delta_2$;\\\\    \>{\tt if }$(\Delta_1\neq 0)$ {\tt then} \={\tt if} \=$(d\leq 0)$\ \={\tt then}\, $ f_1\la  f_1-q\,  x^{-d}\, f_2 $\\\\
  \> \> \>\> {\tt else}
   $\lceil$\= $T\la f_1$; $f_1 \la  x^d\, f_1-q\, f_2$; $f_2\la T$; \\\\
   \>\> \> \>\>$\Delta_2\la  \Delta_1$; $d \la  -d;\rfloor$\\\\
   \> $f_2\la f_2\,z$; $d  \la  1+d; \rfloor$\\\\
{\tt return }$f.\rfloor$
\end{tabbing} 
  \end{algorithm}

\be  \label{011}{\rm  For $\sbbk=\mathrm{GF}(2)$, $m\leq 0$ and $F=x^m$ we have $\vv=m$ and $f=(x^{1-m},z)$. If $G=1\sm F=x^{m-1}+x^mz^{-1}$ then $q=1$ and $d=m\leq 0$, giving  $(x^{1-m}-x^{-m}z,z^2)$ for $\I_G$.}
\eex

\be \label{ADex4} {\rm For $\sbbk=\mathrm{GF}(2)$ and $F=x^{-6}z^{-1}+x^{-4}z^{-3}+x^{-3}z^{-4}+z^{-7}$ we obtain the viable pair $f=(x^4+xz^3+z^4,(x^3+x^2z+xz^2+z^3)z^2)$ in the following table, where $B$ is the boolean statement $(q\neq0)\,\&\, (d>0)$, with values $0,1$:
\begin{center}
\begin{tabular}{|r|r|c|c|c|l|}\hline
$i$&$F^{(i)}$ & $q$  & $d$ &$B$    &$f$  \\\hline\hline
$0$&$1$   &$-$  &$0$  &0& $(x,z)$  \\\hline
$-1$&$z^{-1}$   & $0$ &$1$ &0& $(x,z^2)$   \\\hline
$-2$&$z^{-2}$   & $0$ &$2$  &1 &$(x,z^3)$  \\\hline
$-3$&$x^{-3}+z^{-3}$   &$1$  &$-1$  &0& $(x^3+z^3,xz)$    \\\hline
$-4$&$x^{-4}+x^{-3}z^{-1}+z^{-4}$   & $1$ &$0$  &0& $(x^3+x^2z+z^3,xz^2)$ \\\hline
$-5$&$x^{-4}z^{-1}+x^{-3}z^{-2}+z^{-5}$   &$1$  &$1$  &1& $(x^3+x^2z+xz^2+z^3,xz^3)$ \\\hline
$-6$&$x^{-6}+x^{-4}z^{-2}+x^{-3}z^{-3}+z^{-6}$   & $1$ & $0$ &0& $(x^4+x^3z+x^2z^2,(x^3+x^2z+xz^2+z^3)z)$ \\\hline
$-7$&$x^{-6}z^{-1}+x^{-4}z^{-3}+x^{-3}z^{-4}+z^{-7}$   &$1$  & $1$  &1& $(x^4+xz^3+z^4,(x^3+x^2z+xz^2+z^3)z^2)$. \\\hline
\end{tabular}
\end{center}
}
\eex 

\subsection{A Worst-Case Analysis}

\bp \label{analysis} Let $F\neq 0$ and $m=|F|$. Ignoring  linear and constant terms, Algorithm \ref{calPA} requires at most $m^2/2$ multiplications in $\sbbk$ to compute a viable $f$ for $\I_F$\,.
\ep
 \bpr  Finding $(x^{1-\vv},z)$ requires no multiplications. For $m\leq i\leq \vv-1$, let $f^{(i)}$ be the viable pair obtained in the loop of Algorithm \ref{calPA} for $F^{(i)}$. From Lemma \ref{prebasic},   $\Delta(f_1^{(i)}; F^{(i)})=[f_1^{(i)}\cdot F^{(i)}]_{|f_1^{(i)}|+i}$  requires at most $|f_1^{(i)}|$ multiplications in $\sbbk$ since $f_1^{(i)}$ is monic.  Updating $f_1^{(i)}$ requires at most $|f_2^{(i)}|$ more multiplications.   Now $|f_1^{(i)}|+|f_2^{(i)}|=2-i$, so altogether  at most 
$$\sum_{i=1-\vv}^m (2-i)\leq\sum_{i=0}^m (2-i)=2(1-m)+m(m-1)/2$$ multiplications in $\sbbk$ are required by Algorithm \ref{calPA} to find a viable $f$ for $\I_F$\,. Ignoring constant and linear terms, this gives a total of at most $m^2/2$.
\epr
\subsection{Essential Inverse Forms}
In this subsection we develop the notion of an 'auxiliary  triple'  for an 'essential inverse form'. This is similar to an auxiliary triple for an essential sequence  in \cite{N15}. This subsection forms an important part of constructing a Gb for $\I_F$ in the next section.  
We close with an application to the uniqueness/non-uniqueness of $f_1,f_2$ for an inverse form $F$. 
\bd We say that $F$ is essential if $\lambda_F>\lambda_0$ (and in particular, $|F|\leq-1$). 
\ed

Next a trichotomy of inverse forms, based on \cite[Proposition 2.10]{N15}.
\bp\label{trichotomy} Either $F=1$, $F$ is geometric or $F$ is essential.
\ep
\bpr Recall that $\lambda$ is non-decreasing. Since $F\neq0$, $\vv$ is well-defined. If $\vv<0$ then $\lambda_F\geq 1-\vv>0=\lambda_0$ and $F$ is essential, so suppose that $\vv=0$ and $F_0=1$. Now if $m\leq-1$, $F^{(-1)}$ is always geometric, so if $F$ is not geometric then $F^{(i)}$ is geometric for some $i$ where $m+1\leq i\leq -1$, but $F^{(i-1)}$ is not.  Corollary \ref{geom} implies that $\lambda_F=\lambda_{m}\geq  \lambda_{i-1}=-(i-1)=1-i>1=\lambda_0$ and so $F$ is essential.
\epr
We now motivate the definition of an auxiliary triple $T'_F$ for an essential $F$, with $m\leq-1$. The triple will consist  of an integer in $(m,0]$, a former form and a non-zero discrepancy. \be \label{IRSex} {\rm Let $F=x^m$ with $m\leq-1$. Then $F$ is essential since $F^{(m+1)}=\cdots=F^{(0)}=0$ and $\lambda_m=1-m>\lambda_{m+1}=\cdots=\lambda_0=0$ from Propositions \ref{001}, \ref{Lambda}. Further, $1\in\I_{F^{(m+1)}}$ and $\Delta(1;F)=[1\circ F]_{0+|F|}=1$. Our triple will be $T'_F=(m+1,1,1)\in(m,0]\times\R\times\sbbk^\times$.}
\eex
\be \label{geomex} {\rm Let $m\leq-1$, $F$ be geometric, so that $f=(x-rz,z^{1-m})$ is viable for $F$ by Corollary \ref{geomviable}. Suppose that $G=a\sm F$ is not geometric. Then $(x^{-m}f_1- q\,z^{1-m},f_1 z)$ is viable for $\I_{G}$ by Corollary \ref{geom}, where $q=a-r^{1-m}\neq0$. So $\lambda_G=1-m>1=\lambda_0$ and $G$ is essential. Secondly, $f_1\in\I_F$ but $f_1\not\in\I_G$ since $G$ is not geometric; in fact $\Delta(f_1;G)=q$. Our triple for $G$ will be $T'_G=(m, f_1,q)\in(m-1,0]\times\R\times\sbbk^\times$.}
\eex

\bd \label{tripled} {\rm We say that $F$ is {\em essential with triple} $T'=T'_F\in(m,0]\times\R\times\sbbk^\times$ if $F$ is essential and
 
(i)   $T'_1=\min_{i\leq0}\{i: \lambda_i<\lambda_m\}$

(ii)    $T'_2\in\I_{F^{(T'_1)}}\cap\Phi$  and $|T'_2|=\lambda_{T'_1}$

(iii)  $T'_3=\Delta(T'_2;F^{(T'_1-1)})\in\sbbk^\times$.}
\ed  

Note that (iii) implies that $T'_1-1\leq \vv$. To simplify the notation, we will often write $T'=(m',f_1',\Delta_1')$ and $F'=F^{(m')}$ so that $f_1'\in \I_{F'}\setminus \I_{F^{(m'-1)}}$.
The triple $T'=T'_F$  is  \ul{undefined} if $F=1$ or $F$ is geometric since the integer $m'=T'_1$ is undefined. 

Let $f$ be viable for $\I_F$\,. Our next goal is to relate $f_2$ and $f_1'$ when $F$ is essential with triple $T'$. 
This requires the following lemma, which also shows that annihilator ideals  are related to  ideal quotients.
\bl\label{IQ}If  $m<l\leq\vv$ and $|F|=m$ then

(i) $z^{l-m}\circ F=F^{(l)}$

(ii) if $f_1'\circ F^{(l)}=0$ then $(f_1'z^{l-m})\circ F=0$

(iii) $\I_{F^{(l)}}=(\I_F:z^{l-m})=\{\varphi\in\R: \varphi\,z^{l-m}\in\I_F\}$.
\el
\bpr  (i) Since $l-m\geq1$ and $z^{l-m}\circ z^i=0$ for $0\geq i\geq m-l+1$ 
 $$z^{l-m}\circ(F_mx^m+\cdots+F_{l-1}x^{l-1}z^{m-l+1}+F_lx^lz^{m-l}+\cdots+x^\vv z^{m-\vv})=F^{(l)}.$$
 (ii) Immediate consequence of (i). (iii) By Part (i),  
$(\varphi\, z^{l-m})\circ F=\varphi\circ F^{(l)}$  for any $\varphi\in\R$ and  so $\varphi\, z^{l-m}\in\I_F$ if and only if $\varphi\in \I_{F^{(l)}}$.
\epr

Since $m<m'\leq0$, it follows from Lemma \ref{IQ} that if $F$ is essential with $(m',f_1',\Delta_1')$ then $f_1'z^{m'-m}\in\I_F\cap z\,\R$. 

\bl \label{ess} Suppose that $F$ is essential with triple $T'$, $f$ is viable for $\I_F$ and $f_2=f_1'z^{m'-m}$. Let $G=a\sm F$ and  $\Delta(f_1;G)\neq 0$. If
 $$U'=\left\{\begin{array}{ll}
T'& \mbox{ for } d\leq0\\
(m,f_1,\Delta(f_1;G))&\mbox{ otherwise}
\end{array}\right. $$
then $G$ is essential with $U'$ and if $g$ is as in Theorem \ref{template}  then  $(f_1\ominus f_2\,,g_1'\,z^{(m-1)'-(m-1)})$ is viable for $\I_G$, where $(m-1)'=U'_1$ and $g_1'=U'_2$\,.
\el
\bpr  We have $\lambda_0<\lambda_F$\,. Hence $\lambda_0<\lambda_F\leq \lambda_G$ and $G$ is also essential. First we show that $\Delta(f_2;G)=\Delta(f_1';F^{(m'-1)})=\Delta_1'$ which is non-zero by definition.  From Lemma \ref{IQ},  $z^{m'-m}\circ G=z^{m'-1-(m-1)}\circ G=G^{(m'-1)}=F^{(m'-1)}$ as $m\leq m'-1$. Since $|f_2|=|f_1'|+m'-m$ 
$$[f_2\circ G]_{|f_2|+m-1}=[f_1'z^{m'-m}\circ G]_{|f_1'|+m'-1}=[f_1'\circ G^{(m'-1)}]_{|f_1'|+m'-1}=[f_1'\circ F^{(m'-1)}]_{|f_1'|+m'-1}=\Delta_1'.$$
As $\Delta(f_2;G)=\Delta_1'\neq0$ it follows that $f_2\not\in\I_G$. 
 
 Next we have to show that $G$ is essential with $U'$. 
  If $d\leq 0$, $\lambda_{m-1}=\lambda_G=\lambda_F=\lambda_m$ and   $U'_1=m'=\min_{i\leq0}\{i:\lambda_i<\lambda_m\}=\min_{i\leq0}\{i:\lambda_i<\lambda_{m-1}\}$. By definition, $F'=F^{(m')}=G^{(m')}=G'$, $g_1'=(T'_G)_2=f_1'\in\I_{F'}\cap\Phi=\I_{G'}\cap\Phi$  and $|g_1'|=|f_1'|=\lambda_{F'}=\lambda_{G'}$. Thirdly,  $\Delta(g_1';G)=\Delta(f_1';G)=\Delta_1'\in\sbbk^\times$. Thus $G$ is essential with $U'$ and we have 
 $$g_2=f_2\,z=f'_1\,z^{m'-m}\,z=g_1'\,z^{(m-1)'-(m-1)}$$
  since $g_1'=U'_2=T'_2=f_1'$ and $(m-1)'=m'$. 
  
  Now let $d>0$. Then $\lambda_{m-1}=\lambda_G=d+\lambda_F>\lambda_F=\lambda_m$, so $U'=(m;f_1,\Delta(f_1;G))$ and $\min_{i\leq0}\{i:\lambda_i<\lambda_{m-1}\}=m$. We now have $g_1'=f_1\in\I_F\cap\Phi=\I_{G'}\cap\Phi$ and $|g_1'|=|f_1|=\lambda_F=\lambda_{G'}$. Thirdly $\Delta(g_1'; G^{(U_1'-1)})=\Delta(f_1;G)$, which is non-zero by assumption. Finally $$g_2=f_1\,z=g_1'\,z^{(m-1)'-(m-1)}$$ since $g_1'=(T'_G)_2=f_1$ and $(m-1)'=m$. 
   The fact that $(f_1\ominus f_2\,,g_2)$ is viable for $\I_G$ was proved in Theorem \ref{template}.
  \epr
  Part (ii) of the next result is our  version of \cite[Equation (13)]{Ma69} for inverse forms.
\bt \label{linked!} Let $F$ be essential with auxiliary triple $T'=(m',f_1',\Delta_1')$. The viable $f$ for $\I_F$ of Theorem \ref{indthm} satisfies $f_2=f_1'z^{m'-m}$.  Further,  $\lambda_F+\lambda_{F'}=2-m'$ where $F'=F^{(m')}$.
\et
\bpr  If $F_0=0$ then $F'=F^{(m')}=0$ for some $m'\leq0$ with $F^{(m'-1)}=x^{m'-1}$ and $(x^{2-m'},z)$  viable for $\I_{F^{(m'-1)}}$ by Proposition \ref{001}. Thus $F^{(m'-1)}$ is essential with  $(m',1,1)$ and $f_2=z=f_1'\,z^{m'-(m'-1)}$. Hence we can apply Lemma \ref{ess} inductively to deduce the first part in this case. Now suppose that $m\leq-1$, $F_0=1$ and let  $r=F_{-1}$\,. Since $F$ is essential, it is not geometric. Let $F'=F^{(m')}$ be geometric  with  $f=(x-rz,z^{1-m'})$ viable for $\I_{F'}$ and $\Delta(f_1;F^{(m'-1)})\neq0$. Then $(x^{-m'}f_1- q\,z^{1-m'},f_1 z)$ is viable for $\I_{F^{(m'-1)}}$ by Corollary \ref{geom} and $f_2=f_1z=f_1z^{m'-(m'-1)}$.  Now apply Lemma \ref{ess} inductively.

For the second part, $2-m=|f_1|+|f_2|=|f_1|+|f_1'|+m'-m=\lambda_F+\lambda_{F'}+m'-m$ by Proposition \ref{Lambda}, so  $\lambda_F+\lambda_{F'}=2-m'$.
\epr

The next proposition is an important complement to Theorem \ref{linked!} since  if   $F_0=1$, either (i) $F=1$ or (ii) $m\leq-1$ and $F^{(i)}$ is geometric for some $i$, $m\leq i\leq -1$.
\bd If $F=1$ or $F$ is geometric, we define $f_1'=1$,  $\lambda_1=|f_1'|=0$  and $m'=\min\{i:\lambda_i<\lambda_m\}$.
\ed
\bp\label{dovetail} If $F=1$ or $F$ is geometric and $f$ is  constructed as in Theorem \ref{indthm} then $m'=1$, $f_2=f_1'z^{m'-m}$ and 
$\lambda_m+\lambda_{m'}=2-m'$.
\ep
\bpr  If $F=1$ then $f=(x,z)$, $\lambda_m=1$ and 
$m'=\min\{i:\lambda_i<\lambda_m\}=1$ since $\lambda_1=0$. We now have  $f_1'z^{m'-m}=z=f_2$ and $\lambda_m+\lambda_{m'}=1=2-m'$. Similarly, if $F$ is geometric  Corollary \ref{geomviable} implies that  $f=(x-r\,z,z^{1-m})$ where $r=F_{-1}$ and  $\lambda_m=\cdots=\lambda_0=1$ by Proposition \ref{Lambdageom}. Thus   $m'=\min\{i:\lambda_i<\lambda_m\}=1$ as $\lambda_1<1$, $f_1'z^{m'-m}=z^{1-m}=f_2$ and $\lambda_m+\lambda_{m'}=1=2-m'$. 
\epr

We conclude this section with an application to uniqueness/non-uniqueness of $f_1,f_2$.
\bc Let $F\neq0$ and $f$ as in Theorem \ref{indthm} be viable for $\I_F$\,. The following are equivalent:
\begin{tabbing}
\hspace{1cm}\= (a) $\lambda_F<1-\frac{m}{2}$\\
\>(b) $|f_1|<|f_2|$\\
\>(c) $f_1$ is unique: if $\varphi\in\I_F\cap\Phi$ and $|\varphi|=|f_1|$ then $\varphi=f_1$\\
\>(d)  $f_2$ is not unique: there is a monic form $\varphi\in\I_F^\times$ with  $z|\varphi$, $|\varphi|=|f_2|$ and $\varphi\neq f_2$.
\end{tabbing}
\ec
\bpr  (a) $\Leftrightarrow$  (b). Proposition \ref{Lambda} gives $2\lambda_F=2|f_1|=|f_1|+(2-m-|f_2|)< 2-m$ if and only if $|f_1|<|f_2|$. 

(b) $\Leftrightarrow$  (c). We first show that $\neg$(c) $\Rightarrow$ $\neg$(b). Suppose that $\varphi\in\I_F\cap\Phi$, $|\varphi|=|f_1|$ but $\varphi\neq f_1$\,. 
 We will show that $d=|f_2|-|f_1|\leq 0$. We can write $\varphi=a_1f_1+a_2f_2$ where $a_1\neq0$ since $z\nmid\varphi$. We have
$|f_1|=|\varphi|=|a_1|+|f_1|$ so $a_1\in\sbbk^\times$ and $a_1=1$ since $\varphi,f_1\in\Phi$ and $z|f_2$. If $a_2=0$ then $\varphi=f_1$\,, so $a_2\neq 0$ and by Corollary \ref{ha}, we have $0\geq -|a_2|=|a_1|-|a_2|=d$ as claimed.  Next we show that $\neg$(b) $\Rightarrow$ $\neg$(c). Let $d\leq 0$ and $\varphi=f_1-f_2\,z^{-d}$.  Then $\varphi\neq0$ since $z\nmid f_1$.  It is elementary that $\varphi$ is a monic form and that $|\varphi|=|f_1|$. 

(c) $\Leftrightarrow$  (d). Assume (c). Since (b) $\Leftrightarrow$  (c), we have $d>0$ and hence $m\leq-1$. Put $\varphi=f_1z^d-f_2$.  If $F$ is geometric then by Corollary \ref{geom}, $\varphi=(x-rz)z^d-z^{1-m}$ and $\LM(\varphi)=xz^d$.  Otherwise Theorem \ref{linked!} gives $\varphi=f_1z^d-f_1'z^{m'-m}$ and $\LM(\varphi)=x^{\lambda_m}z^d$ since $|f_1|=\lambda_m>\lambda_{m'}=|f_1'|$. So $\varphi$ is monic, $\varphi\neq f_2$, $|\varphi|=|f_2|$ and (c) $\Rightarrow$ (d). 

We conclude by showing that  $\neg$(c) $\Rightarrow$ $\neg$(d) i.e. if $d=|f_2|-|f_1|\leq 0$ then $f_2$ is unique. Since (b) $\Leftrightarrow$  (c) this will show that (d) $\Rightarrow$  (c). Let $\varphi\in\I_F^\times$, $z|\varphi$ and $|\varphi|=|f_2|$. To see that $\varphi=f_2$, first write $\varphi=a_1f_1+a_2f_2$ for some forms $a_1,a_2$  via Corollary \ref{ha}. If $a_1\neq0$ then
$|f_2|=|\varphi|=|a_1|+|f_1|$, so $|a_1|=d\leq0$ i.e. $a_1\in\sbbk$. But if $a_1\in\sbbk^\times$ then $z\nmid \varphi$. Thus $a_1=0$, $\varphi=a_2f_2$\,, $a_2\neq0$ and $|f_2|=|\varphi|=|a_2|+|f_2|$. Thus $a_2\in\sbbk^\times$. In fact $a_2=1$ since $\varphi$ and $f_2$ are monic. We have shown that $\varphi=f_2$ and the proof is complete. \epr

 \section{Groebner Bases}
 Our derivation of a minimal homogeneous Gb for $\I_F$ is more technical than proving that $f$ is viable for $\I_F$ and proceeds via a number of stages. We
 
 (i)  inductively construct a 'form vector' $\F$ for $\I_F$

 (ii)  use  auxiliary triples from Subsection 4.7 to relate $\F$  to the vectors for $\I_{F^{(i)}}$ where $m<i\leq \vv$; this uses an auxiliary vector of degrees $\D=\D_F$ 
 
 (iii)  derive a syzygy for $(\F_1,\F_2,\F_3)$ from the construction of Theorem \ref{template} and show that $\rem_{\F\,}\Spol(f_1,f_2)=0$ 
 
 (iv)  relate the $\Spol(\F_j,\F_k)$ to previous S polynomials for $1\leq j<k\leq |\F|$, the length of $\F$, 
and  deduce that $\rem_{\F\,}\Spol(\F_j,\F_k)=0$ for $1\leq j<k\leq |\F|$
 
(v) prove that $\F$ is a minimal homogeneous Gb for $\I_F$\,,  Theorem \ref{Gb!}.
 
In the final Subsection \ref{RGbsub} we show how to modify the construction of $\F$ and Algorithm  \ref{calPA} to obtain an algorithm which outputs the RGb $\ol{\F}$ of $\I_F$\,. 

 All Gr\"obner bases (Gb's) are with respect to grlex, written $\succ$.  We will say that a finite subset $\mathsf{G}$ of $\R^\times$   is  a Gb for $I=\langle \mathsf{G}\rangle$ if each element of $\mathsf{G}$ is  monic and   $\langle\LT(I)\rangle=\langle \LT(\mathsf{G})\rangle$.

Let $\mathsf{G}$ be a row vector of non-zero elements of $\R$.  We write $\rem\varphi=\mathrm{rem}\,_\mathsf{G}\,\varphi$ for the remainder on dividing $\varphi\in\R^\times$ by $\mathsf{G}$, and $\mathrm{rem}\,_\mathsf{g}\,\varphi$ if $\mathsf{G}=\mathsf{\{g\}}$. {\em This remainder is unique} as we always work with $\mathsf{G}$ in the fixed order of its components. Further, $\rem\varphi=0$ means that $\varphi=\sum_i a_i \mathsf{G}_i$ where  $\ee(\varphi)\succeq \ee(a_i\mathsf{G}_i)$ if $a_i\neq 0$. If $M$ is a monomial and $\rem \varphi=0$ then $\rem M\,\varphi=0$ and it follows from the uniqueness of $\mathsf{G}$-remainders  that if $\rem \varphi=\rem \varphi'=0$ then $\rem(\varphi+\varphi')=0$. 

The {\em syzygy polynomial} (on the leading monomials) of $\mathsf{G}_i\,,\mathsf{G}_j$ is 
$$\Spol(\mathsf{G}_i,\mathsf{G}_j)=\left(\frac{\mathsf{L}}{\LT(\mathsf{G}_i)}, 
-\frac{\mathsf{L}}{\LT(\mathsf{G}_j)}\right)\cdot\, (\mathsf{G}_i,\mathsf{G}_j)=\frac{\mathsf{L}}{\LT(\mathsf{G}_i)}\mathsf{G}_i 
-\frac{\mathsf{L}}{\LT(\mathsf{G}_j)}\mathsf{G}_j$$
where $\mathsf{L}=\lcm(\LM(\mathsf{G}_i),\LM(\mathsf{G}_j))$; $\Spol(\mathsf{G}_i,\mathsf{G}_j)$ is a form if $\mathsf{G}_i$ and $\mathsf{G}_j$ are. 
If $\rem_\mathsf{G} \Spol(\mathsf{G}_i,\mathsf{G}_j)=0$ for $1\leq i< j\leq |\mathsf{G}|$  then $\mathsf{G}$ is a Gb and conversely, \cite[Theorem 2.6.6]{IVA}, which is Buchberger's criterion. 

First note that there are two simple cases: (i) if $F=x^m$, then $(x^{1-m},z)$ is a Gb and (ii) if $G$ is geometric, then $(x-r\,z,z^{1-m})$ is likewise. A non-zero discrepancy may still  yield a Gb; see Example \ref{011} for instance, where $(x^{1-m}-z^{1-m},z^2)$ yields a Gb.

 \be\label{geomGb} \rm{Let $m\leq-1$ and $F$ be geometric, so that  $f=(x-r\, z,z^{1-m})$ is viable and a Gb for $\I_F$\,. Suppose that  $G= a\sm F$ is not geometric. Then $d=|f_2|-|f_1|=-m>0$, $q=\Delta(f_1;G)\neq 0$ and $g=(x^{-m}f_1-q\,z^{1-m},\ f_1\,z)$ is viable for $\I_G$ by Corollary \ref{geom}. We have  $\Spol(g_1,g_2)=\Spol(x^{-m}f_1-q\,z^{1-m},\ f_1z)= -q\, z^{2-m}$,  so that $g$ is not a Gb for $\I_G$. However, $z^{2-m}=f_2\,z$ and this example shows that if  we append $f_2\,z$ i.e. the 'intermediate or discarded form' $f_2$ multiplied by $z$,  then $\G=(g_1,g_2,f_2\,z)$ is a Gb for $\I_G$\,. For $m=-3$ and $r=0$, we obtain
 $\G=(x^3-z^3,xz,z^4)$; see Figure 2.}
 \eex

 \subsection{The  Form Vector $\F$}\label{G_F}
In this subsection we inductively construct a 'form vector' for $\I_F$\,, inserting the most recent viable pair {\em on the left}. In this way we extend Theorem \ref{indthm} to accumulate a form vector $\F$  with $\F_1=f_1$ and $\F_2=f_2$\,. We have $\langle \F\rangle=\I_F$ since $\I_F=\langle f_1,f_2\rangle\subseteq\langle \F\rangle\subseteq \I_F$\,.

We proceed as follows.  First our inductive basis: if $F=x^m$ then $\F=(x^{1-m},z)$ and $\ee(\F)=\{(1-m,0),(0,1)\}$.  See the set of exponents $\ee(\F)$ in Figure 1 for $m=-2$. 
 
As above, there are  inductive steps I, II: let $f$ be viable for $\I_F$, the form vector $\F$ be defined,  $c=|\F|$ be the {\em length} of $\F$ and  $\F_i$ be monic for $1\leq i\leq c$. If  $G= a\sm F$ and $\Delta(f_1;G)=0$,  we put  $\G_1=\F_1$ and $(\G_2,\ldots,\G_c)=(\F_2\,z,\ldots,\F_c\,z)$ so that $|\G|=|\F|$, as per Proposition \ref{Delta=0}. Clearly the elements of $\G$ are monic.
Inductive step II uses Theorem \ref{template}.

 \bd \label{templateGb}{\rm Suppose that $\F$ has been defined, $c=|\F|\geq2$ and  $\F_i$ is monic for $1\leq i\leq c$.  For  $G= a\sm F$ and $q\neq0$, define  the $\R$ vector $\G$ by $\G_1=f_1\ominus f_2$ and for $2\leq i\leq c$
  $$\G_i= \left\{\begin{array}{ll}
 \F_i\,z&\mbox{ if }d\leq 0 \\
 \F_{i-1}\,z& \mbox{ otherwise}\end{array}\right.
$$
so that   $|\G|=|\F|$ if $d\leq0$ and  $|\G|=|\F|+1$ if  $d>0$.}
\ed

The components of $\G$ are monic since  $\G_1=f_1\ominus f_2$ of Theorem \ref{template} is monic and the components of $\F$ are monic. If $F=x^m$ or $F$ is geometric then $|\F|=2$. 

 \be\label{GbADex4} {\rm  For $F=x^{-6}z^{-1}+x^{-4}z^{-3}+x^{-3}z^{-4}+z^{-7}$  we obtain   the following table, which should be correlated with  Example \ref{ADex4}. Here  $B=(q\neq0)\,\&\,(d>0)$ as before, taking values 0,1. It tells us how  $\F$ changes  to become $\G$ for $\I_G$ where $G=a\sm F$. Let $|\F|=c$. When $B=0$, $\G_1$ is $\F_1$ or is new, the rest of $\G$ is $(\F_2 z,\ldots,\F_cz)$ and  $|\G|=|\F|$.  But when $B=1$,  $\G_1$ is new,  the rest of $\G$ is $(\F_1 z,\ldots,\F_cz)$ and $|\G|=|\F|+1$. Note that the components of $\F$ are neither increasing nor decreasing with respect to $\succ$. Figure 3  corresponds to $m=-3$.}
\begin{center}
\begin{tabular}{|r|c|l|}\hline
$F$ & $B$    &$\F$  \\\hline\hline
$1$   &$0$& $(x,z)$  \\\hline
$z^{-1}$   &$0$& $(x,z^2)$   \\\hline
$z^{-2}$   &$1$ &$(x,z^3)$  \\\hline
$x^{-3}+z^{-3}$   &$0$& $(x^3+z^3,xz,z^4)$    \\\hline
$x^{-4}+x^{-3}z^{-1}+z^{-4}$     &$0$& $(x^3+x^2z+z^3,xz^2,z^5)$ \\\hline
$x^{-4}z^{-1}+x^{-3}z^{-2}+z^{-5}$     &$1$& $(x^3+x^2z+xz^2+z^3,xz^3,z^6)$ \\\hline
$x^{-6}+x^{-4}z^{-2}+x^{-3}z^{-3}+z^{-6}$    &$0$& $(x^4+x^3z+x^2z^2,(x^3+x^2z+xz^2+z^3)z,xz^4,z^7)$ \\\hline
$x^{-6}z^{-1}+x^{-4}z^{-3}+x^{-3}z^{-4}+z^{-7}$     &$1$& $(x^4+xz^3+z^4,(x^3+x^2z+xz^2+z^3)z^2,xz^5,z^8)$. \\\hline
\end{tabular}
\end{center}
\eex

To obtain an upper bound on $|\F|$, we appeal to Theorem \ref{template}. 
\bc \label{card} We have  $|\F|\leq \lambda_F+1$.
\ec
\bpr  For $F=x^m$, $(x^{1-m},z)$ is viable for $\I_F$ and $|\F|=2\leq 2-m=\lambda_F+1$.
Suppose  that the result is true for $F$, $f$ is viable for $\I_F$ and  $G= a\sm F$ has   vector $\G$. If $\Delta(f_1;G)=0$ then $|\G|=|\F|\leq \lambda_F+1=\lambda_G+1$. 
Now let $\Delta(f_1;G)\neq0$. If $d\leq 0$, then $|\G|=|\F|\leq \lambda_F+1=\lambda_G+1$; if $d>0$ then  $|\G|=|\F|+1\leq (\lambda_F+1)+1\leq (\lambda_F+d)+1=\lambda_G+1$.
\epr

We will call   $\lambda_m,\ldots,\lambda_0$ the {\em $\lambda$ profile of $F$}. 
For instance, in Example \ref{ADex4}, the $\lambda$ profile of $F$ is  $4,4,3,3,3,1,1,1$. 
An easy induction shows that  $|\F|-1$ is the number of  decrements in $\lambda_m,\ldots,\lambda_0$ (we also count $\lambda_0=1$ as a decrement)\,.  
In analogy with sequences, we will say that $F$ has a {\em perfect $\lambda$ profile (P$\lambda$P)} if for $m\leq i\leq 0$, $\lambda_i=\lfloor 1-\frac{i}{2}\rfloor$. Equivalently, $\lambda_0=1$ and for $m\leq i\leq -1$, $\lambda_i=\lambda_{i+1}$ if  $i$ is odd and $\lambda_i=\lambda_{i+1}+1$ if $i$ is even. Proposition \ref{stable} implies that if $F$ has a P$\lambda$P then $\Lambda_F=(1-\frac{m}{2})^2$ if $m$ is even and $\Lambda_F=\frac{(1-m)(3-m)}{4}$ if $m$ is odd.

\bp  If $F$ has a P$\lambda$P then $|\F|=\lambda_F+1$.
\ep
\bpr Let $F$ have a P$\lambda$P. If $m=0$, $|\F|=2=\lambda_F+1$ as $\I_F=\m$. Inductively, suppose that $|\F|=\lambda_F+1$ if $F$  has a P$\lambda$P. Let $G= a\sm F$ have a P$\lambda$P. Then $F$ also has a P$\lambda$P, so  $|\F|=\lambda_F+1$. If $m-1$ is odd then $\lambda_F$ has increased, so $|\G|=|\F|+1$ and $\lambda_G=\lambda_F+1$ since $G$ has a P$\lambda$P, so $|\G|=|\F|+1=(\lambda_F+1)+1=\lambda_G+1$. But if $m-1$ is even then $\lambda_G=\lambda_F$ and $|\G|=|\F|$ as $\lambda_F$ has not increased, so  $|\G|=|\F|=\lambda_F+1=\lambda_G+1$. 
\epr

 \subsection{The Degree Vector  $\D$}
 Our second step is to apply the previous subsection to obtain  closed-form expressions for $\F$ and $\ee(\F)$ using a row vector $\D$. The iterative construction of $\D$ is similar to that of  $\F$.  
  
 Let $m=|F|\leq 0$ as usual. The first two components of $\F$ are the viable pair for $\I_F$ by construction. When  $F$ is essential with triple $T'=(m',f_1',\Delta_1')$, we rewrite $f_2$ of Theorem \ref{linked!}  as   $f_2^{(m)}=f_1^{(m')}z^{m'-m}$ instead of  $f_2=f_1'z^{m'-m}$.  We now inductively generalise  $m'$ to all of $\F$ using  $F^{(i)}$ for $m<i\leq\vv$, {\em writing  the viable pair for $\I_{F^{(i)}}$ as $f^{(i)}=(f_1^{(i)},f_2^{(i)})$}. We begin with a detailed example.
 \be\label{Table3}  {\rm We   want to write 
 $\F=(x^4+xz^3+z^4,(x^3+x^2z+xz^2+z^3)z^2,xz^5,z^8)$ of Example \ref{GbADex4} as $$(f_1^{(\D_1)}z^{\D_1-m}, f_1^{(\D_2)}z^{\D_2-m},f_1^{(\D_3)}z^{\D_3-m},f_1^{(\D_4)}z^{\D_4-m}).$$ 
We have $|\F|=4$ and $\lambda_{-7}=4>\lambda_{-5}=3>\lambda_{-2}=1$ so $m'=-5$. Firstly, $\D_1$ is the current degree $m$. Notice that the power of $z$ dividing $f_2^{(\D_1)}=f_1^{(\D_2)}\,z^{\D_2-m}$ is $2=m'-m$, as expected from Theorem \ref{linked!}. Next $f_1^{(\D_3)}z^{\D_3-m}=xz^5$, so $\D_3=-2$ and $5=-2-m$. 
For $\D_4$\,, it is convenient to put $\D_4=1=\vv+1$ and $f_1^{(\D_4)}=1$ (as per Proposition \ref{dovetail}) so that $f_1^{(\D_4)}\,z^{\D_4-m}=z^8=z^{1-m}$.  This gives  
$\D=(-7,-5,-2,1)$. The required $\D$ are:}

\begin{center}
\begin{tabular}{|r|l|r|}\hline
$m$     &$f$ & $\D$ \\\hline\hline
$0$   & $(x,z)$  & $(0,1)$ \\\hline
 $-1$ & $(x,z^2)$ & $(-1,1)$  \\\hline
  $-2$ &$(x,z^3)$ & $(-2,1)$ \\\hline
$-3$  & $(x^3+z^3,xz,z^4)$   & $(-3,-2,1)$ \\\hline
$-4$  & $(x^3+x^2z+z^3,xz^2,z^5)$  & $(-4,-2,1)$\\\hline
$-5$     & $(x^3+x^2z+xz^2+z^3,xz^3,z^6)$  & $(-5,-2,1)$\\\hline
$-6$   & $(x^4+x^3z+x^2z^2,(x^3+x^2z+xz^2+z^3)z,xz^4,z^7)$ & $(-6,-5,-2,1)$ \\\hline
$-7$    & $(x^4+xz^3+z^4,(x^3+x^2z+xz^2+z^3)z^2,xz^5,z^8)$ & $(-7,-5,-2,1)$. \\\hline
\end{tabular}
\end{center}
  \eex

We would like to have a similar description when $m=\vv<0$ or when $F$ is geometric. First let $F=x^m$. Then $\F=(x^{1-m},z)=(f_1^{(m)}, f_2^{(m)})
 =(f_1^{(m)}z^{\D_1-m},f_1^{(\D_2)}z^{\D_2-m})$ so we take $f_1^{(\D_2)}=1$ and $\D_2=m+1=\vv+1$. 
 
Similarly  when  $F$ is geometric. We need $\D_2=1=\vv+1$ and $f_1^{(1)}=1$ as well in this case so that 
$$\F=(x-r\,z,z^{1-m})=(f_1^{(m)}, f_2^{(m)})
 =(f_1^{(\D_1)}z^{\D_1-m}, f_1^{(\D_2)}z^{\D_2-m}).$$
 
Thus for our inductive basis $F=x^m$, we put $\D=(m,m+1)=(m,\vv+1)$ and $f_1^{(\D_2)}=1$. 
  Inductively, let  $f$ be viable for $\I_F$\,, $\D$ be defined and $|\D|=|\F|$. If $G= a\sm F$ and $\Delta(f_1;G)=0$, we define $\calE=\calE_G$  by
$ \calE=(m-1,\D_2,\ldots,\D_{|\F|})$ as per  Proposition \ref{Delta=0}. We have 
$|\calE|=|\D|=|\F|=|\G|$.
 \bd \label{templateN}{\rm Suppose that $\F$ has been defined.  For  $G= a\sm F$ and $q\neq0$, define $\calE=\calE_G$ by $\calE_1=m-1$ and for $2\leq j\leq |\F|$
  $$\calE_j= \left\{\begin{array}{ll}
 \D_j&\mbox{ if }d\leq 0 \\
 \D_{j-1}& \mbox{ otherwise.}\end{array}\right.
$$
}
\ed

In other words, $\calE=(m-1,\D_2,\ldots,\D_{|\F|})$ if $d\leq0$ and $\calE=(m-1,\D_1,\ldots,\D_{|\F|})$ otherwise. The reader may check that   this inductive definition  yields $\D$ for Example \ref{Table3}. Note also that if $F=x^m$ or $F$ is geometric then  $|\F|=2$ and  $\D_2=\D_1'$.
 \be \label{Table2} {\rm Suppose that $m\leq-1$ and $F$ is geometric but $G=a\sm F$ is not, so that 
 $\lambda_{m-1}=1-m>1$ from Corollary \ref{geom}. We have $\D=(m,1)$ and $d>0$ so 
$\calE=\calE_G=(m-1,m,1)$ as in Figure 2. However, if $G$ were geometric, we would have $\calE=(m-1,1)$.}
\eex
\begin{figure}\label{fig1}
\caption{$\ee(\F)$ for $F=x^{-2}$ and $\D=(-2,-1)$.}
\begin{center}
\setlength{\unitlength}{0.75cm}
\begin{picture}(1,2)
\thicklines
\put(0,2){\line(0,-2){2}}
\put(0,0){\line(4,0){4}}
\put(0,1){\circle*{0.25}}
\put(1,1){\circle{0.2}}
\put(2,1){\circle{0.2}}
\put(3,0){\circle*{0.25}}
\put(0.05,1.05){$^{(0,1)=\ee(f_2)=(0,\,\D_2-m)}$}
\put(3.05,0.05){$^{(3,0)=\ee(f_1)=(3,\,\D_1-m)}$}
\put(0,1){\line(3,0){3}}
\put(3,1){\line(0,-1){1}}
\end{picture}
\end{center}
\end{figure}

\begin{figure}\label{fig2}
\caption{$\ee(\G)$ for $G=x^{-3}-z^{-3}$ and $\calE=(-3,-2,1)$ for Example \ref{geomGb}.}
\begin{center}
\setlength{\unitlength}{0.75cm}
\begin{picture}(1,5)
\thicklines
\put(0,5){\line(0,-5){5}}
\put(0,0){\line(4,0){4}}
\put(1,1){\circle*{0.25}}
\put(2,1){\circle{0.2}}
\put(1,2){\circle{0.2}}
\put(3,0){\circle*{0.25}}

\put(0,4){\circle*{0.25}}
\put(1,3){\circle{0.2}}
\put(0.05,4.05){$^{(0,4)=\ee(\G_3)=(\lambda_{\calE_3},\,\calE_3-m)}$}
\put(1.05,1.05){$^{(1,1)=\ee(\G_2)=(\lambda_{\calE_2},\,\calE_2-m)}$}
\put(3.05,0.05){$^{(3,0)=\ee(\G_1)=(\lambda_{\calE_1},\,\calE_1-m)}$}

\put(0,4){\line(1,0){1}}
\put(1,4){\line(0,-3){3}}
\put(1,1){\line(2,0){2}}

\put(3,1){\line(0,-1){1}}
\end{picture}
\end{center}
\end{figure}

The vector $\D$ satisfies the following properties. 
\bp \label{calN} We have $m=\D_1<\cdots<\D_j<\cdots<\D_{|\F|}=\vv+1$\,. If $F$ is essential then 

(i) $\lambda_{\D_1}>\cdots>\lambda_{\D_j}>\cdots >\lambda_{\D_{|\F|}}=0$

(ii) $\lambda_{\D_j-1}>\lambda_{\D_j}$

(iii)  $(\D_j-1)'=\min\{i: \lambda_{\D_j-1}> \lambda_i\}=\D_{j}$   for $2\leq j\leq |\F|$.
\ep
 
\begin{figure}\label{fig3}
\caption{$\ee(\F)$ and  $\D=(-7,-5,-2,1)$ for Example \ref{ADex4}.}
 \begin{center}
\setlength{\unitlength}{0.75cm}
\begin{picture}(1,9)
\thicklines
\put(0,9){\line(0,-9){9}}
\put(0,0){\line(5,0){5}}
\put(4,1){\circle{0.2}}
\put(3,3){\circle{0.2}}
\put(3,4){\circle{0.2}}
\put(2,5){\circle{0.2}}
\put(1,6){\circle{0.2}}
\put(1,7){\circle{0.2}}
\put(0.05,8.05){$^{(0,8)=\ee(\F_4)=(\lambda_{\D_4},\D_4-m)}$}
\put(1.05,5.05){$^{(1,5)=\ee(\F_3)=(\lambda_{\D_3},\D_3-m)}$}
\put(3.05,2.05){$^{(3,2)=\ee(\F_2)=(\lambda_{\D_2},\D_2-m)}$}
\put(4.05,0.05){$^{(4,0)=\ee(\F_1)=(\lambda_{\D_1},\D_1-m)}$}

\put(0,8){\line(1,0){1}}
\put(0,8){\circle*{0.25}}
\put(1,8){\line(0,-3){3}}
\put(1,5){\line(2,0){2}}
\put(1,5){\circle*{0.25}}
\put(3,5){\line(0,-3){3}}
\put(3,2){\line(1,0){1}}
\put(3,2){\circle*{0.25}}
\put(4,2){\line(0,-2){2}}
\put(4,0){\circle*{0.25}} 
\end{picture}
\end{center}
\end{figure}

\bt \label{exp} If $F\neq0$ has form vector $\F$,  vector $\D$ and $c=|\F|=|\D|$ then 
$$\F=(f_1^{(\D_1)}z^{\D_1-m},\ldots,f_1^{(\D_c)}z^{\D_c-m})$$ and  $\ee(\F_i)=(\lambda_{\D_i},\D_i-m)$ for $1\leq i\leq |\F|$.
 \et
\bpr  
 This is a simple verification if $F$ is geometric or $F=x^m$.  We prove the result for essential $F$ by induction on $m\leq-1$.
Inductively, let $F$ be essential, $c=|\F|$, $\D=(\D_1,\ldots,\D_c)$ and $\F=(f_1^{(\D_1)}z^{\D_1-m},\ldots,f_1^{(\D_c)}z^{\D_c-m})$. Suppose that  $G= a\sm F$,  $n=m-1$ and  $d=|\G|$. If $\Delta(f_1;G)=0$ then  $\calE=(n,\D_2,\ldots,\D_c)$ and by construction
\begin{eqnarray*}
 \G&=& (f_1^{(n)},f_1^{(\D_2)}z^{\D_{2}-m}\,z,\ldots, f_1^{(\D_c)}z^{\D_c-m}\,z)\\&=&(f_1^{(\calE_1)}z^{\calE_1-n},f_1^{(\calE_{2})}z^{\calE_{2}-n},\ldots, f_1^{(\calE_c)}z^{\calE_c-n}).\end{eqnarray*}
 Now suppose that $\Delta(f_1;G)\neq 0$ and  $f_1^{(n)}=f_1^{(m)}\ominus f_2^{(m)}$,  $f_2^{(n)}=(f_1^{(m)}\,z)\curlyvee (f_2^{(m)}\,z)\in\I_G$ are defined as in Theorem \ref{template}.  If $d\leq 0$, the proof is similar to the previous case. On the other hand, if $d>0$   then  $d=c+1$, $\calE_1=n$, $(\calE_{2},\ldots,\calE_d)=(\D_1,\D_2,\ldots,\D_c)$ and 
\begin{eqnarray*}
 \G
 &=& (f_1^{(n)},f_1^{(\D_1)}z^{\D_1-m}z,\ldots,f_1^{(\D_c)}z^{\D_c-m}z )\\
 &=&(f_1^{(\calE_1)}z^{\calE_1-n},f_1^{(\calE_{2})}z^{\calE_{2}-n},\ldots,f_1^{(\calE_d) }z^{\calE_d -n}).
 \end{eqnarray*}
Secondly, $z\nmid f_1^{(\D_i)}$ and $\ee(\F_i)=(|f_1^{(\D_i)}|,\D_i-m)=(\lambda_{\D_i},\D_i-m)$. 
\epr
{\em Note to the reader:}  

(i) Since $\lambda_{\D_{i-1}}>_\mathrm{lex}\lambda_{\D_i}$ for $2\leq i\leq c=|\F|$, we are justified in writing $\F$ as a lex-ordered $c$-tuple $(f_1^{(\D_1)}z^{\D_1-m},\ldots,f_1^{(\D_c)}z^{\D_c-m})$. 

(ii) In Figures 1--4,   $\ee(\F)$ is in boldface. The number of points of $\N^2$ strictly below the 'frontier of minimal cardinality of $\I_F$' coincides with  $\Lambda_F$\,, which agrees with Corollary  \ref{minGb} below.

\subsection{A Syzygy and  $\rem\Spol(f_1,f_2)$}
 It is easy to see that  if $F$ is geometric or $F=x^m$  and $f$ is viable for $\I_F$ then $\rem\Spol(f)=0$. 
For $|\F|\geq 3$,  we derive a  syzygy from Theorem \ref{indthm} and show that  $\rem\Spol(f_1,f_2)=0$. 

For fixed $e\geq 1$ and $v\in\R^e$,  the  {\em $\R$-module of syzygies} of $v$ is  
$\Syz(v)=\{w\in\R^e:  w\cdot v=0\}$ where $w\cdot v=\sum_{j=1}^e\, w_j\,v_j$\,. For example if $F$ is geometric,  $\D=(m,1)$  and $(z^{\D_2-m}, -x+r\,z)\in\Syz(x-r\,z,z^{1-m})$ as $(z^{\D_2-m}, -x+r\,z)\cdot(x-r\,z,z^{1-m})=z^{1-m}(x-r\,z)+(-x+r\,z)z^{1-m}=0$. If  $m\leq0$ and $F=x^m$ then  $\D=(m,m+1)$ and $(z^{\D_2-m}, -x^{1-m})\in\Syz(x^{1-m},z)$.  
\bl\label{finally} Let $|\F|\geq 3$ and $p=\lambda_m-\lambda_{m'}>0$. Then (i) there  are explicit $q_{m'}\in\sbbk^\times$ and $h_F\in z\,\R$ such that
$$(z^{m'-m},-x^p+h_F,q_{m'})\in\Syz(\F_1, \F_2,\F_3)$$
 and (ii)  if $ h_F\neq 0$ then $\ee( h_F\,\F_2)\succ\ee(\F_3)$.
\el
\bpr  (i) If $F=x^m$ or $F$ is geometric then $|\F|=2$ so we know that $F$ is essential with $T'=(m',f_1',\Delta_1')$ and that $p>0$. Let $F'=F^{(m')}$, $\lambda=\lambda_m$,  $\lambda'=\lambda_{m'}$, $m''=\D_3$ and $\lambda''=\lambda_{m''}$\,. We will determine $\F_1,\F_2,\F_3$ explicitly.

Since  $(f^{(m')}_1,f^{(m')}_2)$ is viable for $\I_{F'}$ and $2-m'-\lambda'=\lambda$ from Theorem \ref{linked!}, we have
$$d_{m'}=d_{f'}=|f_2^{(m')}|-|f_1^{(m')}|=(2-m'-\lambda')-\lambda'=\lambda-\lambda'=p>0$$  and $q_{m'}=q_{f'}=\Delta(f_1^{(m')};G')/\Delta(f_2^{(m')};G')\in\sbbk^\times$ where $G'=F_{m'-1}\sm F'=F^{(m'-1)}$.  Theorem \ref{template} applied to $F'$ gives
\begin{eqnarray}\label{firsteqn} f^{(m'-1)}_1=f^{(m')}_1\ominus_{d_{m'},\,q_{m'}} f^{(m')}_2=x^{d_{m'}}f^{(m')}_1-q_{m'}\,f^{(m')}_2=x^pf^{(m')}_1-q_{m'}\,f^{(m')}_2\in\I_{F^{(m'-1)}}
\end{eqnarray}
and $f_2^{(m'-1)}=f_1^{(m')}z$.
  If $m\leq m'-2$, Theorem \ref{template} implies that  for some $d_{m'-1}\leq0$ and $q_{m'-1}\in\sbbk$
$$f^{(m'-2)}_1=f^{(m'-1)}_1\ominus_{d_{m'-1},\,q_{m'-1}}f_2^{(m'-1)}=f^{(m'-1)}_1-q_{m'-1}x^{-d_{m'-1}}(f_1^{(m')}\,z)\in\I_{F^{(m'-2)}}\,.$$ 
Iterating for $\lambda_{m'-2}=\cdots=\lambda$ gives
$$f^{(m'-i-1)}_1= f^{(m'-i)}_1-q_{m'-i}\,x^{-d_{m'-i}}f_1^{(m')}\,z^i\in\I_{F^{(m'-i-1)}}\mbox{ for }1\leq i\leq m'-m-1$$
and using Equation (\ref{firsteqn})
\begin{eqnarray}\label{secondeqn}\F_1=f_1^{(m)}=f^{(m'-1)}_1- h_Ff_1^{(m')}=x^pf^{(m')}_1-q_{m'}\,f^{(m')}_2- h_Ff_1^{(m')}
\end{eqnarray}
where $ h_F=\sum_{i=1}^{m'-m-1}q_{m'-i}\,x^{-d_{m'-i}}\,z^i\in z\,\R$ (and  $ h_F=0$ if 
$m=m'-1$). We have (a) $\F_2=f_2^{(m)}=f_1^{(m')}z^{m'-m}$. Now $f_2^{(m')}=f_1^{(m'')}z^{m''-m'}$ from  Theorem \ref{exp} and (b) $f_2^{(m')}z^{m'-m}=(f_1^{(m'')}z^{m''-m'})z^{m'-m}=f_1^{(m'')}z^{m''-m}=\F_3$.   Therefore Equation (\ref{secondeqn}) and (a), (b)  yield
\begin{eqnarray*}
(z^{m'-m},-x^p+ h_F,q_{m'})\cdot(\F_1,\F_2,\F_3)&=&z^{m'-m}\F_1-x^p\F_2+h_F\F_2+ q_{m'}\F_3\,\\
&=&z^{m'-m}[x^pf^{(m')}_1-q_{m'}\,f^{(m')}_2- h_F\,f_1^{(m')}]\\
&&-x^pf_1^{(m')}z^{m'-m}+ h_Ff_1^{(m')}z^{m'-m}+\,q_{m'}\F_3\\
&=&-q_{m'}z^{m'-m}f^{(m')}_2+q_{m'}\F_3=0
\end{eqnarray*}
and $(z^{m'-m},-x^p+h_F,q_{m'})\in\Syz(\F_1, \F_2,\F_3)$ as claimed.

(ii) If $ h_F\neq 0$, Equation (\ref{secondeqn}) implies that $\lambda=|h_Ff_1^{(m')}|$.
From the definition of $h_F$, $|h_F|=-d_{m'-i}+i$ where $d_{m'-i}\leq0$ and $1\leq i\leq m'-m-1$. 
Thus $\lambda=-d_{m'-i}+i+\lambda'$ and  $1\leq i= \lambda-\lambda'+d_{m'-i}\leq \lambda-\lambda'$. Now  $\F_2=f_1^{(m')}z^{m'-m}$ and so for some $i$, $1\leq i\leq \lambda-\lambda'$
$$\ee( h_F\,\F_2)=(-d_{m'-i},i)+(\lambda',m-m')=(\lambda-\lambda'-i,i)+(\lambda',m-m')=(\lambda-i,m-m'+i).$$ As $\lambda-i\geq \lambda'>\lambda''$ we conclude that  $\ee( h_F\,\F_2)\succ(\lambda'',m'-m)=\ee(\F_3)$.
\epr

\begin{figure}\label{fig4}
\caption{$\ee(\G)$ and $\calE_G=(-5,-4,0)$ for Example \ref{01101}.}
\begin{center}
\setlength{\unitlength}{0.75cm}
\begin{picture}(1,6)
\thicklines
\put(0,6){\line(0,-6){6}}
\put(0,0){\line(5,0){5}}
\put(2,1){\circle*{0.25}}
\put(4,0){\circle*{0.25}}
\put(2,2){\circle{0.2}}
\put(2,3){\circle{0.2}}
\put(2,4){\circle{0.2}}
\put(1,5){\circle{0.2}}
\put(3,1){\circle{0.2}}
\put(0,5){\circle*{0.25}}
\put(0.05,5.05){$^{(0,5)=\ee(\G_3)=(\lambda_{\calE_3},\,\calE_3-n)}$}
\put(2.05,1.05){$^{(2,1)=\ee(\G_2)=(\lambda_{\calE_2},\,\calE_2-n)}$}
\put(4.05,0.05){$^{(4,0)=\ee(\G_1)=(\lambda_{\calE_1},\,\calE_1-n)}$}

\put(0,5){\line(2,0){2}}
\put(2,5){\line(0,-4){4}}
\put(2,1){\line(4,0){2}}

\put(4,1){\line(0,-1){1}}
\end{picture}
\end{center}
\end{figure}

\be\label{01101}{\rm Consider $\sbbk=\GF(2)$ and $F=x^{-4}+x^{-2}z^{-2}+x^{-1}z^{-3}$. We obtain   $(x^2,z)$, $(x^2+xz,z^2)$, $(f_1,z^3)$ and $f=(f_1,z^4)$ for $\I_F$\,,  where $f_1=x^2+xz+z^2$. Since $\vv=-1$,  $\D_F=(-4,0)$. 
Now let  $n=-5$ and $G=0\sm F$. Since $d=2$ and $\Delta_1=1$,   $g_1=x^2f_1+f_2=x^4+x^3z+x^2z^2+z^4$ and $(g_2,g_3)=(f_1 z,f_2z)$ giving $\G=(x^4+x^3z+x^2z^2+z^4,x^2z+xz^2+z^3,z^5)$ for $\I_G$ and vector $\calE=(-5,-4,0)$. See Figure 4. In   Theorem \ref{finally} applied to $\G$, $n'=-4$ and $p=\lambda_{-5}-\lambda_{-4}=2$. This gives $h_F=0$, so that 
$(z^{n'-n},x^p,1)=(z,x^2,1)\in\Syz(\G_1,\G_2,\G_3)$ as the reader may verify.  } 
\eex

\bt\label{spol} If  $f$ is viable for $\I_F$ then 
$\rem\Spol(f)=0$.
\et
\bpr If $F$ is geometric, then $f=(x-r\,z,z^{1-m})$ and $\rem\Spol(f)=\rem(-r\,z^{2-m})=0$; if $F=x^m$ then   $f=(x^{1-m},z)$ and $\Spol(f)=0$. Thus   we can assume that $|\F|\geq 3$. Let $ h_F$ and $q_{m'}$ be as in Lemma \ref{finally}, $\lambda=\lambda_m$, $m'=\D_2$, $\lambda'=\lambda_{m'}$ and $p=\lambda-\lambda'$\,. Then $\LM(f_1)=x^{\lambda}$,  $\LM(f_2)=x^{\lambda'}z^{m'-m}$ and $\lambda>\lambda'$, so that $\lcm(\,\LM(f_1),\LM(f_2)\,)=x^\lambda z^{m'-m}$. Lemma \ref{finally} yields
$$\Spol(f)=( z^{m'-m},-x^p)\cdot f=- h_F\,\F_2-q_{m'}\F_3\,.$$ 
If $ h_F=0$ we are done and if $ h_F\neq 0$ Lemma \ref{finally} implies that $\ee\,\Spol(f)=\ee( h_F\,\F_2)\succ \ee(q_{m'}\F_3)$ and hence $\rem\Spol(f)=0$.
\epr
\subsection{A Minimal Gb}
We  finally show that $\F$ is a minimal Gb for $\I_F$ using two technical lemmas on S polynomial remainders. As an application, we determine $\dim_\sbbk(\R/\I_F)$.

Recall that for $1\leq j<k\leq  |\F|$, we have $\D_j<\D_k\leq1$ and $\lambda_{\D_j}>\lambda_{\D_k}$\,.
\bl \label{Gbindstep} For $1\leq j<k\leq |\F|$,  put $K=\D_k$, $J=\D_j$ and $J'=\D_{j+1}$. 

(i) 
$\Spol(\F_j,\F_k)=z^{K-J}\F_j-x^{\lambda_J-\lambda_K}\F_k$ 

(ii) if   $|\F|\geq 3$ then
$$\Spol(\F_1,\F_{j+1})=z^{J'-J}\Spol(\F_1,\F_{j})+x^{\lambda_m-\lambda_{J}}\,\Spol(\F_{j},\F_{j+1})$$
and if $\rem\Spol(\F_1,\F_{j}),\rem\Spol(\F_{j},\F_{j+1})=0$ then $\rem\Spol(\F_1,\F_{j+1})=0$.
\el
\bpr We have $ m\leq J<K\leq 1$ and $\lambda_J>\lambda_K$\,.  Hence $$M=\lcm(\LM(\F_j),\LM(\F_k))=\lcm(x^{\lambda_J}z^{J-m},x^{\lambda_K}z^{K-m})=x^{\lambda_J}z^{K-m}$$
 $M/\LM(\F_j)=z^{K-J}$ and  $M/\LM(\F_k)=x^{\lambda_J-\lambda_K}$. This now gives $\Spol(\F_j,\F_k)$. From (i),  $\Spol(\F_1,\F_{j})=z^{J-m}\F_1-x^{\lambda_m-\lambda_{J}}\F_{j}$ and
$ \Spol(\F_j,\F_{j+1})=z^{J-J'}\F_j-x^{\lambda_{J}-\lambda_{J'}}\F_{j+1}$. 
Expanding and cancelling $z^{J-J'}x^{\lambda_m-\lambda_{J'}}\F_{j+1}$ we obtain
\begin{eqnarray*}\Spol(\F_1,\F_{j+1})&=& z^{J'-m}\F_1-x^{\lambda_m-\lambda_{J'}}\F_{j+1}=z^{J'-J}[z^{J-m}\F_1] -x^{\lambda_m-\lambda_{J}}[x^{\lambda_{J}-\lambda_{J'}}\F_{j+1}]\\
&=&z^{J'-J}[\Spol(\F_1,\F_{j})+x^{\lambda_m-\lambda_{J}}\F_{j}]-x^{\lambda_m-\lambda_{J}}[z^{J'-J}\F_{j}-\Spol(\F_j,\F_{j+1})]\\
&=&z^{J'-J}\Spol(\F_1,\F_{j})+x^{\lambda_m-\lambda_{J}}\Spol(\F_j,\F_{j+1})\end{eqnarray*}
and by hypothesis, each summand has remainder zero. Hence $\rem\Spol(\F_1,\F_{j+1})=0$.
\epr
The second is an analogue of Theorem \ref{exp} for our S forms.

\bl \label{guts}  For fixed $j$,   $2\leq j<k\leq |\F|$, let $F'=F^{(\D_j)}$  with vectors $\F^\prime,\D^\prime$. Then

(i) $\Spol(\F_j,\F_k)=\Spol(\F'_1,\F'_{k-j+1})z^{\D_j-m}$

(ii) if $\rem_{\F'}\Spol(\F'_1,\F'_{k-j+1})=0$ then  $\rem_\F\Spol(\F_j,\F_k)=0$.
\el
\bpr Put $c=|\F|\geq3$, $n=\D_1'$ and $c'=|\F'|$ so that  $\F'_{l'}=f_1^{(\D'_{l'})}z^{\D'_{l'}-n}$ for $1\leq l'\leq c'$ from Theorem \ref{exp}. Since $F=F_m\sm\cdots\sm F_{j-1}\sm F'$ we have $\D_i=\D'_{i-j+1}$  and therefore $f_1^{(\D_i)}=f_1^{(\D'_{i-j+1})}$ for $j\leq i\leq c$ and $c'=c-j+1$.
Hence for $j\leq i\leq c$
$$\F_i=f_1^{(\D_i)}z^{\D_i-m}=f_1^{(\D'_{i-j+1})}z^{\D_i-m}=(\F'_{i-j+1}z^{n-\D'_{i-j+1}})
\,z^{\D_i-m}=\F'_{i-j+1}z^{\D_j-m}$$
since $n=\D_j$\,.
 Thus 
$\Spol(\F_j,\F_k)=\Spol(\F'_1\,z^{\D_j-m},\F'_{k-j+1}\,z^{\D_j-m})$ which is easily seen to be $\Spol(\F'_1,\F'_{k-j+1})z^{\D_j-m}.$
(ii) By hypothesis, we can write $\Spol(\F'_1,\F'_{k-j+1})=\sum_{l'=1}^{c'}a_{l'}\,\F'_{l'}$  where $\ee\,\Spol(\F'_1,\F'_{k-j+1})\succeq \ee(a_{l'}\,\F'_{l'})$ if $1\leq l'\leq c'$ and $a_{l'}\neq 0$. Now  $\F'_{l'}z^{\D_j-m}=\F_{l'+j-1}$ for $1\leq l'\leq c'$, so from (i) 
\begin{eqnarray*}\Spol(\F_j,\F_k)&=&\left(\sum_{l'=1}^{c'}a_{l'}\,\F'_{l'}\right)z^{\D_j-m}=\sum_{l'=1}^{c'}a_{l'}\,(\F'_{l'}\,z^{\D_j-m})=
\sum_{l'=1}^{c'}a_{l'}\,\F_{l'+j-1}=\sum_{l=j}^c\,a_{l-j+1}\,\F_l\,.
\end{eqnarray*}
Suppose that $j\leq l\leq c$ and $a_{l-j+1}\neq0$. Then  $1\leq l'=l-j+1\leq c'$ and from (i) 
$$\ee\,\Spol(\F_j,\F_k)=\ee\,(\Spol(\F'_1,\F'_{k-j+1})z^{\D_j-m})\succeq
\ee(a_{l'}\,\F'_{l'}z^{\D_j-m})=\ee\,(a_{l-j+1}\,\F_l).$$
From the  expression for $\Spol(\F_j,\F_k)$, we therefore have  $\rem_\F\,\Spol(\F_j,\F_k)=0$. 
\epr

Recall that   $\mathsf{G}$ is  a {\em minimal  Gb for $\langle \mathsf{G}\rangle$} if it is a Gb and for $1\leq i\neq j\leq |\mathsf{G}|$, $\LM(\mathsf{G}_i)$ does not divide $\LM(\mathsf{G}_j)$. 
We now prove our version of \cite[Theorem 3]{AD}.
\bt \label{Gb!}  The vector $\F$  is  a minimal homogeneous  Gb for $\I_F$\,. 
\et
\bpr  We know that $\F=(x^{1-m},z)$ is a Gb for $\I_F$ if $F=x^m$ and $\F=(x-r\,z,z^{1-m})$ is a Gb for $\I_F$ if $F$ is geometric.
We need to show that if $|\F|\geq3$ then $\rem_\F\,\Spol(\F_j,\F_k)=0$ for $1\leq j<k\leq |\F|$.  First of all, $\rem\Spol(\F_1,\F_2)=0$ by Theorem \ref{spol}. Suppose for the moment that $\rem\Spol(\F_j,\F_{j+1})=0$ for $2\leq j< |\F|$.
Then   by  Lemma
\ref{Gbindstep}, $\rem\Spol(\F_1,\F_{3})=0$ and again, $\rem\Spol(\F_1,\F_{4})=0$. Iterating, $\rem\Spol(\F_1,\F_{j+1})=0$ for $2\leq j<|\F|$.  

To see that  $\rem\Spol(\F_j,\F_{j+1})=0$ for $2\leq j< |\F|$, we assume inductively that   $F'=F^{(\D_j)}$ and $\F'$   is a Gb  for $\I_{F'}$ with  $|\F'|=|\F|-j+1<|\F|$. Since $(\F'_1,\F'_2)$ is viable for $\I_{\F'}$\,, $\rem_{\F'}\,\Spol(\F'_1,\F'_2)=0$  by 
 Theorem \ref{spol}. Hence  by Lemma \ref{guts}, $\rem_\F\,\Spol(\F_j,\F_{j+1})=0$.

Now let $2\leq j<k\leq |\F|$. Inductively, $\rem_{\F'}\,\Spol(\F'_1,\F'_{k-j+1})=0$ since $\F'$ is a Gb for $\I_{\F'}$ and so $\rem_\F\,\Spol(\F_j,\F_{k})=0$ by Lemma \ref{guts}. We now have   $\rem_\F\,\Spol(\F_j,\F_k)= 0$  for $1\leq j<k\leq |\F|$ and by  \cite[Theorem 2.6.6]{IVA}, $\F$ is a Gb, homogeneous as each $\F_j$ is a form. 
  
 For minimality, let $1\leq j<k\leq |\F|$, so that $\D_j<\D_k$ and $\lambda_{\D_j}>\lambda_{\D_k}$\,. From Theorem \ref{exp}, $\ee(\F_j)=(\lambda_{\D_j},\D_j-m)$. Hence if $\LM(\F_j)|\LM(\F_k)$ then $\lambda_{\D_j}\leq \lambda_{\D_k}$ and if $\LM(\F_k)|\LM(\F_j)$ then $\D_k-m\leq \D_j-m$, both of which are impossible.  
\epr 

\bc  Let $\F$, $\D$ be the vectors of forms and (total) degrees for $F$. Then

(i) for any $\varphi\in\R^\times$ there is a unique remainder on dividing $\varphi$ by $\F$, no matter how the elements of $\F$ are ordered

(ii) if $\mathsf{G}$ is any minimal grlex Gb for $\I_F$ then $\ee(\mathsf{G})=\{(\lambda_{\D_i},\D_i-m):1\leq i\leq |\F|\}$ and in particular, $|\mathsf{G}|=|\F|\leq \lambda_F+1$. 
\ec
\bpr Apply  Theorems \ref{exp}, \ref{Gb!},  \cite[Proposition 2.1.6]{IVA} and \cite[Exercise 7, p. 94]{IVA}. 
\epr

Recall that $\Lambda_F=\sum_{i=m}^\vv\lambda_{F^{(i)}}$.
\bc \label{minGb} If  $f$ is viable for $\I_F$ then $\dim_\sbbk(\R/\I_F)=\Lambda_F=|f_1|\,|f_2|$.
\ec
\bpr    It is an elementary  exercise that the first equality is true if $F$ is geometric or $F=x^m$.  Inductively,  assume that the result is true for $F$. It is well-known that   $\B_F=\{x^i\,z^j+\I_F\ :  (i,j)\not\in\ee(\I_F)\}$  is a $\sbbk$ basis for $\R/\I_F$, see \cite[Proposition 5.3.4]{IVA} and that $\ee(\I_F)=\ee(\F)+\N^2$ since $\F$ is a Gb for $\I_F$.   Let  $G= a\sm F$ and $T_\F$ be the translate $\N^2+(\lambda_F,1)$. We have 
$\ee(\G)+\N^2= ((\lambda_G,0)+\N^2)\,\cup\, (\ee(\F\,z)+T_\F)$ for all $d\in\Z$ by Theorem \ref{exp}. Denoting disjoint union of sets by '$\sqcup$'
\begin{eqnarray*}\N^2\setminus\ee(\I_G) &=&
([0,\lambda_G-1]\times\N)\,\cap\,[(\N\times\{0\})\sqcup(\ee(\F\,z)+T_\F)]\\
&=&([0,\lambda_G-1]\times\{0\})\,\sqcup \, [T_\F\setminus(\ee(\F\,z)+T_\F)].
\end{eqnarray*}
The inductive hypothesis now gives $|\B_G|=\lambda_G+|\B_F|=\lambda_G+\Lambda_F=\Lambda_G$\,.
The second equality  follows  from Corollary \ref{ha} since $f$ is  viable for $\I_F$\,. 
\epr

In Figures 1--4,  projecting  onto the first coordinate illustrates  $|\F|\leq \lambda_m+1$, which was proved in  Corollary \ref{card}. For $1\leq j\leq |\F|-1$, the LC increments $\lambda_{\D_j}-\lambda_{\D_{j+1}}$ appear as  lengths of  horizontal line segments and on the vertical line segments,  $\lambda_{\D_j}$ appears $(\D_{j+1}-m)-(\D_j-m)=\D_{j+1}-\D_{j}$ times, as expected from  Theorem \ref{exp}. 
   
\subsection{The Reduced Gb $\ol{\F}$}\label{RGbsub}
It is evident that we can extend Algorithm \ref{calPA} to accumulate the  form vector $\F$ of $F$, which will be a minimal Gb for $\I_F$ by Theorem \ref{Gb!}.  In this subsection we show how to modify  this vector version of Algorithm \ref{calPA} to obtain the RGb of $\I_F$\,. 
 Algorithm 1 of \cite{AD} was not extended to compute the RGb of $\I_s$  in \cite{AD}. 

Recall that (i) a minimal Gb $\mathsf{G}$ is {\em reduced} if for all $\mathsf{g}\in\mathsf{G}$, no monomial of $\mathsf{g}$ is in $\langle \LT(\mathsf{G}\setminus\mathsf{g})\rangle$ and (ii) 
 $\langle \mathsf{G}\rangle$ has a unique (grlex) reduced Gb,  which we write as $\ol{\mathsf{G}}$. The standard method for obtaining $\ol{\mathsf{G}}$ is to successively replace $\mathsf{G}$ by $(\rem_{\mathsf{G}\setminus\mathsf{g}}\,\mathsf{g})\cup(\mathsf{G}\setminus\mathsf{g})$,  \cite[p. 92]{IVA}.  We say that a monomial $M$ {\em can be reduced by} $\mathsf{G}$ or $\mathsf{G}$ {\em reduces} $M$ if $\LM(\mathsf{g})|M$ for some $\mathsf{g}\in\mathsf{G}$.

 Thus if $F$  has form vector $\F$, we can always construct the RGb $\ol{\F}$ of $\I_F$ according to the  method of \cite{IVA}. However, we will see that this method can be considerably improved in  our case: it will suffice to replace $f_1$ by $\rem_{f_2}\,f_1$\,. 
 
 For example, let  $F=x^{-1}+z^{-1}$. Then $\F=(x-z,z^2)$, which is reduced. Let $a\neq 1$ and $G=a\sm F=ax^{-2}+x^{-1}z^{-1}+z^{-2}$. From Corollary \ref{geom} with $q=a-1\neq0$,  $g=(xf_1-q\,z^2,f_1z)=(x^2-xz-qz^2,xz-z^2)$  and $\G=(x^2-xz-qz^2,xz-z^2,z^3)$ as in Example \ref{geomGb}. We have $\rem _{g_2}\,g_1=x^2-az^2$ and $(x^2-az^2,xz-z^2,z^3)$ is reduced.
 
Likewise, for Example \ref{ADex4} with $F=x^{-6}+x^{-4}z^{-2}+x^{-3}z^{-3}+z^{-6}$
 $$f=(x^4+x^3z+x^2z^2,(x^3+x^2z+xz^2+z^3)z)$$  and  $\rem _{f_2}\,f_1=x^4+xz^3+z^4$. Now $\rem _{f_2}\,f_1=f^{(8)}_1$ and $\Delta(\rem _{f_2}\,f_1;0\sm F)=0$. We will see that the form vector $\F$ for $F$ is an RGb.

 The proof of the following lemma adapts an inequality from  \cite[Lemma 7]{AD}. 
 \bl\label{j=2}
 If  $f$ is viable for $\I_F$\,,  $|\F|\geq 3$ and for some $j>1$, $\F_j$ reduces a monomial $M$ of $f_1$  then $j=2$. 
 \el
\bpr
 Since  $m\geq|\F|\geq 3$, $F$ is necessarily essential. We will show that $j\geq 3$ gives a contradiction. We have $z|M$ since $\LM(\F_j)|M$ and $j>1$.   As $f_1$ is a form of total degree $\lambda_m$\,, 
$M=x^{\lambda_m-p}z^p$ for some $p$, $1\leq p\leq \lambda_m$ by Theorem \ref{exp}. If $\D$ is the degree vector then $\LM(\F_j)=x^{\lambda_{\D_j}}z^{\D_j-m}$ by Theorem \ref{exp}. 
Hence if $\LM(\F_j)|M$, we have $\lambda_{\D_j}\leq \lambda_m-p$ and
$\D_j-m\leq p$.  From Theorem \ref{linked!} we have $\lambda_m=2-m'-\lambda_{m'}$. 

Now $m'=\D_2<\D_j$ since $j\geq 3$, so $m'\leq \D_j-1$ and hence $\lambda_{m'}\geq\lambda_{\D_j-1}$ as $\lambda$ is non-decreasing.
Also from Proposition \ref{calN}, $(\D_j-1)'=\min\{i:  \lambda_{\D_j-1}>\lambda_i\}=\D_j$. Hence  $\lambda_{\D_j-1}=2-(\D_j-1)'-\lambda_{(\D_j-1)'}=2-\D_j-\lambda_{\D_j}$ from Theorem \ref{linked!} and $-\lambda_{m'}\leq -\lambda_{\D_j-1}= -(2-\D_j-\lambda_{\D_j})$. Therefore
\begin{eqnarray*}
\lambda_{\D_j}&\leq& \lambda_m-p=(2-m'-\lambda_{m'})-p\\
&\leq& 2-m'-\lambda_{\D_j-1}-p=2-m'-(2-\D_j-\lambda_{\D_j})-p=-m'+\D_j+\lambda_{\D_j}-p\\
&<&-m+\D_j+\lambda_{\D_j}-p=\lambda_{\D_j}+(\D_j-m-p)\leq \lambda_{\D_j}
\end{eqnarray*}
the last line since $m<m'$ and $\D_j-m\leq p$. 
\epr

\bt \label{RGb} (Inductive RGb) Let  $f$ be viable for $\I_F$, $\F$ be the form tuple for $F$ and assume that $\F$ is reduced. Let $G= a\sm F$,  $g$ be viable for $\I_G$ and $\G$ be the form tuple for $G$. Then 

(i)  if $\Delta(f_1;G)= 0$ then $\G$ is reduced 

(ii)  $\G$  is  reduced if and only if $g_1=\rem_{g_2}\,g_1$

(iii)  if $|g_2|> |g_1|$ then $\G$ is reduced.
\et
\bpr  Let $\Delta_1=\Delta(f_1;G)$ and $c=|\F|$. (i) If $\Delta_1=0$ then $\G=(\F_1,\F_2z,\ldots,\F_{c}z)$. Since $\langle \LM(\F_i)\,z:i>1\rangle\subset\langle \LM(\F_i):i>1\rangle$ and $\F$ is reduced, a monomial of $\G_1=\F_1$ cannot be reduced.  Likewise, if $2\leq i\leq |\G|=|\F|$\,,  $\langle \LM(\F_1),\{\LM(\F_j)z: j\neq i\}\rangle\subset\langle \LM(\F_1),\LM(\F_j): j\neq i\}\rangle$ so no monomial of $\G_i=\F_iz$ can be reduced.

 (ii) It suffices to show that if $\G$ is not reduced then a monomial of $g_1$ lies in $\langle\LT(g_2)\rangle$.  From (i) we have $\Delta_1\neq 0$, so either (a) $d\leq 0$, $\LM(g_1)=\LM(f_1)$  and $\G=(g_1,\F_2z,\ldots,\F_{c}z)$ or (b) $d>0$,  $\G=(g_1,\F_1z,\ldots,\F_{c}z)$ and $\LM(g_1)=x^d\,\LM(f_1)$. As before, no term of $\F_i\,z$ lies in  $\langle \LM(\F_j)z: j\neq i\rangle$ since $\F$ is reduced. Suppose that for some $i$, $2\leq i\leq |\G|$\,, we could reduce a term of $\F_i\,z$  by $\LT(g_1)$. Since $\R$ has unique factorisation and $z\nmid g_1$, a term of $\F_i$ would be reducible by either $\LM(g_1)=\LM(f_1)$ or by $x^d\,\LM(f_1)$ i.e. be reducible by $\LM(f_1)$ and so $\F$ would not be reduced. Thus if $\G$ is not reduced, a monomial $M$ of $g_1$ can be reduced by $\G_j$ for some $j>1$. If $|\G|=2$ we are done and if $|\G|\geq 3$ then Lemma \ref{j=2} implies that $j=2$.

(iii) From (ii) if $\G$ is not reduced then $g_1$ can be reduced by $g_2$ and so  $|g_2|\leq |g_1|$.
\epr

\be\label{ADEx1} {\rm (\cite[Example 1]{AD}) Let $\sbbk=\Q$ and $F=x^{-1}+2z^{-1}$ so $f=(x-\frac{1}{2}z,z^2)$ is viable for $\I_F$ and $\F=f$ is reduced. If $G=2\sm F=2x^{-2}+x^{-1}z^{-1}+2z^{-2}$, we obtain $\Delta_1=\frac{3}{2}$, $\Delta_2=2$, $q=\frac{\Delta_1}{\Delta_2}=\frac{3}{4}$ and $d=1$ so $g_1=xf_1-q f_2=x^2-\frac{1}{2}\,xz-\frac{3}{4}z^2$, $(g_2,g_3)=(f_1z,f_2z)=(xz-\frac{1}{2}z^2,z^3)$ and  $\G=(g_1,g_2,g_3)$ is a Gb for $\I_G$\,. Since  $g_1$ reduces by $g_2$ we have $\ol{\G}=(\rem _{g_2}\,g_1, g_2,g_3)=(x^2-z^2, xz-\frac{1}{2}z^2,z^3)$.}
\eex
\be {\rm Let $\sbbk=\GF(2)$ and $G=x^{-4}z^{-1}+x^{-2}z^{-3}+x^{-1}z^{-4}$   as in Example \ref{01101}, where we found $\G=(x^4+x^3z+x^2z^2+z^4,x^2z+xz^2+z^3,z^5)$ which is not reduced. We have  $d=-1<0$ and $\ol{\G}=(\rem _{g_2}\,g_1, g_2,g_3)=(x^4+xz^3+z^4,x^2z+xz^2+z^3, z^5)$.}
\eex
\bc\label{RGbT} For any  $F$ we can construct  the unique RGb $\ol{\F}$ of $\I_F$\,. 
\ec
\bpr  It is immediate that  $\F$ is reduced  if $F$ is geometric or $F=x^m$. Suppose inductively that  $F$ is essential, the form vector $\F$ for $\I_F$  is reduced and $G= a\sm F$\,. If $\Delta(f_1;G)=0$ then $\G$ is already reduced by Theorem \ref{RGb}, so assume that $\Delta(f_1;G)\neq 0$. If $\G$ is not reduced then $d\leq 0$ and we replace $g_1$ by  $\rem_{g_2}\,g_1$ as per Theorem \ref{RGb}. Then $\ol{\G}$ is reduced by Theorem \ref{RGb}. Finally,  $(\rem_{g_2}\,g_1,g_2)$ is a suitable input for Theorem \ref{template} (which does not require that  the input viable pair be  constructed via  Theorem \ref{template}):  it is viable for $\I_G$ and $g_2=g_1'\,z^{m'-m}$ still obtains. This completes the induction. 
\epr

We conclude with the RGb-version of  Algorithm \ref{calPA}, computing $\ol{\F}$ by accumulating all of $\F$ and including   '{\tt if} $(d\leq 0)$ {\tt then} $\F_1\la \rem _{\F_2}\,\F_1$' at the end of each iteration. 

 \begin{algorithm}\label{RGbA}(RGb for $\I_F$)
\begin{tabbing}
\noindent {\tt Input}:\ \ \ \ \ \ \=  Inverse form $F\neq0$.\\

\noindent {\tt Output}: \> The unique reduced (grlex) Gr\"obner basis $\ol{\F}$ of $\I_F$\,.\\\\

$\lceil$  $i\la 1$; {\tt repeat}  $i\la i-1$   {\tt until} $(F_i\neq0)$; $\vv\la i$; $ \F\la (x^{1-\vv},z)$; $\Delta_2\la 1$;\ $d \la  \vv$;\ $c\la|\F|$; \\\\

$G\la x^\vv$;\\
{\tt for} \= $i\la \vv-1$ {\tt downto }$|F|$ {\tt do}\\\\

  \>$\lceil$\ $G\la F_i\sm G$; $\Delta_1\la \Delta(\F_1;G); q\la \Delta_1/\Delta_2$\,;\\\\  
   \>{\tt if}\ $(\Delta_1\neq 0)$\ {\tt then} \={\tt if} \=$(d\leq 0)$\ \={\tt then}\ $  \F_1\la  \F_1-q\,  x^{-d}\, \F_2$\\\\
  \> \> \>\> {\tt else}
   $\lceil$ \= $\mathcal{T}_1\la \F_1$; $\F_1 \la  x^d\, \mathcal{T}_1-q\, \F_2$\,;   \\\\
 \> \> \>\>\>   $\Delta_2\la  \Delta_1$\,; $d \la  -d$; \\\\
  \> \> \>\>\>  $(\F_2,\ldots,\F_{c+1})\la \mathcal{T}$; $c\la c+1; \rfloor$\\\\
       \> $(\F_2,\ldots,\F_c)\la (\F_2,\ldots,\F_c)z$; $d  \la  1+d$;\\\\
       \>  {\tt if} $(d\leq 0)$ {\tt then} $\F_1\la \rem _{\F_2}\,\F_1;\rfloor$ \\\\
     {\tt return }$\ol{\F}\,.\rfloor$
\end{tabbing} 
  \end{algorithm}
\section{Finite Sequences}
We apply Sections 4,  5  to finite sequences. In particular, we derive a version of Algorithm \ref{calPA} for this case, namely Algorithm \ref{calPAs}. Then we recompute  two examples of ideal intersection using \cite[Section 5.3]{UAG}. Finally we show that dehomogenisation induces a one-to-one correspondence between the outputs of Algorithm \ref{calPAs} and the first components of \cite[Algorithm 4.12, normalised]{N15}.
\subsection{Notation and Applications}
We will write $s=s_0,\ldots,s_{n-1}$ for a  typical non-trivial finite sequence over $\sbbk$ with $n\geq 1$ terms\,\footnote{\ Writing $s=s_m,\ldots,s_0$ would be  more appropriate for a sequence $s$ with $1-m\geq1$ terms, but we follow conventional notation.} so that $s_i\in\sbbk$ for $0\leq i\leq n-1$.
For us, the inverse form of  $s$ is 
$$ F^{(s)}=\sum_{i=1-n}^0 s_{-i}x^iz^{1-n-i}\in\M^\times.$$
  Thus  $ F^{(0,\ldots,0,1)}=x^{1-n}$ if $s_{n-1}=1$ and $ F^{(1,0,\ldots,0)}=z^{1-n}$ where there are $n-1$ zeroes. It is clear that there is a one-to-one, order-reversing correspondence of (non-zero) inverse forms of total degree $m\leq0$ and non-trivial sequences of length $1-m\geq1$.   
  
    If $s,t$ are non-trivial we write $s\sim t$ if  $ F^{(s)}\sim F^{(t)}$\,, {\em $\I_s$ for the  annihilator ideal  $\I_{ F^{(s)}}$ and $\vv=\vv( F^{(s)})$.}
As $\I_s$ depends only on the equivalence class $\ol{F^{(s)}}$, we can assume that $s_{-\vv}=1$.
   
   Thus  $-\vv\geq 0$ is the  number of initial zeroes  of $s$,  $s_{-\vv}=1$, $s$ has $1-\vv$ terms and we can regard $s$ as being obtained from $0,\ldots,0,1$  by successively adding the $n+\vv-1\geq0$ terms $s_{1-\vv},\ldots,s_{n-1}$.  For $a\in\sbbk$ and $s$ non-trivial, we write $s,a$ for the sequence
   $s_0,\ldots,s_{n-1},a$. 
   
   A sequence  $s=0,\ldots,0,1$ is  called  an {\em  impulse response sequence (IRS)}. Thus $s$ is an IRS  of $n$ terms if and only if $ F^{(s)}=x^{1-n}$. In this case   $\I_s=\langle x^n,z\rangle$ by Proposition \ref{001}  and $\F^{(s)}=(x^n,z)$ by construction.
 From Theorem \ref{indthm} we have

\bc For any non-trivial $s$ we can compute a viable $f$ for $\I_s$\,.
\ec

 It is immediate that if $n>1$ then $s$ is geometric (as in high school) if and only if $ F^{(s)}$ is geometric. Further,  $s$ is geometric if and only if  $n>1$, $s_0=1$ and $(x-r\,z,z^n)$ is viable for $\I_s$\,, where $r=s_1$ and by construction $\F^{(s)}=(x-r\,z,z^n)$. 

For  non-trivial $s$, Algorithm \ref{calPA} applies as is, with $d=|f_2|-|f_1|$ unchanged. But if $s$ is trivial, neither loop of Algorithm \ref{calPA}  terminates. To avoid this  we (i)  initialise via $f\la(1,0)$, $\Delta_2\la 1$ and $d\la 1$ (ii) put  $\Delta(1;0)=0$ and $\Delta(1;x^{1-n})=1$  (iii) terminate after $n$ iterations. 
\begin{algorithm}\label{calPAs}(Algorithm \ref{calPA} modified for  sequences)
\begin{tabbing}
\noindent {\tt Input}: \ \ \=  Sequence $s=s_0,\ldots,s_{n-1}$.\\

\noindent {\tt Output}: \> Viable ordered pair $f$ for $ \I_s$ if $s$ is non-trivial and $f=(1,0)$ otherwise.\\\\

$\lceil$ $f\la (1,0)$;  $\Delta_2\la 1$;\ $d \la  1$;\\\\

{\tt for} \= $i\la 0$ {\tt to }$n-1$ {\tt do}\\\\

   \>$\lceil \ \Delta_1\la \Delta(f_1;F^{(s_0,\ldots,s_i)}); q\la \Delta_1/\Delta_2$;\\\\    \>{\tt if }$(\Delta_1\neq 0)$ {\tt then} \={\tt if} \=$(d\leq 0)$\ \={\tt then}\, $ f_1\la  f_1-q\,  x^{-d}\, f_2 $\\\\
  \> \> \>\> {\tt else}
   $\lceil$\= $T\la f_1$; $f_1 \la  x^d\, f_1-q\, f_2$; $f_2\la T$; \\\\
   \>\> \> \>\>$\Delta_2\la  \Delta_1$; $d \la  -d;\rfloor$\\\\
   \> $f_2\la f_2\,z$; $d  \la  1+d; \rfloor$\\\\
{\tt return }$f.\rfloor$
\end{tabbing} 
  \end{algorithm}
  If $s$ is trivial, $f=(1,0)$. If $s_0,\ldots,s_{i-1}=0,\ldots,0,1$  we obtain $d=i$, $f=(x^i\cdot 1- 1\cdot0,1)=(x^i,1)$, $f_2\la z$, giving $f=(x^i,z)$ and then $d=1-i\leq0$. If $i\leq n-1$, the next iteration gives $f=(x^i,z^2)$ if $\Delta_1=0$ and $f=(x^i-\Delta_1\, x^{i-1}z,z^2)$ otherwise.
Algorithm \ref{calPAs}  has been implemented in COCOA, \cite{COCOA}.
\be \label{101} {\rm  Let $\sbbk=\mathrm{GF}(2)$, $n\geq 3$ and $s=1,0,\ldots,0,1$. The pair $(x^{n-1}+z^{n-1},xz)$ is viable for $\I_s$\,. This example appeared in \cite[Example 2]{AD} for $n=8$.}
\eex

\be {\rm For $\sbbk=\GF(2)$ and $s=1,0,0,1,1,0,1,0$ we obtain $f=(x^4+xz^3+z^4,(x^3+x^2z+xz^2+z^3)z^2)$ from  Example \ref{ADex4}. This sequence is  \cite[Example 4]{AD}.}
\eex
\subsection{Annihilating Polynomials}
If $s$ is non-trivial then $| F^{(s)}|=1-n$ and Corollary \ref{basic} yields the following  characterisation of forms in $\I_s^\times$.
\bp \label{char} If  $s$ is non-trivial and $\varphi\in\R^\times$ is a  form  then $\varphi\in\I_s^\times$ if and only if $$\sum_{j=0}^{|\varphi|}\varphi_j\, s_{j-i}=0\ {\mathrm{for}}\ |\varphi|+1-n\leq i\leq 0.$$
\ep

Let $\psi\in\sbbk[x]$ be monic. We recall from  \cite[Definition 2.1]{N15} that $\psi$ is an {\em annihilating polynomial} of $s=s_0,\ldots,s_{n-1}$ if   $ \sum_{j=0}^{|\psi|}\psi_j\, s_{j-i}=0$ for $|\psi|+1-n\leq i\leq 0$. In particular, if $|\psi|\geq n$ then $\psi$ is vacuously an annihiliating polynomial of $s$; for example, $x^n$ is one.
It is notationally convenient to allow $0\in\sbbk[x]$ to be an annihilating polynomial. We write $\Ann_s$ for the {\em set of monic annihilating polynomials} of $s$. 
 Thus 
$$\Ann_s^\times=\{\psi\in\sbbk[x]^\times: \psi{\rm \ monic}\,,\sum_{j=0}^{|\psi|}\psi_j\, s_{j-i}=0\mbox{ for }|\psi|+1-n\leq i\leq 0\}\neq\emptyset.$$  
\bc \label{1-1corresp}For any $s$, we have a 1-1 correspondence $^\wedge : \Ann_s^\times\rightleftarrows \I_s\cap \Phi : \ ^\vee$.
\ec
\bpr  We know that $^\wedge : \sbbk[x]^\times\rightleftarrows  \Phi : \ ^\vee$ are  degree-preserving mutual inverses from Subsection \ref{homogsubsection}. Let $\psi\in\Ann_s^\times$ and put $\varphi=\psi^\wedge\in\Phi$.  We want to show that $\varphi\in\I_s$. Now $|\varphi|=|\psi|$ and if $|\psi|+1-n>0$ then  $|\varphi|+1-n>0$ and $\varphi\in\I_s$ by Lemma \ref{prebasic}, so  assume that $|\psi|+1-n\leq 0$. We want to show that $\varphi\in\I_s$ so let $|\varphi|+1-n\leq i\leq 0$. Now $\psi_j=\varphi_j$ for $0\leq j\leq |\varphi|=|\psi|$ so
$$[\varphi\cdot  F^{(s)}]_i= \sum_{j=0}^{|\varphi|}\,\varphi_j\, F^{(s)}_{i-j}=\sum_{j=0}^{|\varphi|}\,\varphi_j\, s_{j-i}=\sum_{j=0}^{|\psi|}\psi_j\, s_{j-i}$$ 
 which is zero since  $|\psi|+1-n\leq i\leq 0$   and $\psi\in\Ann_s^\times$. Hence  $\varphi\in\I_s$ by Proposition \ref{char}. Now let $\varphi\in\I_s\cap\Phi$ and put $\psi=\varphi^\vee$. Then $|\psi|=|\varphi|$. As before, we can assume that $|\varphi|+1-n\leq 0$. We want to show that $\psi\in\Ann_s^\times$ so let  $|\psi|+1-n\leq i\leq 0$. Then 
$$ \sum_{j=0}^{|\psi|}\,\psi_i\, s_{j-i}=\sum_{j=0}^{|\varphi|}\varphi_i\, s_{j-i}=[\varphi\cdot  F^{(s)}]_i$$
which is zero  by Proposition \ref{char} as  $|\varphi|+1-n\leq i\leq 0$. Thus $\psi\in\Ann_s^\times$.
\epr

Let $n>1$.  It is easy to check that if $s$ is geometric then  $x-s_1\in\Ann_s$\,. Conversely, let $s_0=1$ and  $x-r\in\Ann_s$ for some $r\in\sbbk$. From the definition of $\Ann_s$, $s_i=r s_{i-1}$ for $1\leq i\leq n-1$ i.e.  $s$ is geometric with $r=s_1$\,.
 
We recall from  \cite{N15} that the {\em linear complexity} of  $s$ can be defined as 
 $$\LC_s=\min\{|\psi|: \psi\in\Ann_s^\times\}$$
and $\mu\in\Ann_s^\times$ of degree $\LC_s$  was called a {\em  minimal polynomial (MP)} of $s$; such a  polynomial always exists since $\Ann_s^\times\neq\emptyset$ and $\sbbk$ is a field.  For example, $\mu=1$ is an MP of a trivial sequence; if $s$ is geometric  then $\mu=x-s_1$ is an MP of $s$  since $\mu\in\Ann_s^\times$\,, and $s_0=1$ implies that any element of $\Ann_s^\times$ has degree at least 1.\\

We now relate  $\ell_s$ and $\I_s$\,.

\bp \label{LC} For any sequence $s$

(i) $\LC_s=\lambda_{ F^{(s)}}$

(ii)  if  $s$ is  non-trivial  and $f$ is viable for $\I_s$ then  $\LC_s=|f_1|$.
\ep
\bpr (i) By definition, $\lambda_{ F^{(s)}}=\min\{|\varphi|:\ \varphi\in\I_s\cap\Phi\}$. If $s$ is trivial, then $\LC_s=0=\lambda_{ F^{(s)}}$\,. Suppose $s$ is non-trivial. If $\varphi\in \I_s\cap\Phi$ then $\varphi^\vee\in\Ann_s^\times$ by Proposition \ref{1-1corresp} so $|\varphi|=|\varphi^\vee|\geq \LC_s$ and $\lambda_{ F^{(s)}}\geq \LC_s$\,. Conversely,  if $\psi\in \Ann_s^\times$  then $\psi^\wedge\in\I_s\cap\Phi$ by Proposition \ref{1-1corresp},  $|\psi|=|\psi^\wedge|\geq \lambda_{ F^{(s)}}$ and so $\LC_s\geq  \lambda_{ F^{(s)}}$\,.
(ii) We have $\lambda_{ F^{(s)}}=|f_1|$ from Proposition \ref{Lambda}, so this follows from (i).
\epr

For example, if $s$ is an IRS then $\LC_n=n$  since $(x^n,z)$ is viable for $\I_s$; if $n>1$ and $s=1,0,\ldots,0$ then $\LC_s=1$ as $(x,z^n)$ is viable for $\I_s$\,. Indeed, $1,0\ldots,0$ is geometric. 
If $f$ is viable for $\I_s$  then $f_1^\vee$  is a minimal polynomial for $s$, for $f_1^\vee\in\Ann_s$ by Corollary \ref{1-1corresp} and $|f_1^\vee|=|f_1|=\LC_s$\,. 

To  simplify notation we will also write  $\LC_i$ for $\LC_{s_0,\ldots,s_{i-1}}$ where $1\leq i\leq n$\,. Since $s$ is geometric if and only if $ F^{(s)}$ is geometric, Proposition \ref{Lambdageom} gives

   \bp\label{simplegeom} The sequence $s$ is geometric $\Leftrightarrow$ $n>1$ and $\LC_n=\cdots=\LC_1=1$. 
\ep

We can also prove Proposition \ref{simplegeom} directly as it is well-known (and easy to see) that for any $s$, the function $i\mapsto \LC_i$ is non-decreasing for $1\leq i\leq n$\,.
From Theorem \ref{Gb!} if $s$ is non-trivial, the inverse form $ F^{(s)}$ has a form vector $\F^{(s)}$ which is a minimal Gb for $\I_s$\,, and if $s$ is an IRS or geometric then  $|\F^{(s)}|=2$ and $\F^{(s)}$ is reduced. 
\bc (\cite[Lemma 7]{AD}) Let $s$ be non-trivial with  form vector $\F^{(s)}$ and $|\F^{(s)}|\geq3$. If   $2\ell_n\leq n$ then $\F^{(s)}$ is reduced.
\ec
\bpr  We show that if $\F^{(s)}$ is not reduced then $2\ell_n>n$. From Theorem \ref{indthm}, there is a viable $f$ for $\I_s$ and by  Lemma \ref{j=2}, if $\F^{(s)}$ is not reduced then $f_1$ can be reduced by $f_2$ and hence $|f_1|\geq |f_2|$. Proposition \ref{LC} now implies that $ \ell_n=|f_1|\geq |f_2|=n+1-|f_1|=n+1-\ell_n$ and so $2\ell_n>n$.
\epr

\subsection{Several Sequences}
 For completeness, we recompute the two examples of ideal intersection   given in \cite{AD}. Let $s,t$ be non-trivial finite sequences, 
  $f$ be viable for $\I_s$ and  $g$ be viable for $\I_t$\,. 
We can obtain generators for $\I_s\cap\I_t$ as in \cite[Section 5.3]{UAG}:  put
$$M_{s,t}=\langle[1,\ 1]^\T, [f_1,\ 0]^\T, [f_2,\ 0]^\T, [0,\ g_1]^\T, [0,\ g_2]^\T\rangle.$$
We obtain generators for $\I_s\cap\I_t$ by  finding the generators of the syzygy module of  $M_{s,t}$ and then taking first coordinates. The following examples were computed using COCOA.
\be {\rm (\cite[Example 3]{AD})
 If $s=1,0,0,1,1,1$ and $t=1,0,0,0,1,0,0,1$, then Algorithm \ref{calPAs} gives $f=(x^3 + x^2z + z^3, xz^3)$ and  $g=(x^4 + xz^3 + z^4,  xz^4)$ so that  $$M_{s,t}=\langle[1, 1], [x^3 + x^2z + z^3, 0], [xz^3, 0], [0, x^4 + xz^3 + z^4], [0, xz^4]\rangle$$
\begin{eqnarray*}\Syz(M_{s,t})&=&\langle[xz^4, 0, z, 0, 1], [x^4z + xz^4 + z^5, xz + z^2, x, z, 0],\\ && [x^6 + x^3z^3 + x^2z^4, x^3 + x^2z + xz^2, x^2 + z^2, x^2, 0] \rangle
\end{eqnarray*}
and $\I_s\cap\I_t=\langle x^6 + x^3z^3 + x^2z^4, x^4z + xz^4 + z^5,xz^4\rangle=\langle \gamma_1,\gamma_2,\gamma_3\rangle$ say. 
From Corollary \ref{1-1corresp}, $\gamma_1^\vee=x^6+x^3+x^2\in\Ann_s\cap\Ann_t$\,. If $\psi\in (\Ann_s\cap\Ann_t)^\times$ then $\psi^\wedge\in\I_s\cap\I_t\cap\Phi$ and $\psi^\wedge=\sum a_i\gamma_i$  where  $a_1\neq 0$ and $z\nmid a_1$\,. Thus $|\psi|=|\psi^\wedge|=|a_1|+6\geq 6$. 

We remark that in {\em loc. cit.} the authors obtain  
$x^6+x^3\in \Ann_s\cap\Ann_t$. However, $[(x^6+x^3)\cdot(1+x^{-3}+x^{-4}+x^{-5})]_0=1$ so that by definition $x^6+x^3\not\in\Ann_s$\,.
}
\eex
\be {\rm (\cite[Example 4]{AD}) If $s=1,0,0,1,1,1,0,1$ and $t=1,0,0,1,1,0,1,0$ we obtain 
\begin{eqnarray*}\I_s\cap\I_t&=&\langle x^3 + x^2z + z^3, x^4,xz^5,z^8\rangle\cap\langle x^4 + xz^3+z^4,x^3z^2+x^2z^3+xz^4+z^5,xz^5,z^8\rangle\\
&=&\langle x^6+x^5z+x^2z^4+z^6,x^3z^3+x^2z^4+z^6,xz^5,z^8\rangle.
\end{eqnarray*}
As in the previous example, $x^6+x^5+x^2+1\in\Ann_s\cap\Ann_t$ and if $\psi\in(\Ann_s\cap\Ann_t)^\times$ then $|\psi|\geq 6$.  }
\eex

\brs {\rm (i) For $\I_s\cap\I_t$\,, we  compute syzygies in $\R^5$. If the Gb's  of $\I_s\,,\I_t$ computed in  \cite{AD} by Algorithm 1, {\em loc. cit.} have cardinality $c_s\,,c_t$ respectively then $\I_s\cap\I_t$ is computed using  syzygies in $\R^{c_s+c_t+1}$.
(ii) It is clear that we may use this method to find generators of $\I_F\cap\I_G$ for  inverse forms $F,G$.}
\ers

\subsection{A One-to-One Correspondence}
We will now apply the results so far and in particular Subsection 6.3 to show that dehomogenisation induces a one-to-one correspondence between  a viable pair  and the normalised output of \cite[Algorithm 4.12]{N15}, recalled here  as Algorithm \ref{calPA1}.

We will say that {\em $s$ is essential} if $F^{(s)}$ is essential. In particular, $n>1$. If $s$ is essential, a triple $T'$ for $s$ will mean a triple for $F^{(s)}$, except that we replace the first component  $T'_1$  by  $1-T'_1$ for consistency with \cite{N15}.  
From Example \ref{IRSex} we see that if $s$ is an IRS then $s$ is essential with triple $T'_s=(n-1,1,1)$. Suppose that  $s$ is geometric with $r=s_1$, but $t=s,a$ is not. From Example \ref{geomex}, Corollary \ref{geom} and Proposition \ref{LC},     $T'_t=(n,x-r\,z, a- r^n)$ is a  triple for $t$.
  Rewriting Definition \ref{tripled} for $F^{(s)}$ gives 
\bp \label{triples} If $s$ is essential, $T'$ is a triple for $s$ and $n'=1-T'_1$ then 

(i) $1\leq n'=\max_{1\leq i\leq n-1}\{i:\ell_i<\ell_s\}<n$

(ii) $T_2'\in\I_{s_0,\ldots,s_{n'-1}}\cap\Phi$  and $|T'_2|=\ell_{n'}$

(iii) $\Delta(T_2';F^{(s')})\in\sbbk^\times$ where $s'=s_0,\ldots,s_{n'}$

(iv) $\ell_s+\ell_{s'}=n'+1$.
\ep
\bpr Parts (i)-(iii) are straightforward applications of Proposition \ref{LC}. For part (iv),  Proposition \ref{LC} gives  $\ell_s+\ell_{s'}=\lambda_F+\lambda_{F'}$, which is $n'+1$ by Theorem \ref{linked!}.
\epr

Next we discuss \cite[Algorithm 4.12, normalised]{N15}. Firstly,  a sequence was called essential in \cite{N15} if $\LC_s>\LC_1$. This is consistent with $F^{(s)}$ being essential by Proposition \ref{LC}, and again if $n>1$, a non-trivial sequence is either geometric or essential. We will say that $\mu'$ is an {\em auxiliary polynomial} of $s$ if either (i) $s$ is trivial and $\mu'=0$ or (ii)  $s=1$ or $s$ is geometric and $\mu'=1$ or (iii)  $s$ is essential and $\mu'$ is an MP for $s_0,\ldots,s_{n'-1}$ where $T'$ is a triple for $s$ and $n'=T'_1$. 

The normalised version of \cite[Algorithm 4.12]{N15} returns 
 $(\mu,\mu')\in \sbbk[x]^2$ where $\mu$ is an MP  for $s$ and  $\mu'$ is an auxiliary polynomial for $s$. It uses $\Delta(1;0,\ldots,0,1)=1$ (see \cite[Note (iv), p. 448]{N15}) and from \cite[Definition 2.5]{N15},  the discrepancy for $\psi\in\sbbk[x]^\times$ and $s,a$ is 
 $$\Delta(\psi; s,a)=[\psi\cdot(ax^{-n}+s_{n-1}x^{1-n}+\cdots+1)]_{|\psi|-n}=\sum_{j=0}^{|\psi|} \psi_j\, s_{n+j-|\psi|}.$$

We have also made some trivial notational changes for easy comparison with Algorithm \ref{calPAs}: Algorithm \ref{calPA1} uses $n+1-2\LC_n\in\Z$ as in \cite[Definition 4.1]{N15} rather than $d=|f_2|-|f_1|$ when $f$ is viable for $\I_s$\,. But if $f$ is viable for $\I_s$ then  $n+1-2\LC_n=n+1-2\lambda_{F^{(s)}}=(n+1-|f_1|)-|f_1|=|f_2|-|f_1|=d$ by Propositions \ref{Lambda} and \ref{LC}, so we  use $d$ in both algorithms.

 \begin{algorithm}\label{calPA1}(\cite[Algorithm 4.12, normalised]{N15})
\begin{tabbing}
\noindent {\tt Input}: \ \ \=  Sequence $s=s_0,\ldots,s_{n-1}$\,.\\

\noindent {\tt Output}: \> $(\mu,\mu')\in\sbbk[x]^2$, $\mu$ an MP and $\mu'$ an   auxiliary polynomial  of $s$.\\\\

$\lceil\ (\mu,\mu')\leftarrow (1,0)$; $\Delta'\leftarrow1$;\ $d \leftarrow 1$;\\\\

{\tt for} \= $i\leftarrow0$ {\tt to }$n-1$ \\\\

   \>$\lceil \ \Delta\leftarrow\Delta(\mu;s_0,\ldots,s_i)$;  $q\la \Delta/\Delta'$;\\\\
    \>{\tt if}$\ \Delta\neq 0$ {\tt then}\ \={\tt if} \=$d\leq 0$\ \={\tt then}\, $\mu\leftarrow  \mu-q\,  x^{-d}\, \mu'; $\\\\
  \> \> \> {\tt else}\ \=
   $\lceil\psi\leftarrow\mu$; $\mu \leftarrow x^d\, \mu-q\, \mu'$; 
    $\mu'\la \psi$; $\Delta'\la \Delta$;\ $d \leftarrow -d;\rfloor$\\\\
   \>\  $d  \leftarrow d+1;\  \rfloor$\\\\
{\tt return }$(\mu,\mu').\rfloor$
\end{tabbing} 
\end{algorithm}

The reader cannot fail to notice the overall similarity of Algorithms \ref{calPAs} and \ref{calPA1} even though their derivations are different. Furthermore ignoring constant terms  and terms linear in $n$,  Algorithm \ref{calPA1} also requires at most $n^2/2$ multiplications in $\sbbk$ to compute $(\mu,\mu')$  by \cite[Corollary 4.17]{N15}.
Next the main theorem of this subsection. 
 \bt \label{dehomog} Let $s$ be non-trivial and $f$, $(\mu,\mu')$ be the pairs  output by Algorithm \ref{calPA},  Algorithm \ref{calPA1} respectively. There is a one-to-one correspondence  $$f\stackrel{\vee '}\longmapsto (f_1^\vee,f_2^\vee) \mbox{ and }(\mu,\mu')\stackrel{\wedge '}\longmapsto (\mu^\wedge,(\mu')^\wedge\, z^{n+1-|\mu|-|\mu'|}).$$
 In particular, $(f_1')^\vee=\mu'$ and $\I_s=\langle \mu^\wedge,(\mu')^\wedge\, z^{n+1-|\mu|-|\mu'|}\rangle$.
\et
\bpr If $n=1$ the  outputs are $(x,z)$, $(x,1)$ respectively and $n+1-|x|-|1|=1$. Let $n>1$. If 
 $s$ is geometric, the  outputs are $f=(x-r\,z,z^n)$, $(\mu,\mu')=(x-r,1)$ respectively. Further, $n+1-|\mu|-|\mu'|=n$. We omit the remaining trivial verifications for the one-to-one correspondence. For the remainder of the inductive proof, we assume that $s$ is essential with triple $T'$ and $n'=T'_1$.
 
We begin with Algorithm \ref{calPA}.  First the inductive basis. If $s_0=0$ we can assume that 
 $s$ is an IRS as it is essential by hypothesis. From direct computation  Algorithm \ref{calPA1} returns $(\mu,\mu')=(x^n,1)$ and from Proposition \ref{001}, we have $f=(x^n,z)$. We have $\vee'(x^n,z)=(x^n,1)$ and since $n+1-|\mu|-|\mu'|=1$,  $\wedge'(x^n,1)=(x^n,z)$. 
 
 For the other inductive basis, suppose that  $s_0=1$, $r=s_1$, $n>2$, $n'=n-2$ and $s'=1,\ldots,r^{n-2}$  but $s=s',s_{n-1}$ is not geometric i.e. $s_{n-1}\neq r^{n-1}$.  For Algorithm \ref{calPA1} with input $(x-r,1)$ 
 $$\Delta=[(x-r)\cdot(s_{n-1}x^{n-1}+r^{n-2}x^{n-2}+\cdots +1)]_{1-(n-1)}=s_{n-1}-r^{n-1}\neq 0.$$
Since  $\Delta'=1$ and $d=n-2>0$,  the next iteration gives $(\mu,\mu')=(x^{n-2}(x-r)- \Delta,x-r)$.  
 From Corollary \ref{geom},  $(x^{n-2}f_1- \Delta\,z^{n-1},f_1z)$ is viable for $\I_s$\,. Now $$\vee'(x^{n-2}f_1- \Delta\,z^{n-1},f_1z)=(x^{n-2}(x-r)- \Delta,x-r)=(\mu,\mu').$$ We also have $n+1-|\mu|-|\mu'|=n+1-(n-1)-1=1$ so that $\wedge'(\mu,\mu')$ is as required. This completes the proof for the inductive bases.
 
  For the inductive step, suppose that $f$ is viable for $\I_s$ and we have shown that the respective outputs correspond for $s$. Let $t=s,a$ and $e=|f_1|$. Then $|f_1^\vee|=e$ and $\Delta=\Delta(f_1^\vee;t)=[f_1^\vee\cdot (ax^{-n}+s_{n-1}x^{1-n}+\cdots+s_0)]_{e-n}$. 
   Therefore 
  $$\Delta(f_1;t)=[f_1\circ F^{(t)}]_{e-n}=\sum_{i=0}^e [f_1]_i\cdot t_{n+i-e}=\sum_{i=0}^e [f_1^\vee]_i\cdot t_{n+i-e}=\Delta.$$
 Dehomogenising is just evaluation at $z=1$, so the updating of $\mu$ and $f_1$ in each algorithm is consistent.   We also saw in Lemma \ref{ess} that $\Delta_2=\Delta(f_2;t)=\Delta(f_1';s',s_{n-1})=\Delta'$. Thus $\Delta/\Delta'=q$  and the updating of $\Delta_2$ and $\Delta'$ is consistent.   We note that dehomogenising the statement '$f_2\la f_2\,z$' in Algorithm \ref{calPAs} makes it redundant and $d$ is updated in precisely the same way in each algorithm.

For good measure, we include a proof of the inductive step beginning with the outputs of Algorithm \ref{calPA1}
for $s$  essential with $T_1'=n'$ and $s'=s_0,\ldots,s_{n'}$. 
 Write $\LC=\LC_s$ and $\LC'=\LC_{s'}$ so that $n'=\LC+\LC'-1$ from Proposition \ref{triples}. Put $\psi=\mu'$. 
Suppose inductively that if $(\mu,\psi)$ is the output for $s$ then  $f=\wedge'(\mu,\psi)=(\mu^\wedge,\psi^\wedge\,z^{n+1-\LC-\LC'})$ agrees with the output of Algorithm \ref{calPA}.  We want to show that we obtain the values of Theorem \ref{template} for $t=s,a$. 

If $d\leq 0$ then $\mu$ is updated to $\mu-qx^{-d}\psi$, $\psi$ is unchanged, $\ell_t=\ell$, $t'=s'$ 
and $n+2-\LC_t-\LC_{t'}=n-n'+1$ since $1-\ell-\ell'=-n'$. Hence $f_2=\psi^\wedge\,z^{n+1-\LC-\LC'}=\psi^\wedge\,z^{n-n'}$  and $d+\ell-\ell'=(n+1-2\ell)+\ell-\ell'=n-n'$. Thus 
\begin{eqnarray*}\wedge'(\mu-qx^{-d},\psi)
&=&(\,[\mu-qx^{-d}\psi]^\wedge,\psi^\wedge\, z^{n+2-\LC_t-\LC_{t'}})\\
&=&(\,[\mu(x/z)-q\,(x/z)^{-d} \psi(x/z)]z^{\LC_t},\psi(x/z)z^{\LC_{t'}}\,z^{n+2-\LC_t-\LC_{t'}})\\
&=&(\mu^\wedge -q\, x^{-d}\psi^\wedge z^{d+\LC-\LC'},\psi^\wedge\, z^{n-n'+1})=(f_1-q\,x^{-d}\psi^\wedge\, z^{n-n'},f_2\, z)\\
&=&(f_1-q\,x^{-d}f_2,f_2\, z)
\end{eqnarray*}
which agrees with Theorem \ref{template}.
If $d>0$, $\mu$ is updated to $x^d\mu-q\psi$, $\psi$ is updated to $\mu$, we have $\ell_t=d+\ell$, $\ell_{t'}=\ell$, $n+2-(d+\ell)-\ell=n+2-(n+1-2\ell)-2\ell=1$ so that
\begin{eqnarray*}\wedge'(x^d\mu-q\psi,\mu)
&=&(\,[(x^d\mu-q\psi]^\wedge,\mu^\wedge \, z^{n+2-\ell_t-\ell_{t'}})\\
&=&(\,[(x/z)^d\mu(x/z)-q\, \psi(x/z)]z^{d+\LC},\mu(x/z)z^\LC\, z^{n+2-(d+\LC)-\LC})\\
&=&(x^d\mu^\wedge -q\,\psi^\wedge\, z^{d+\LC-\LC'},\mu^\wedge\, z)\\
&=&(x^df_1-q\,\psi^\wedge z^{n-n'},\mu^\wedge\, z)=(x^df_1-q\,f_2,f_1\, z)
\end{eqnarray*} 
 which again agrees with Theorem \ref{template}.  
 \epr 

 \bc \label{deriv}If $s$ is  non-trivial and $f$ is computed by Algorithm \ref{calPA} then $f_1^\vee$ is an MP of $s$.
\ec
\bpr Since $f$ is viable for $\I_s$, $f_1^\vee\in\Ann_s$ by Corollary \ref{1-1corresp} and $|f_1^\vee|=|f_1|=\LC_s$ by Proposition \ref{LC} i.e. $f_1^\vee$ is as claimed.
\epr

The significance of  Corollary \ref{deriv} is  that we can  compute an MP  of a non-trivial $s$ via Algorithm \ref{calPA}, which has a less technical  derivation, and then dehomogenising the output. 
\section{Appendix}
We show that our annihilator ideal coincides with the ideal of \cite{AD}.
\subsection{An Evaluation Homomorphism}
It will be convenient to rephrase the $\R$ module structure of $\M=\sbbk[x^{-1},z^{-1}]$ in terms of an evaluation (ring) epimorphism $\varepsilon:\LL=\sbbk[x^{-1},z^{-1},x,z]\ra\M$ given by $\varepsilon\,|\,\M$ is the identity and $x,z\mapsto 0$. As $\varepsilon$ is a ring homomomorphism

  \bp  \label{Rmodulemap} For $\varphi\in\R$ and $F\in\M$,  $\varphi\circ F=\varepsilon(\varphi\cdot F)$ makes $\M$ into an $\R$ module. 
\ep

The reader may check that if $p,q,u,v\geq 0$ then $\varepsilon(x^p\,z^q\cdot x^{-u}\,z^{-v})=x^p\,z^q\circ x^{-u}\,z^{-v}$.
Let $I=\ker\varepsilon$.   There is an induced isomorphism $\varepsilon':\LL/I\ra\M$ given by $\varepsilon'(F+I)=\varepsilon(F)$ for $F\in\LL$.  The quotient $\LL/I$ is an $\R$ module via $\varphi\ast (F+I)=\varphi\cdot F+I$ for $\varphi\in\R$ and $F\in\LL$. We include the following fact for completeness.
 \bp \label{Eiso}The induced isomorphism $\varepsilon':(\LL/I,\ast)\ra(\M,\circ)$
 is an isomorphism of $\R$ modules.
 \ep
 \bpr It suffices to check that $\varepsilon'$ is a homomorphism of $\R$ modules on monomials i.e. that
$\varepsilon'(x^p\,x^q\ast (x^u\,z^v+I))=x^px^q\circ\varepsilon'(x^u\,z^v+I)=x^p\,x^q\circ\varepsilon(x^u\,z^v)$. 
The left-hand side is 
$\varepsilon(x^{p+u}\,z^{q+v})$. The right-hand side is
$\varepsilon(x^p\,x^q\cdot\varepsilon(x^u\,z^v))=
\varepsilon(x^{p+u}\,z^{q+v})$ if $u\leq 0$ and $v\leq 0$, and zero otherwise. But the left-hand side 
$\varepsilon(x^{p+u}\,z^{q+v})$ is also zero  if $u> 0$ or  $v> 0$.
\epr

\subsection{The Module of Inverse Power Series}
\label{ADsubsect}
As in \cite[Introduction]{Northcott}, the ring  $\sbbk[[x^{-1},z^{-1}]]$ of {\em inverse power series} can also be  defined as an $\R$ module using  Equation (\ref{module}). 
We will we make $\sbbk[[x^{-1},z^{-1}]]$ into an $\R$ module using an evaluation homomorphism. This enables us to show that our  annihilator ideal coincides with the annihilator ideal of \cite{AD}.

For us, a non-zero {\em Laurent series} in $x^{-1},z^{-1}$  is  $$S=\sum_{-\infty<i\leq k,\,-\infty<j\leq l} S_{i,j}x^iz^j\in\sbbk((x^{-1},z^{-1}))^\times$$
  where  $S_{i,j}\in\sbbk$, $k,l\in\Z$ and $S_{k,l}\neq 0$.  Define  $\zeta:\sbbk((x^{-1},z^{-1}))\ra \sbbk[[x^{-1},z^{-1}]]$ to be the unique evaluation ring homomorphism with $\zeta|\sbbk[[x^{-1},z^{-1}]]$ the identity and $x,z\mapsto 0$. Note that $\zeta\,|\,\M=\varepsilon$, the evaluation homomorphism of the previous subsection.
    
  Using $\zeta$, we can rephrase the $\R$ module structure on $\sbbk[[x^{-1},z^{-1}]]$ as follows. For $\varphi\in\R$, $S\in\sbbk[[x^{-1},z^{-1}]]$, $\varphi\cdot S\in\sbbk((x^{-1},z^{-1}))$ and 
$\varphi\circ S = \zeta(\varphi\cdot S)$. As $\zeta$ is a homomorphism of rings, $(\sbbk[[x^{-1},z^{-1}]],\circ)$ is an $\R$ module and $(\M,\circ)$ is an $\R$ submodule of $(\sbbk[[x^{-1},z^{-1}]],\circ)$. In \cite{Macaulay}, the element $\varphi\circ S\in\sbbk[[x^{-1},z^{-1}]]$ is called the {\em $\varphi$-derivate} of  $S$.

 Let $J=\ker\zeta$. 
  The quotient  $\sbbk((x^{-1},z^{-1}))/J$ is  also an $\R$ module via $\varphi\ast (S+J)=\varphi\cdot S+J$ for $\varphi\in\R$ and $S\in\sbbk((x^{-1},z^{-1}))$.  Elements of the $\R$ module $\sbbk((x^{-1},z^{-1}))/J$ play the role of   inverse power series  in \cite{AD}.
For  $S\in\sbbk((x^{-1},z^{-1}))$, we also have the annihilator ideal of $S+J$ with respect to $\ast$ namely
 $$(0:_{\,\ast} S+J)=\{\varphi\in\R: \varphi\ast (S+J)=0\}$$
which we now relate to $(0:_{\,\circ} F)$ of the previous section, where $F \in\M$.

We have an induced isomomorphism $\zeta':\sbbk((x^{-1},z^{-1}))/J\cong\sbbk[[x^{-1},z^{-1}]]$ of rings given by $\zeta'(S+J)=\zeta(S)$ for $S\in \sbbk[[x^{-1},z^{-1}]]$ and as in Proposition \ref{Eiso}, $\zeta'$ is an isomorphism of $\R$ modules. 

Recall that $\LL=\sbbk[x^{-1},z^{-1},x,z]$ denotes the ring of Laurent polynomials in $x,z$. 
Let  $q:\LL\ra\sbbk((x^{-1},z^{-1}))/J$ be given by $q(L)=L+J$ and  let $\iota:\M\subset\sbbk[[x^{-1},z^{-1}]]$ be the inclusion. It is straightforward to check that $\iota\,\varepsilon=\zeta'q$ by evaluating both sides on a monomial of $\LL$. Hence $I\subseteq J$ and we have an induced map $q':\LL/I\ra\sbbk((x^{-1},z^{-1}))/J$ given by $q'(L+I)=q(L)+J$. It is trivial that $q'$ is a homomorphism of $\R$ modules. We now have
\bp\label{diag} If $I=\ker \varepsilon$, $J=\ker \zeta$ and $\iota:\M\subset\sbbk[[x^{-1},z^{-1}]]$ then

(i) the following diagram of $\R$ modules and induced $\R$ module maps commutes:
$$\xymatrix{
(\LL/I,\ast)\ar[d]_{\varepsilon'}&\ar[r]^{q'}&&(\sbbk((x^{-1},z^{-1}))/J,\ast)\ar[d]^{\zeta'}\\
(\M,\circ)&\ar[r]^{\iota}        &&(\sbbk[[x^{-1},z^{-1}]],\circ)}
$$
where  the vertical maps are $\R$ module isomorphisms

(ii) for $F\in\M$ we have $(0:_{\,\circ} F)=(0:_{\,\ast}\iota F+J)$.
\ep

\bpr  (i) This follows from   $\iota\,\varepsilon=\zeta'\,q$. (ii) Let $\I=(0:_{\,\circ} F)$ and $\J=(0:_{\,\ast}\iota F+J)$. 
From the definitions, $\varphi\in\I$ if and only if $\varphi\cdot F\in\ker \varepsilon=I$ and $\varphi\in\J$ if and only if $\varphi\cdot \iota F+J\in\ker \zeta=J$. We already know that $\ker(\varepsilon)\subseteq \ker(\zeta)$, so that $\I\subseteq\J$. We have  $\varepsilon'(F+I)=F$ and $\zeta'(\iota F+J)=\iota F$. 
If $\varphi\in\J$ then Part (i)  implies that 
$$0=\varphi \ast (\iota F+J)=\varphi \ast(\zeta')^{-1}\iota F=\varphi\ast q'(\varepsilon')^{-1} F=\varphi \ast q'(F+I)=\varphi\cdot F+J$$
i.e. $\varphi\cdot F\in J=\ker \zeta$. But $\varphi\cdot F\in \LL$ and $\zeta|\LL=\varepsilon$ and so $\varphi\cdot F\in\ker\varepsilon$ i.e. $\varphi\circ F=0$. \epr

We have defined  an evaluation map $\zeta:\sbbk((x^{-1},z^{-1}))\ra\sbbk[[x^{-1},z^{-1}]]$ with kernel $J$ and an  action of $\R$ on $\sbbk((x^{-1},z^{-1}))/J$ namely $\varphi\ast (S+J)=\varphi\cdot S+J$ for $S\in\sbbk((x^{-1},z^{-1}))$. Let $\iota:\M\subset\sbbk[[x^{-1},z^{-1}]]$ be the inclusion, $s$ be a finite sequence and $F^{(s)}$ be its inverse form. In \cite{AD}, the {\em generating form} of $s$  is $\iota  F^{(s)}$   and the annihilator ideal of $s$ is $(0:_{\,\ast} \iota F^{(s)}+J)=\{\varphi\in\R:\varphi \ast (\iota F^{(s)}+J)=0\}$.
\bc For any sequence $s$, $\I_s=(0:_{\,\ast} \iota F^{(s)}+J)$. 
\ec
\bpr  
Taking $F= F^{(s)}$ in Proposition \ref{diag} gives $\I_s=(0:_{\,\circ} F^{(s)})=(0:_{\,\ast}\iota F^{(s)}+J)$.
\epr

\end{document}